\documentstyle[12pt,twoside,epsfig,rotating,amstex,cite,times]{article}
\newcommand{\F}{$ F_{2}(x,Q^2)\,$} 
\newcommand{\FL}{$ F_{L}(x,Q^2)\,$}
\newcommand{\V}{$ V(x,Q^2)\,$}
\newcommand{\Vc}{$ V\,$}
\newcommand{\A}{$ A(x,Q^2)\,$}
\newcommand{\Ac}{$ A\,$}              
\newcommand{\Fc}{$ F_{2}~$}

\newcommand{\gv}{GeV$^2\,$}
\newcommand{\as}{$\alpha_s\,$}
\newcommand{\amz}{$\alpha_s(M_Z^2)\,$} 
\newcommand{\bs}{\overline{s}}
\newcommand{\bu}{\overline{u}}
\newcommand{\bd}{\overline{d}}
\newcommand{\bq}{\overline{q}}    
\newcommand{\FLc}{$ F_{L}\,$} 
\newcommand{\xg}{$xg(x,Q^2)\,$}
\newcommand{\xgc}{$xg\,$}

\newcommand{\pdsi}{$(\partial \sigma_r / \partial \ln y)_{Q^2}\,$}
\newcommand{\pdff}{$(\partial F_{2} / \partial \ln  Q^{2})_x\,$ }
\newlength{\dinwidth}                                                          
\newlength{\dinmargin}                                                         
\begin{document}              

\noindent                                                                 
\begin{titlepage}                                                    

\begin{flushleft}

DESY-00-181  \hfill  ISSN 0418-9833 \\

December 2000

\end{flushleft}

\vspace{2.2cm}

\begin{center}

\begin{Large}
{\bf
 Deep-Inelastic Inclusive {\boldmath $ep$}
 Scattering \\
  at Low  {\boldmath $x$} and a Determination of  {\boldmath $\alpha_s$} \\}
\vspace*{2.cm}                                                                 
H1 Collaboration
\\                                                          
\end{Large}  
\vspace*{2.cm}                                                                 
{\bf 
Abstract:}                                                                
\begin{quotation}                                                              
  \noindent
  A precise measurement of the inclusive deep-inelastic $e^+ p$
  scattering cross section is reported in the kinematic range $1.5 \ 
  \le Q^2 \ \le \ 150$\,\gv and $ 3 \cdot 10^{-5} \le x \le 0.2$.  The
  data were recorded with the H1 detector at HERA in 1996 and 1997,
  and correspond to an integrated luminosity of $20$\,pb$^{-1}$.  The
  double differential cross section, from which the proton structure
  function \F and the longitudinal structure function \FL are
  extracted, is measured with typically 1\% statistical and 3\%
  systematic uncertainties.  The measured derivative $(\partial
  F_2(x,Q^2) / \partial \ln Q^2)_x $ is observed to rise continuously
  towards small $x$ for fixed $Q^2$.  The cross section data are
  combined with published H1 measurements at high $Q^2$ for a
  next-to-leading order DGLAP QCD analysis.  The H1 data determine the
  gluon momentum distribution in the range $3 \cdot 10^{-4} \le x \le
  0.1$ to within an experimental accuracy of about 3\% for $Q^2
  =20$\,\gv.  A fit of the H1 measurements and the $\mu p $ data of
  the BCDMS collaboration allows the strong coupling constant $\alpha
  _s$ and the gluon distribution to be simultaneously determined.  A
  value of $\alpha _s (M_Z^2) =0.1150 \pm 0.0017 (exp)
  ^{+0.0009}_{-0.0005}~(model)$ is obtained in NLO, with an
  additional theoretical uncertainty of about $\pm 0.005$, mainly due
  to the uncertainty of the renormalisation scale.
\end{quotation} 
\vspace{0.2cm}
To be submitted to Eur. Phys. J. C
\end{center}
\cleardoublepage                     
\end{titlepage}
%
%
\noindent
C.~Adloff$^{33}$,              
V.~Andreev$^{24}$,             
B.~Andrieu$^{27}$,             
T.~Anthonis$^{4}$,             
V.~Arkadov$^{35}$,             
A.~Astvatsatourov$^{35}$,      
I.~Ayyaz$^{28}$,               
A.~Babaev$^{23}$,              
J.~B\"ahr$^{35}$,              
P.~Baranov$^{24}$
% ...

\noindent
{\it
 $ ^{1}$ I. Physikalisches Institut der RWTH, Aachen, Germany$^{ a}$ \\
% ...
$ ^{4}$ Inter-University Institute for High Energies ULB-VUB, Brussels;
          Universitaire Instelling Antwerpen, Wilrijk; Belgium$^{ c}$ \\
% ...
$ ^{23}$ Institute for Theoretical and Experimental Physics,
          Moscow, Russia \\
 $ ^{24}$ Lebedev Physical Institute, Moscow, Russia$^{ e,h}$ \\
% ...
$ ^{27}$ LPNHE, Ecole Polytechnique, IN2P3-CNRS, Palaiseau, France \\
 $ ^{28}$ LPNHE, Universit\'{e}s Paris VI and VII, IN2P3-CNRS,
          Paris, France \\
% ...
$ ^{33}$ Fachbereich Physik, Bergische Universit\"at Gesamthochschule
          Wuppertal, Wuppertal, Germany$^{ a}$ \\
% ...
$ ^{35}$ DESY, Zeuthen, Germany$^{ a}$ \\
% ...
$ ^a$ Supported by the Bundesministerium f\"ur Bildung, Wissenschaft,
      Forschung und Technologie, FRG,
      under contract numbers 7AC17P, 7AC47P, 7DO55P, 7HH17I, 7HH27P,
      7HD17P, 7HD27P, 7KI17I, 6MP17I and 7WT87P \\
% ...
$ ^c$ Supported by FNRS-NFWO, IISN-IIKW \\
% ...
$ ^e$ Supported by the Deutsche Forschungsgemeinschaft \\
% ...
$ ^h$ Supported by Russian Foundation for Basic Research
      grant no. 96-02-00019 \\
% ...
}
\newpage
%
%
\section{Introduction}    
%
%
Deep-inelastic lepton-nucleon scattering (DIS) has been pivotal in the
development of the understanding of strong interaction
dynamics. Measurements of the inclusive DIS cross section have been
essential for testing Quantum Chromodynamics (QCD)~\cite{tamps}.
Previous fixed target DIS experiments have
observed scaling violations, i.e. the variation of the structure
functions with $Q^2$, the squared four-momentum transfer between
lepton and nucleon, for fixed values of Bjorken-$x$, which are well
described by perturbative QCD.  The $Q^2$
evolution of the proton structure function \F is related to the gluon
momentum distribution in the proton, \xg, and to the strong
interaction coupling constant, \as.  These can be determined with
precision deep-inelastic scattering cross section data measured over a
wide range of Bjorken-$x$ and $Q^2$.

The first measurements of \Fc at low $x \sim 10^{-3}$ and $Q^2 \sim
20$~\gv at HERA revealed a steep rise of \F towards low $x$ for fixed
$Q^2$~\cite{h1f292, zeus92}.  Strong scaling violations are observed
at low $x$ and are attributed to a high gluon density in the proton.
The validity of the DGLAP evolution equations~\cite{dglap}, which
neglect higher-order terms~\cite{bfkl, ccfm} proportional to
$\alpha_s \cdot \ln(1/x)$, is questionable in the low $x$ range and
therefore has to be tested against data.  At extremely low $x$,
non-linear gluon interaction effects have been considered in order to
damp the rise of the cross section in accordance with unitarity
requirements~\cite{glr}. The study of quark-gluon interaction dynamics
at high parton densities therefore continues to be a challenging
subject.  Knowledge of the parton densities at low $x$ is also
necessary for interpreting measurements at hadron colliders and of cosmic
neutrino interactions.

This paper presents new cross section measurements for the neutral
current process $e^+ p \rightarrow e^+ X$ in the kinematic region $1.5
\leq Q^2 \leq 150{~\rm GeV^2}$ and $3 \cdot 10^{-5}~\leq~x~\leq~0.2$,
obtained from data taken in the years 1996 and 1997 with positrons of
energy $E_e=27.6$\,GeV and protons of energy $E_p=820$\,GeV,
corresponding to a centre of mass energy $\sqrt{s} = 300.9$\ GeV.
Cross section measurements at low $x$ and medium $Q^2$, based on the
1994 HERA data, were previously published by the H1 collaboration
\cite{sf94} and by the ZEUS collaboration \cite{zeusf2}.  The present
measurement uses upgraded detectors to measure and identify the
scattered positron, including new precision tracking for low $Q^2$
scattering. It also benefits from increased luminosity from HERA which
enables an accuracy of typically 3\% to be reached for the DIS cross
section.  Thus it considerably improves the former structure function
measurements of the H1 collaboration~\cite{sf94,sfsvx,sfl} at $Q^2
\leq 150$\,GeV$^2$.  The kinematic range is extended to larger $x$,
yielding an overlap of H1 data with measurements from fixed target
muon-proton scattering experiments for the first time.  The paper
includes a measurement of the derivative \pdff, which serves as a
sensitive test of the dynamics of strong interactions.

The longitudinal structure function \FL is obtained with improved
precision in an extended range of inelasticity $y$ and $Q^2$, as
compared to its first determination at low $x$~\cite{sfl}.  A new
method is introduced to extend the extraction of \FLc to $Q^2$ values
below 10\,\gv which uses the derivative \pdsi of the reduced cross
section $\sigma_r$.

A next-to-leading order (NLO) DGLAP QCD analysis is performed using
inclusive lepton-proton scattering data by introducing a new flavour
decomposition of the structure function \Fc. Hence it is independent
of nuclear binding effects in the deuteron or heavier nuclei. The QCD
analysis of the present low $x$ data and of the recently published
high $Q^2$ H1 data~\cite{hiq} determines the gluon distribution \xg
at low $x$.  The combination of the low $x$ H1 data with large $x$
data from the BCDMS experiment~\cite{BCDMS} enables an accurate,
simultaneous determination of both \xg and \amz.  The present
analyses use all of the available information regarding the
experimental uncertainties of the data sets considered and explore the
QCD model and fit parameter variations in a systematic way.

The paper is organised as follows. Section~2 defines the inclusive
cross section and the methods used to reconstruct the event kinematics.
The detector, the event selection and the
simulation are described in section~3. Section~4 presents the
alignment and calibration methods, and summarises the cross section
measurement.  Section~5 presents the measurement of \pdsi and the
determination of the longitudinal structure function \FLc.  The results
for the proton structure function \Fc and its derivative \pdff 
are given in section~6.  The QCD interpretation of the data is
discussed in section~7, which refers to an appendix presenting details
of the analysis. The paper is summarised in section~8.
%
%
\section{Cross Section and Kinematic Reconstruction \label{kin}}
The inclusive DIS cross section of the
 measured reaction $e^+p \rightarrow e^+ X$
 depends on two independent kinematic
variables, chosen to be $x$ and $Q^2$, and on the centre of mass
energy squared~$s$, with the inelasticity variable
 $y=Q^2/sx$.  In the one-photon exchange
approximation the neutral current double differential cross section,
$d^2\sigma /dxdQ^2$, is given by
\begin{equation}
 \frac{d^2\sigma}{dxdQ^2}  =  \frac{2\pi \alpha^2 Y_+}{Q^4 x} \cdot
\sigma_r 
\end{equation}
where the reduced cross section is defined as
\begin{equation}
 \sigma_r  \equiv       F_2(x,Q^2) - \frac{y^2}{Y_+} \cdot F_L(x,Q^2)
 \label{sig}
\end{equation}  
and $Y_+ = 1+ (1-y)^2$.  Due to the positivity of the cross sections
for longitudinally and transversely polarised photons scattering off
protons, the two proton structure functions \Fc and \FLc obey the
relation $0 \leq F_{L} \leq F_{2}$. Thus the contribution of the
longitudinal structure function \FLc to the cross section can be
sizeable only at large values of the inelasticity $y$, and in most of
the kinematic range the relation $\sigma_r \approx F_2$ holds to a
very good approximation.

The HERA collider experiments allow DIS kinematics to be reconstructed
using the scattered positron, the hadronic final state, or a
combination of the two. This is important for maximum coverage of the
kinematic range and the  control of systematic uncertainties.

In the ``electron method'' the event kinematics are determined
using the measured energy of the scattered positron $E_e'$ and its
polar angle $\theta_e$ according to the relations
\begin{equation}
y_e=1-\frac{E_e'}{E_e}~\sin^2(\theta_e/2),
 \hspace*{2cm}
Q^2_e= \frac{E^{'2}_e \sin^2{\theta_e}}{ 1-y_{e}}.
\label{qy}
\end{equation}
The coordinate system of H1 is defined such that the positive $z$ axis
is in the direction of the incident proton beam.  Polar angles
$\theta$ are defined with respect to the proton beam direction.  While
the electron method is accurate at large $y$, corresponding to low
$E_e'$, the resolution rapidly degrades with $1/y$ as $E_e'$
approaches the positron beam energy $E_e$.

The inelasticity $y$ can also be determined as \cite{jb}
\begin{equation}
       y_h=\frac{\Sigma_{i}(E_i-p_{z,i})}{2E_e} \equiv \frac{\Sigma}{2E_e},
       \label{yh}
\end{equation}
where $E_i$ and $p_{z,i}$ are the energy and longitudinal momentum
component of a particle $i$ in the hadronic final state, the masses
being neglected.  In this analysis the kinematics are also reconstructed
 with the ``$\Sigma$ method'' using the variables~\cite{babe}
\begin{equation} 
  y_{\Sigma} = 
   \frac{\Sigma}{\Sigma+E_e'(1-\cos \theta_e)},
   \hspace*{2cm} 
  Q^2_{\Sigma} = \frac{E^{'2}_e \sin^2{\theta_e}}{ 1-y_{\Sigma}}.
\label{ys}
  \end{equation}
For all reconstruction methods, Bjorken-$x$ is calculated
as $x=Q^2/sy$.  Due to energy-momentum conservation the variable
\begin{equation} 
 E-p_z = \Sigma+E_e'(1-\cos \theta_e)
\label{epz}
  \end{equation}
is approximately equal to $2 E_e$. 
The hadronic variables $y_h$ and $y_{\Sigma}$ are related according to
  \begin{equation}
 y_{\Sigma} =\frac{y_h}{ 1+y_h-y_e} 
   \label{yyy}
\end{equation}
and can be well measured down to low $y \simeq 0.004$.  The variable
$y_{\Sigma}$ is less sensitive to initial state radiation than $y_h$
since the initial energy $E_e$ in the denominator in equation~\ref{yh}
can be calculated using the total energy reconstructed in the detector
which leads to equation~\ref{ys}.

The hadronic scattering angle is defined as
\begin{equation}
    \tan \frac{\theta_{h}}{2} = \frac{\Sigma}{P_{t,h}},
                       \label{thetah}
\end{equation}
where $P_{t,h}$ is the total transverse momentum of the hadronic final
state particles. In the naive quark parton model, $\theta_{h}$ defines
the direction of the struck quark related to $\theta_e$ as
        \begin{equation}
           \tan \frac{\theta_{h}}{2} =
                \frac{y}{1-y} \cdot
           \tan \frac{\theta_{e}}{2}.
                        \label{teh}
        \end{equation}     
For $y > 0.5$ the positron scattering angle $\theta_{e}$ is
smaller than $\theta_{h}$.  This relation, together with the
definition of $y_e$ (equation~\ref{qy}), determines the scattered
positron energy from $\theta_{e}$ and $\theta_{h}$ in the
``double angle method''~\cite{dang}. 
%
%
\section{Experimental Procedure}
\subsection{H1 Detector}
The H1 detector~\cite{h1detec} combines tracking in a
solenoidal magnetic field of 1.15~T
with nearly hermetic calorimetry to investigate high-energy
$ep$ interactions at HERA.  The low $Q^2$ cross section measurement
relies mainly on the central and backward tracking systems, the backward
calorimeter (SPACAL) and  the Liquid Argon (LAr) calorimeter. These
components are briefly described below.
 
The energy of the positron, when scattered into the backward region of
the H1 detector ($153^\circ < \theta_e < 177^\circ$), is measured in
the SPACAL, a lead-fibre calorimeter~\cite{awe, Spacal}.  The SPACAL
has an electromagnetic section with an energy resolution of
$7\%/\sqrt{E/GeV} \oplus 1\%$, which together with a hadronic section
represents a total of two interaction lengths. The SPACAL time
resolution of less than 1~ns allows proton beam induced background to
be vetoed.  The calorimeter has a high transverse granularity which
provides a determination of the transverse coordinates of
electromagnetic clusters with an accuracy of a few millimeters and
positron identification capability.  Identification of the scattered
positron is improved and the polar angle measured with a backward
drift chamber (BDC), situated in front of the SPACAL, and a new
backward silicon strip detector (BST)~\cite{BST}. The BST consists of
four detector planes, arranged perpendicularly to the beam axis which
are equipped with 16 wedge shaped, double metal silicon strip
detectors.  The BST measures the polar angle of tracks with an
internal resolution of about 0.2~mrad from radial coordinates between
5.9~cm and 12.0~cm.

The hadronic final state is reconstructed with the Liquid Argon
calorimeter~\cite{LAr}, the tracking detectors and the SPACAL.  The
LAr calorimeter is built of eight wheels of modules with an octant
structure.  The total depth of the calorimeter varies between 4 and
8.5 interaction lengths depending on the polar angle. In test beam
measurements pion induced hadronic energies were reconstructed with a
resolution of about $50\%/\sqrt{E/GeV} \oplus 2.0\%$ after software
energy weighting~\cite{LArw}.

The interaction vertex is determined with the central tracking
detector consisting of two concentric jet drift chambers (CJC) and two
concentric $z$ drift chambers (CIZ and COZ).  The vertex determination
is complemented by the inner proportional chamber CIP, for $167^\circ
< \theta_e < 171^\circ$, and by the silicon tracker BST, for
$171^\circ < \theta_e < 176.5^\circ$.

The luminosity is determined using the small-angle bremsstrahlung
process $ep~\rightarrow~ep~\gamma$. The final state photon and the
positron, scattered at very low $Q^2$, can be detected in calorimeters
(``electron and photon taggers'') which are situated close to the beam
pipe at distances of 33~m and 103~m from the interaction point in the
positron beam direction. The luminosity is measured with a precision
of 1.5\% using the method outlined in~\cite{lumi}.
\subsection{Data Samples and Interaction Triggers \label{datrig}}
The analysis comprises two different DIS event samples:
\begin{itemize}
\item{Sample $A$ - data taken in the years 1996 and 1997 with
    luminosities of 4.5~pb$^{-1}$ and 13.4~pb$^{-1}$, respectively.
    These two data sets are combined to provide the cross section
    measurement for $Q^2$ values from 15\,\gv to 150\,GeV$^2$ and for
    $Q^2=12$\,GeV$^2$ at $y > 0.17$.}
\item{Sample $B$ - data taken in the autumn of 1997 during a two week period
    dedicated to the lower $Q^2$ region. The data from this special run
    with a luminosity of 1.8~pb$^{-1}$ are used in the $Q^2$ range from
    1.5\,GeV$^2$ to 8.5\,GeV$^2$ and for $Q^2=$12\,GeV$^2$ at low $y <
    0.17$.}
\end{itemize}
DIS events at low $Q^2$ are characterised by a positron scattered into
the backward part of the H1 apparatus.  The event trigger for
 sample $A$ requires  local energy sums in the SPACAL calorimeter
to be above an energy threshold of 6\,GeV. This threshold is lowered
to 5\,GeV in  sample $B$.

Both data samples are contaminated by photoproduction events in which
the scattered positron escapes through the beam pipe and a particle in
the hadronic final state mimics the signature of a scattered positron.
For a fraction of these background events the scattered positron is
detected in the electron tagging calorimeter.  This background is 
significant only at low energies $E_e'$, corresponding to values of $y >
0.6$. 

The region of high $y > 0.75$ is accessed with a dedicated trigger which
requires a compact energy deposition (cluster) of
 more than 2\,GeV of energy  in the SPACAL,
 and a vertex signature in the proportional
chamber system.  The data accumulated with this trigger correspond to
a luminosity of 2.8~pb$^{-1}$ in 1996 and 3.4~pb$^{-1}$ in 1997. 

The SPACAL energy triggers are monitored with independent track
triggers and found to be fully efficient for energies of about 1\,GeV
above threshold. The high $y$ trigger efficiency is determined to be 97\%
using independent calorimeter triggers.  The online data
reconstruction leads to a maximum loss of 0.5\% of  DIS events.
This loss is estimated from monitor data and corrected for.
\subsection{Event Selection \label{evsel}}
The scattered positron is identified with the cluster of maximum
transverse momentum $p_{t}$ in the SPACAL calorimeter, for which
requirements on the cluster shape are satisfied. Electromagnetic
energy deposition leads to clusters of smaller transverse extension
than hadronic energy deposition.  The transverse energy distribution
of positron showers is determined experimentally using QED Compton
events, and from radiative DIS events in which a photon is radiated
from the incoming positron and detected in the photon tagger. The
positron cluster radius~\cite{Spacal,sasha} can thus be measured at
all energies considered in the analysis, and a cluster radius cut of
4~cm is chosen. This cut removes a sizeable fraction of the
photoproduction background while retaining more than 99\% of the DIS
signal.

A positron candidate cluster is required to be associated with a track
segment in the BDC. The efficiency of the BDC is measured to be 98\%
on average with small radius dependent variations. A signal is also
required in either the CIP or the BST where geometrically available.
The efficiency of the CIP is about 98\%. The track efficiency of the
BST in the 1997 special run period is found to be about 93\%.
Efficiencies at low energies are determined using QED Compton
scattering events and radiative DIS events.

Reconstruction of the interaction vertex is necessary to determine the
event kinematics and to suppress beam background events. In the
intermediate $y$ region, hadrons measured in the central tracking
chambers allow vertex reconstruction with an efficiency exceeding
98\%.  However, at very low $y$ and also at very high $y$, where no
hadron may be measured in the central tracking chambers, the vertex
can be defined by the scattered positron if it falls within the
acceptance of the CIP or the BST.
\begin{table}[h] \centering 
\begin{tabular}{|l|c|}
\hline 
                SPACAL energy  &  $ >$ 6.9 GeV              \\
radius of cluster              &  $< 4$ cm                 \\
fraction of energy in the hadronic section  &  $<$ 15\% of $E_{e}'$   \\
cluster-BDC track distance     &  $ < 1.5$ cm      \\
cluster-BST track distance     &  $ < 1.0$ cm       \\
$z$ vertex position            &  $|z| < 30$ cm            \\
$E-p_{z}$                &  $> 35$ GeV              \\
\hline
\end{tabular}
\caption{\label{tabcut} \sl Basic DIS event selection criteria.
The radius of the cluster~\cite{Spacal,sasha}
defines the lateral shape of the energy deposition.
The high $y$ region, corresponding to energies between $E_e' = 3$\,GeV
and 6.9\,GeV,  is accessed with
 additional cuts, as discussed in Section~\ref{sechihy}.}
\end{table}  

Longitudinal momentum conservation in neutral current DIS events gives
the constraint that $E-p_{z}$, summed over the final state particles
is about equal to $2 E_e$ (equation~\ref{epz}). In radiative events a
photon may carry a significant fraction of the $E-p_{z}$ sum. Such
events are thus suppressed by requiring $E-p_{z} > 35$\,GeV.

The criteria applied to select DIS events are summarised in table~\ref{tabcut}.
\subsection{Simulation \label{simp}}
For the calculation of the detector acceptance and efficiency control
about $ 10^7$ inelastic events are simulated. Deep-inelastic events
are generated using the DJANGO~\cite{django} event generator.  This
program is based on HERACLES~\cite{heracles} for the electroweak
interaction and on the Lund Monte Carlo generator program
ARIADNE~\cite{ari}, which includes the generation of diffractive
events.  This generator, when tuned to HERA data~\cite{herafi2},
presently gives the most reliable description of the final state
properties~\cite{herafin}. To describe higher order QCD radiation
processes the ARIADNE program uses the Colour Dipole Model
(CDM)~\cite{CDM}.  For hadron fragmentation the JETSET program is
used~\cite{jetset}. Comparisons  are done using the
generator LEPTO~\cite{lepto}. QED Compton events are generated using the
program COMPTON~\cite{comp}. DIS events are also generated with the
HERWIG event generator~\cite{herwig}, which includes resonant final
state production.  This is important for the description of rejected events at
very low $y$~\cite{vova}.  Photoproduction background is generated
with the PHOJET \cite{phoj} program using the parameterisation of
CKMT~\cite{ckmt} to determine the virtual photon-proton interaction
cross section. The normalisation of the PHOJET event sample is
adjusted to the data measured with the electron tagging calorimeter.
It is found to agree within 20\% with the calculation of the cross
section using the Weizs\"acker-Williams approximation. Using the
leading logarithmic approximation~\cite{job} the  effect of
photon radiation is estimated to be negligible.

The detector response is simulated in detail with a program based on
GEANT3 \cite{geant}.  The Monte Carlo events are subject to the same
reconstruction and analysis chain as the real data. In the comparisons
shown here, the simulated distributions are normalised to the measured
luminosity. In the event generation the DIS cross section is
calculated with the parton distributions of~\cite{grv94} and with the
longitudinal structure function $F_L=0$. A QCD fit to all
the data is used to reweight the simulated cross section.
%
%
\section{Measurement of the Cross Section}
This section presents briefly the methods and results of the
measurement of the deep-inelastic scattering cross section. Further
details of the analysis are described in~\cite{sasha, vova, rainer,
doris}.
\subsection{Detector Alignment}
The coordinate system of the H1 detector is defined by the central
tracking chambers which determine the spatial coordinates of the
interaction vertex. The variation of the vertex position in the $x,
y$ plane along the beam direction is used to determine the inclination
of the beam axis with respect to the $z$ axis.

The central tracking chambers are aligned with respect to each other
using cosmic muon tracks. The alignment of the BDC and of the SPACAL
with respect to the central tracker is done by studying the difference
of the polar angles measured by these detectors as a function of the
azimuthal angle. This results in 1 to 2~mm adjustments of the nominal
detector positions.  QED Compton events, which have the signature of
back-to-back positron and photon clusters, provide a cross check for
the alignment~\cite{sasha} of the SPACAL in the transverse plane to an
accuracy of 0.2~mm.  After internal adjustment of the strip detector
planes, the spatial position of the BST is determined using the event
vertex $z$ coordinate measured with the central tracker.

In the BST angular acceptance range the polar angle is measured both
by the BDC using the interaction vertex, and by the BST track segment.
This allows the alignment procedure to be checked to within an
accuracy of $\Delta \theta$ of 0.1~mrad~\cite{vova}.  From the
residual dependence of $\Delta \theta$ on the azimuthal angle and from
the uncertainties of the alignment procedure, a measurement error of
0.3~mrad is estimated for the angle of the scattered positron.
\subsection{Calibration of the Energy Measurements}
\begin{figure}[tbp]
 \epsfig{file=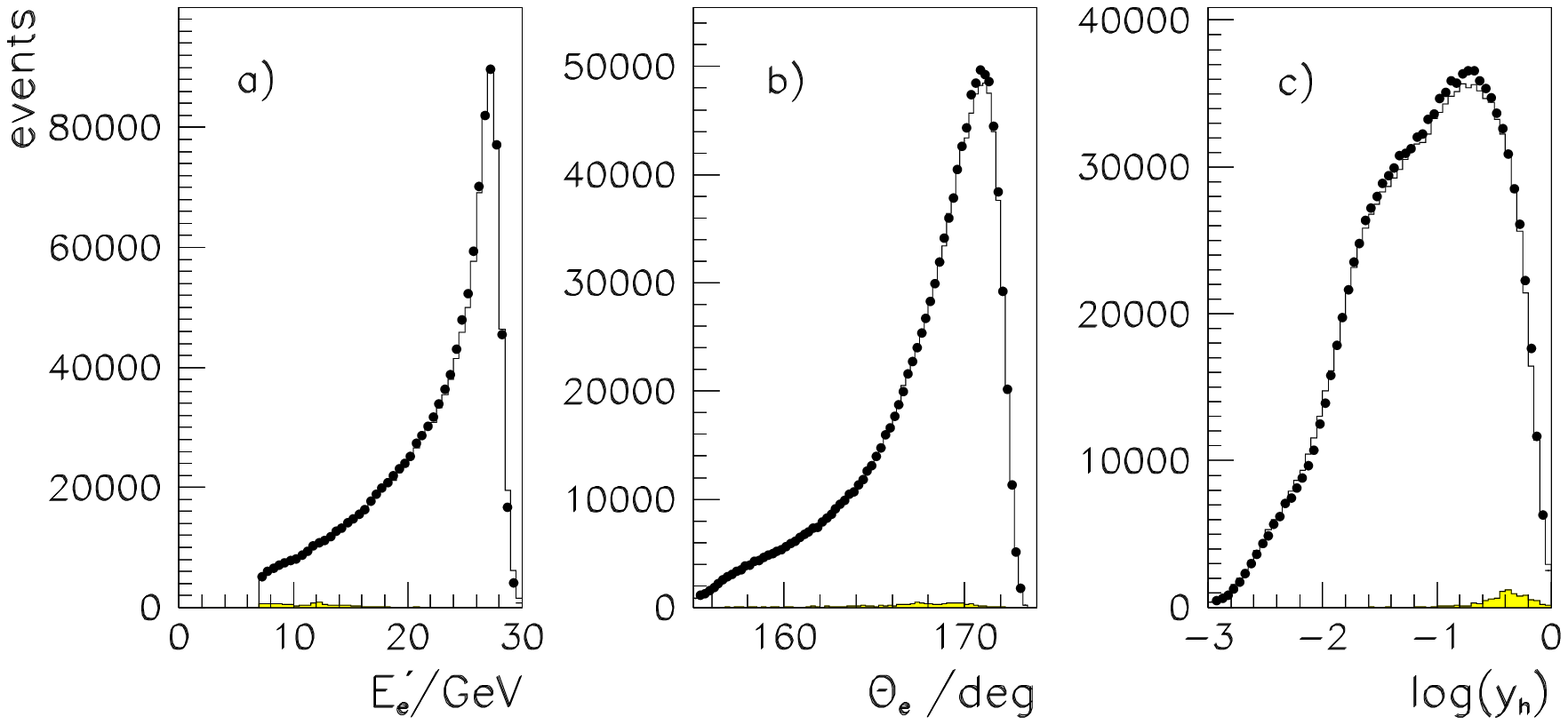,width=15.5cm}
  \caption{ \sl  Distributions 
    of a) the energy, b) the polar angle of the scattered positron, and c)
    $y_h$ for the data sample $A$ taken in 1996/97 (solid
    points).  The  histograms show the simulation of
    DIS and the small  photoproduction background (shaded),
   normalised to the luminosity
    of the data.} \protect\label{contm}
\end{figure}

\begin{figure}[tbp]
%
%
  \epsfig{file=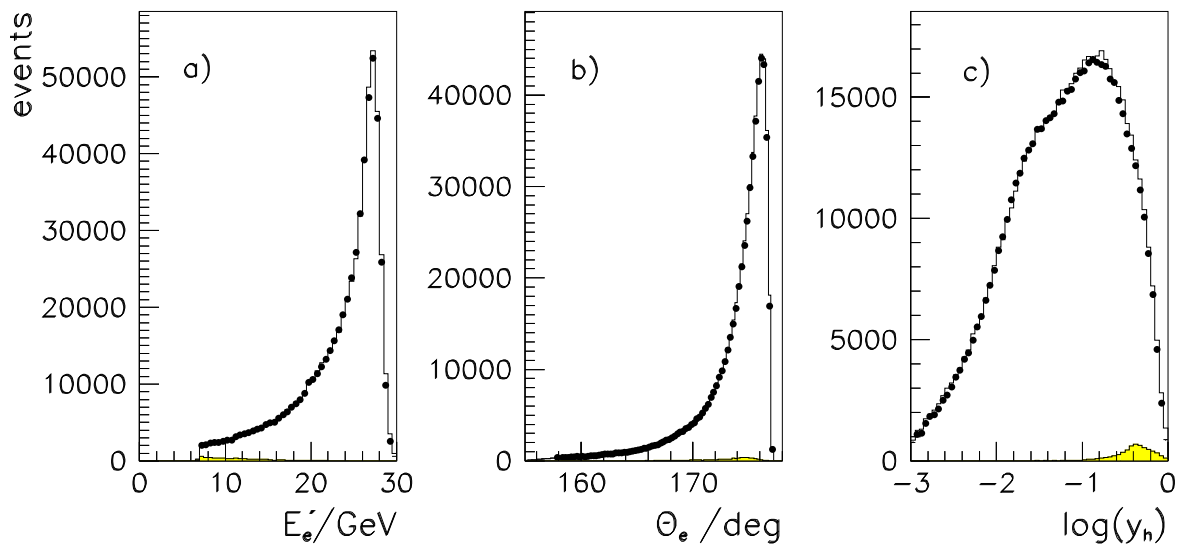,width=15.5cm}
  \caption{ \sl Distributions
    of a) the energy, b) the polar angle of the scattered positron, and c)
    $y_h$ for the low $Q^2$ data sample $B$ taken in 1997. 
    The  histograms represent the simulation of
    DIS and the small  photoproduction background (shaded),
    normalised to the luminosity of the data.
   } \protect\label{contsp}
\end{figure}

The energy of the scattered positron is measured in the 
SPACAL, which has a transverse cell size of $4 \times
4$~cm$^2$ and a Moli\`ere radius of 2.5\,cm.  In a first step the
responses of the SPACAL cells are equalised using cosmic muons.
The energy scale of each cell is determined with DIS events using the
double angle method, which allows the energy of the scattered positron
to be expressed as a function of $\theta_e$ and $\theta_h$. This
method is applied to the data of the various run periods, and also to
the simulated events. Agreement of the energy scales is found at the
level of 0.2\%.  The calibration procedure leads to a systematic error
of the $E_e'$ scale of 0.3\% for most of the SPACAL area and energies
$E_e'$ above 20~GeV.  The energy calibration at lower energies is
performed using QED Compton events. This leads to an estimated maximum
energy scale uncertainty of 2.7\% at 3\,GeV. This uncertainty is
observed to approach linearly the 0.3\% level at maximum energies
$E_e' \simeq E_e$.  The SPACAL response at the lowest energies is
cross checked by studying the $\pi^0$ mass reconstructed from pairs of
photons in the energy range of $0.8-4$\,GeV~\cite{sasha,pizero}.
Material in front of the calorimeter leads to showering and energy
losses, which are corrected for using the backward drift chamber as a
preshower detector~\cite{rainer}.
Figures~\ref{contm}a,b and \ref{contsp}a,b show the energy and polar
angle distributions for data samples $A$ and $B$. These are described
by the simulation of DIS and photoproduction events.

The cross section measurement at low $y$ relies on the measurement of
$y$ using the hadronic final state (equation~\ref{yh}).  The
determination of $y$ is optimised by combining calorimeter energy
deposits with low momentum tracks. The sum over energy clusters in the
calorimeters can be strongly affected by electronic noise, in
particular for low $y < 0.03$.  Thus an additional noise
suppression is performed which excludes isolated depositions of energy
less than 400~MeV (800~MeV) in the central (forward) region of the LAr
calorimeter from the analysis of both data and simulated events. This
leads to a small signal loss but improves the $y$ resolution at low
$y$. The uncertainty of this subtraction procedure is estimated to
correspond to a quarter of the suppressed noise contribution to the
reconstructed hadron energy.

The calibration of the hadronic energy measurement uses the $p_t$
balance between the scattered positron and the hadronic final state.
The energy scales for the electromagnetic and hadronic sections of
wheels and octants~\cite{vova} of the LAr calorimeter result from a
Lagrangian multiplier technique which simultaneously determines all
128 calibration constants.  A systematic uncertainty of 2\% on the
hadronic energy scale in the LAr calorimeter is determined, based on a
binwise comparison of different calibration methods.  The results of
the energy calibration procedure are consistent with those recently
presented~\cite{hiq}.

Figures~\ref{contm}c and \ref{contsp}c show the experimental $y_h$
distributions over three orders of magnitude in $y$. In the
simulation, the $y_h$ reconstruction is found to be accurate to within
a few~\% over a wide range of $y_h$ and down to the small $Q^2$
region. The worse $y$ resolution at low $y$ is accounted for with
an increased bin size, allowing the $y$ range to be extended down to
$y \simeq 0.004$.
 
The response of the SPACAL to hadronic energy flow is calibrated using
longitudinal momentum conservation in the DIS events~\cite{sasha} to
within an uncertainty of 5\%. This scale uncertainty affects the final
cross section data at large $y$ through the $E-p_z$ momentum balance
requirement.
\subsection{Measurement at Large {\boldmath $y$}} \label{sechihy}
For the measurement of the longitudinal structure function it is
essential to reach the maximum possible values of $y$ (see
equation~\ref{sig}). This requires an efficient rejection of
photoproduction background events in which low energy deposits in the
SPACAL can mimic the signature of a deep-inelastically scattered
positron.

\begin{figure}[tbp]
  \epsfig{file=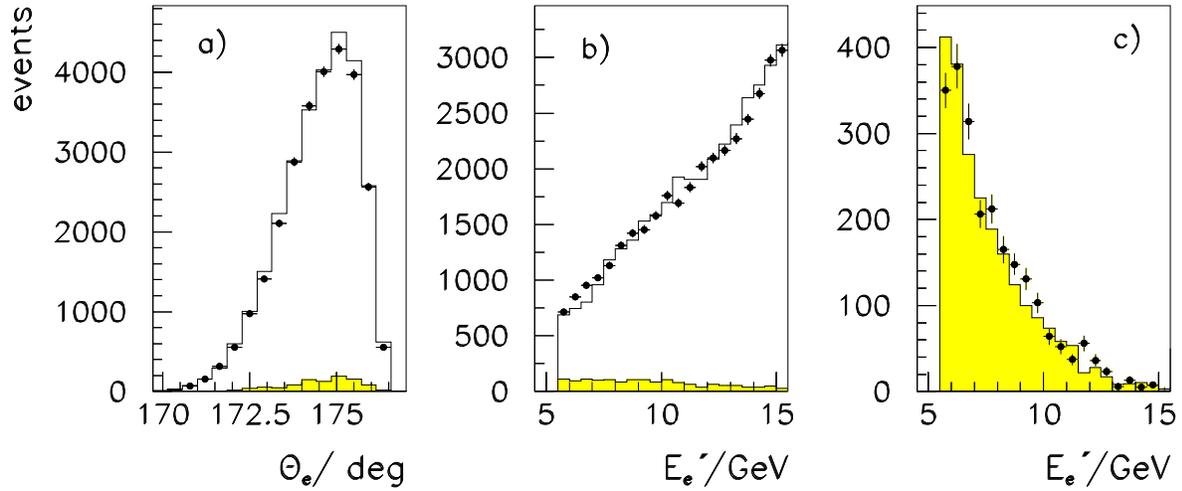,width=15.5cm}
  \caption{ \sl
    Distributions illustrating the cross-section measurement at high
    $y$ ($0.46 < y < 0.82 $)
     and low $Q^2$ ($2 < Q^2 < 5$~GeV$^2$) for
    events in the BST acceptance range. 
     DIS event distributions of a) the polar angle a) and
     b) the SPACAL energy  of the scattered positron candidate.
      c) SPACAL energy distribution for tagged photoproduction events
  fulfilling the
  DIS event selection criteria, apart from the $E-p_z$ requirement.
   Solid points: H1 data; shaded histograms: simulation of
photoproduction
    events; open histograms: added distributions of
    simulated  DIS and  photoproduction events.}
 \protect\label{contbst}
\end{figure}

At $Q^2$ below 10~\gv, the range $y \leq 0.75$ is accessed by
requiring a track signal in the BST. This requirement removes a
sizeable fraction of the background where a cluster in the SPACAL is
due to photons from $\pi^0 \rightarrow \gamma \gamma$ decays.  The
remaining background is due to photon conversion and showering in the
passive material, possible overlap of $\pi^0$ decays with charged
tracks, and misidentified charged pions. This background is subtracted
bin by bin using the PHOJET event simulation. Figures~\ref{contbst}a
and ~\ref{contbst}b illustrate the range of polar angle and energy for
high $y$ events with a track in the BST. The photoproduction
background can be estimated experimentally using a data sample of
events which have a positron detected in the electron tagger.
Figure~\ref{contbst}c shows the energy spectrum of SPACAL clusters for
those events which satisfy the DIS selection criteria, apart from the
$E-p_z$ requirement. This distribution is well described by the
simulation.

At $Q^2 $ above 10~\gv, for $y < 0.75$, the photoproduction background
is subtracted using the PHOJET event simulation. For $y$ above 0.75,
however, experimental information is used by employing the charge
assignment of central tracks associated with SPACAL energy clusters.
This allows the energy range to be extended down to $E_e'=3$\,GeV,
corresponding to $y \leq 0.89$.  For $12 \leq Q^2 \leq 35$\,GeV$^2$
tracks reconstructed in the CJC can be linked to low energy SPACAL
clusters with an efficiency of 95\% (93)\% in 1996 (1997). For such
tracks with energies up to 15\,GeV the charge is determined with an
efficiency of 99.5\%~\cite{sasha}. The sample of candidates with
negative charge is taken to represent the background in the positron
data sample.

\begin{figure}[tbp]
  \epsfig{file=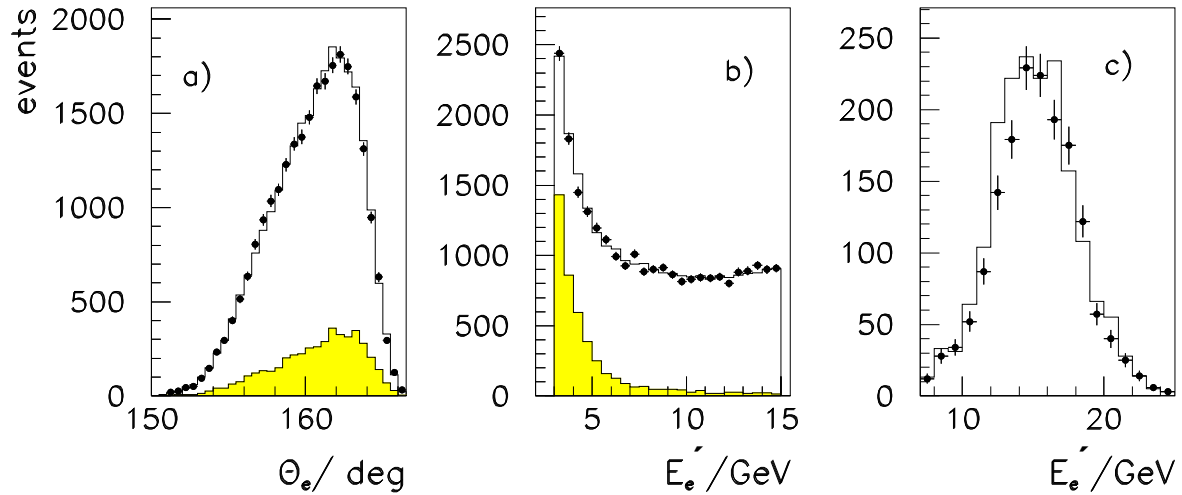,width=15.5cm}
  \caption{\sl
    Distributions illustrating the cross-section measurement
    at high $y$  ($0.46 < y < 0.89 $)
     and large $Q^2$  ($10 < Q^2 < 35$~GeV$^2$). a) Polar angle and b)
    SPACAL energy distributions before subtraction of the
    photoproduction background using the charge measurement by the
    CJC. Solid points: data with positive charge assignment.  Shaded
    histogram: data with negative charge assignment.  Open histogram:
    sum of data with negative charge assignment and DIS
    event simulation, normalized to the data luminosity. c) Spectrum
    of energy measured in the electron tagger for DIS candidate events
    with a linked track of either positive charge (solid points)
    or of negative charge (histogram).} \protect\label{conthyq}
\end{figure}

The statistical subtraction procedure requires the study of any
process which may cause a charge asymmetry. This asymmetry can be
measured using tagged photoproduction events which fulfill the DIS
event selection criteria.  A small charge asymmetry
$(N_+ - N_-)/(N_+ + N_-)$ is
 found with an average of $-4.8$\% with a statistical
accuracy of $1.9$\%, for $0.65 < y < 0.89$. Here $N_+(N_-)$ is the
number of events with positive (negative) charge of the track
associated with the SPACAL cluster.  Comparing the energy distribution
for a sample of negative tracks in $e^+p$ scattering, taken in
1996/1997, with that for a sample of positive tracks in $e^-p$
scattering, taken in 1999, a consistent asymmetry of $(-3.5 \pm
2.5)$\% is measured.  Background simulation studies and measurements
of the ionisation losses in the CJC reveal that this small asymmetry
is due to the antiproton interaction cross section exceeding that for
proton interactions at low energies~\cite{pdg}. Annihilation leads to
larger energy deposits in SPACAL than proton interactions which
introduces an asymmetry for low energies above a given threshold.
This charge asymmetry is taken into account in the measurement of the
positron DIS cross section at high $y$.  Selected control
distributions~\cite{sasha, doris} are shown in figure~\ref{conthyq}
illustrating the good understanding of this kinematic region down to
scattered positron energies of $E_e' = 3$\,GeV.
\subsection{Results \label{sigmeas}}
\begin{figure}[tbp]
  \epsfig{file=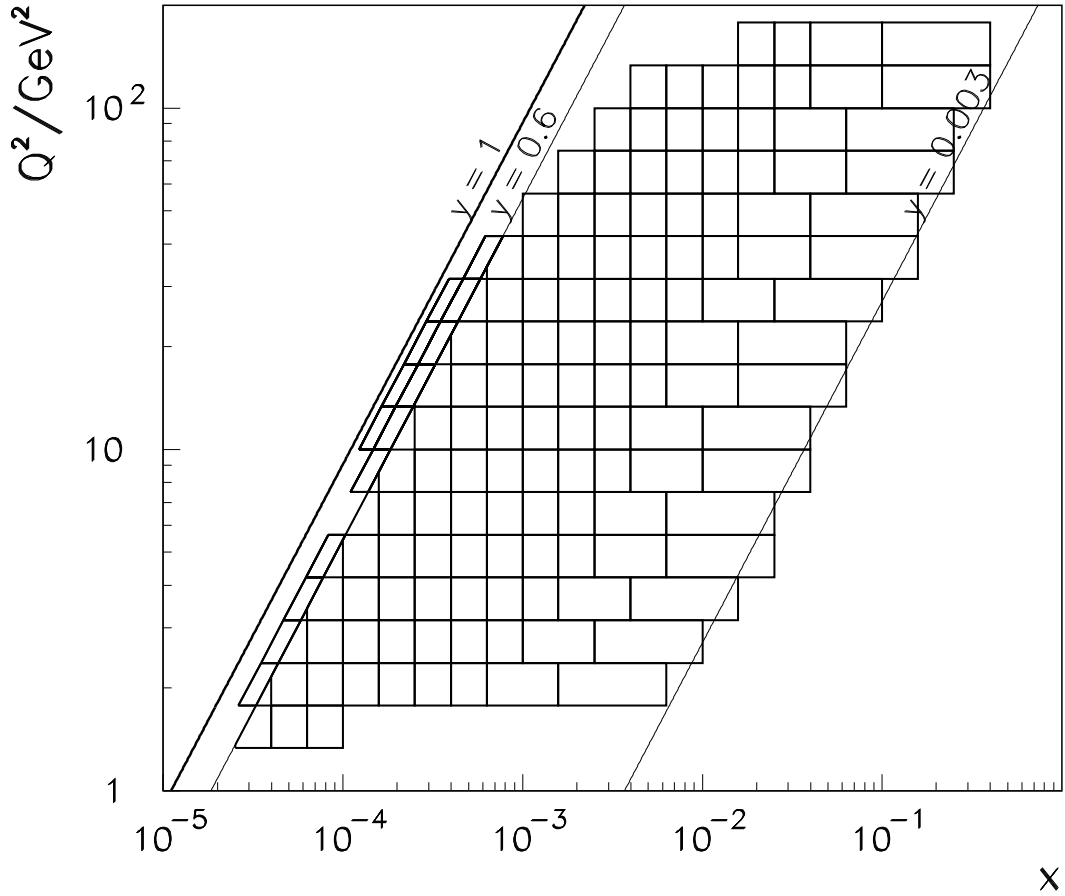,width=14cm}
  \caption{\sl
 Division of the $(x, Q^2)$ plane for the measurement of the
 inclusive DIS cross section. At low $y$ the bin size
 increases as the resolution deteriorates.
 At large $y$
 the data are binned in intervals of $Q^2$ and $y$ 
 in order to account for the 
 $y$ dependent effect of \FLc on the cross-section
 and the variation of the systematics with $y$.
 The triangular regions inside the acceptance do not
 represent valid analysis bins.} \protect\label{bins}
\end{figure}

An iterative bin wise correction procedure is adopted for the
extraction of the double differential cross section $\sigma_{r}$.
This procedure requires that the bin sizes are adapted to the
resolution in the measurement of the kinematic variables.  The data
and the simulated events are binned in a grid in $x$ with five bins
per decade and in $Q^{2}$ with eight bins per decade, as illustrated
in figure~\ref{bins}. At low $y$ the resolution of the measurement
degrades and the bin size is widened.  For $y > 0.6$ the data are
divided in bins of $y$, and the $Q^2$ division is kept.  In this
region the cross section may receive a large negative contribution
from \FLc proportional to $y^2$ and therefore a fine binning in $y$ is
desirable.  Bins are accepted if the purity and stability are bigger
than 30\% with typical values being 60\%. Here the purity (stability)
is defined as the number of simulated events which originate from a
bin and which are reconstructed in it, divided by the number of
reconstructed (generated) events in that bin.

The longitudinal momentum conservation constraint, $E-p_z > 35$\,GeV,
 limits the amount of radiative corrections to at
most 5\% at high $y$.  The program HERACLES \cite{heracles}, which is
used in the DIS event simulation, accounts for first order radiative
corrections to the one photon exchange approximation.  The radiative
corrections are extracted using a high statistics calculation within
the HERACLES Monte Carlo program, and compared with the results of the
numerical program HECTOR~\cite{hechec}, which includes higher order
and hadronic corrections.  The corrections are found to agree within
the statistical accuracy of the radiative event simulation of 0.5\%.

The results of the measurement are summarised in
tables~\ref{tabsiga}-\ref{tabsigd}. At $y > 0.17$ the kinematics are
reconstructed using the quantities $Q^2_e$ and $y_e$. At $y < 0.17$,
where the resolution of $y_e$ degrades, the variables $Q^2_{\Sigma}$
and $y_{\Sigma}$ are used.  The error calculation for the measurement
is discussed below. The full error correlation matrix can be obtained
from the H1 Collaboration~\cite{daterr}.

\begin{figure}[tbp]
  \epsfig{file=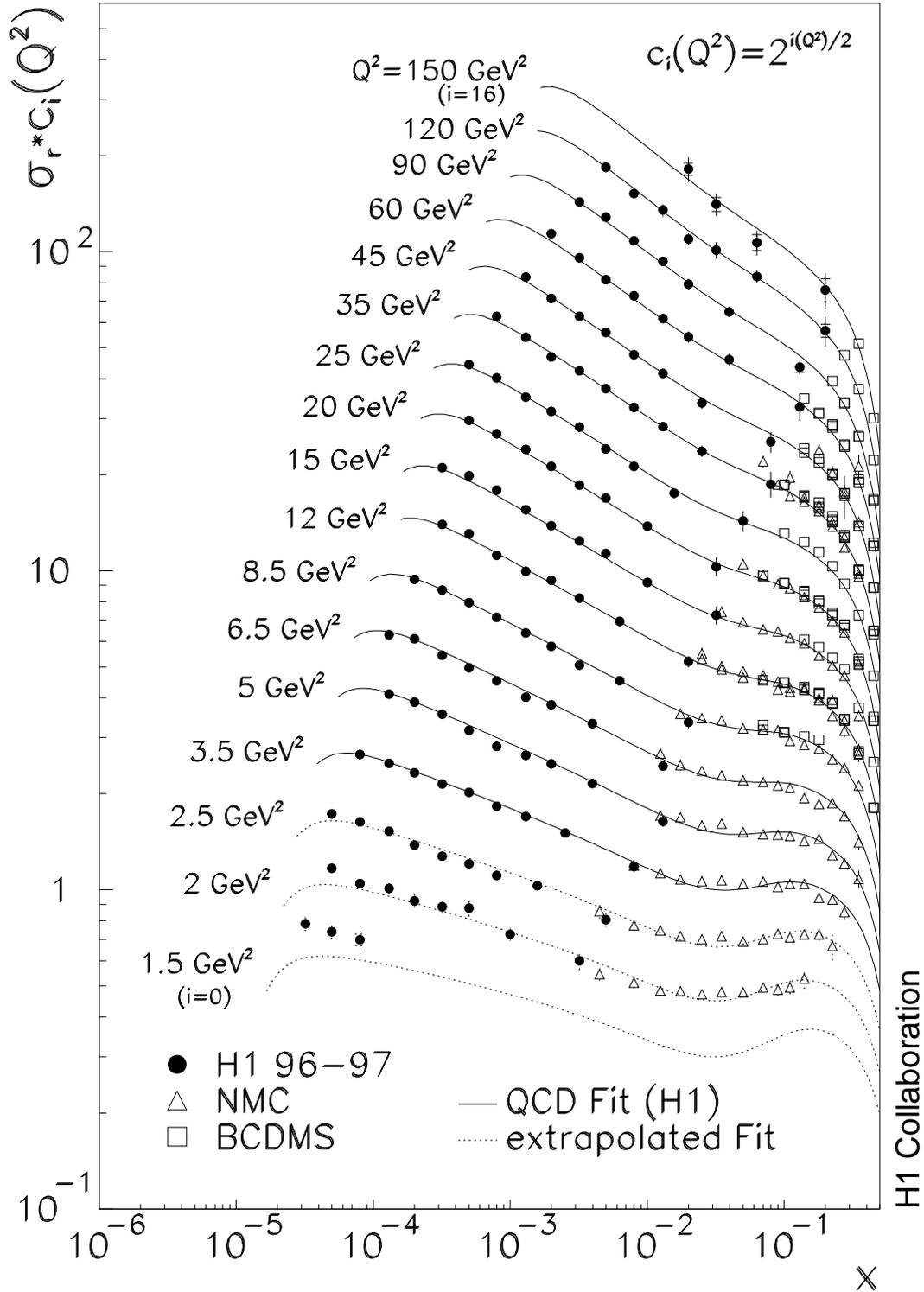,width=14cm}
  \caption{\sl Measurement of the reduced DIS scattering cross section
   (closed points). Triangles (squares) represent data from
    the NMC (BCDMS) muon-proton scattering experiments.  The solid curves
    illustrate the cross section obtained in  a NLO DGLAP QCD fit
    to the present data at low $x$,
    with $Q^2_{min}= 3.5$~\gv,
    and to the H1 data at high $Q^2$. The dashed curves
    show the extrapolation of this fit towards lower $Q^2$.
    The curves are labelled with the $Q^2$ value the data points belong 
    to and scale factors are coveniently chosen to separate the
    measurements.
   } \protect\label{sigqcd}
\end{figure}

The cross section measurement is shown in figure~\ref{sigqcd} as a
function of $x$ for different $Q^2$. Due to the extension of the
measurement towards low $y$, the H1 data overlap with data of fixed
target $\mu p$ experiments. The H1 measurement agrees well with the
fixed target data within the uncertainty of about 7~\%. The cross
section rises towards low $x$.  This rise is observed to be damped at
the smallest values of $x$, which is attributed to the longitudinal
structure function, see section~\ref{flsec}.  The cross section can be
well described by a QCD fit to the data as discussed in
section~\ref{qcdana}.
\subsection{Systematic Errors}
The statistics of the data presented here exceed 10$^4$ events in most
of the bins. The precision for this measurement is dominated by
systematic uncertainties of typically 3\%.  These are classified into
a global normalisation uncertainty, kinematically correlated errors
($\delta_{cor}$), the statistical errors of the data ($\delta_{sta}$)
and uncorrelated errors ($\delta_{unc}$). The uncorrelated errors
contain the statistical uncertainty of the simulation and further
errors due to local systematic effects.

Table~\ref{tabsy1} lists those errors which result in a possible
global change of all data points. The resulting total normalisation
uncertainty of the data is 1.7\%. It is dominated by the error on the
luminosity measurement.
\begin{table}[h] \centering 
\begin{tabular}{|l|c|}
\hline
  source  &  cross section error [\%]  \\
\hline
 online event selection          &  0.5       \\
 BDC efficiency                        &  0.5       \\
 trigger efficiency                  &  0.5       \\
 luminosity measurement     &  1.5       \\
\hline
\end{tabular}
\caption{\label{tabsy1}
  \sl Sources and sizes of normalisation errors.}
\end{table}

Energy calibration and alignment uncertainties cause systematic errors
which depend on the kinematics and introduce correlations between the
measured data points.  These errors are determined using the
simulation program and verified by an analytic calculation.  They are
found to be symmetric to good approximation.  The uncertainty of the
photoproduction background simulation is estimated to be 20\% . The
correlated error sources are listed in table~\ref{tabsy2}.
\begin{table}[h] \centering 
\begin{tabular}{|l|c|c|}
\hline
source &  size of uncertainty &  typical cross section error [\%]    \\
\hline
scattered positron energy scale & 0.3\% at $E_e' \simeq 27.5$\,GeV   & 1  \\
                          & 2.7\%
     at $E_e'=3$\,GeV & 2    \\
scattered positron angle   & 0.3 mrad                & 0.5        \\
hadronic energy scale in LAr &  2\%                   & 2   \\
LAr noise         & 25\% of noise &  max of 5 at lowest $y$ \\
photoproduction background   & 20\% of background  & 3 at large $y$ \\
\hline
\end{tabular}
\caption{\label{tabsy2}
  \sl Sources of correlated systematic errors and their typical effect
  on the cross section measurement accuracy.
}
\end{table}

As a cross check of the positron identification, the scattered
positron is also considered to be the cluster of maximum energy. When
this alternative positron identification criterion is used, the cross
section changes by less than 1\%. This is accounted for in the
systematic error.  Detailed studies using different event generators
with differing simulations of the hadronic final state verify the
reliability of the positron identification procedure within the quoted
systematic uncertainty~\cite{sasha}.

Uncertainties due to radiative corrections, positron identification
and final state simulation details are treated as uncorrelated
systematic errors. The errors introduced by the track based
background subtraction procedure in the high $y$ data analysis
(section~\ref{sechihy}) are summarised in table~\ref{tabhiy}.
\begin{table}[h] \centering 
\begin{tabular}{|l|c|}
\hline
  source  &   error of cross section [\%]   \\
\hline
 positron identification                    &  1    \\
 track charge determination                 &  0.5  \\
 charge asymmetry                           &  1    \\
 CJC-SPACAL track link efficiency           &  2    \\
 hadronic track requirement in CJC          &  1    \\
 high $y$ trigger efficiency                &  1    \\
 radiative corrections                      &  1    \\   
\hline
\end{tabular}
\caption{\label{tabhiy}
\sl 
Additional sources of systematic errors and effect on the cross section
for the ana\-lysis, in the range $12 < Q^2 < 25$\,\gv, in which 
central jet chamber tracks are used to measure the
charge of the positron candidate. }
\end{table}

For each kinematic bin the resulting cross section errors are given in
the cross section tables~\ref{tabsiga}-\ref{tabsigd}.
%
%
\section{Longitudinal Structure Function 
 \boldmath{$F_L(x,Q^2)$} \label{flsec}}
The extraction of the longitudinal structure function is based on the
reduced double differential cross section (equation~\ref{sig}), which
depends on two proton structure functions, \F and \FL.  The
contribution of \FLc is enhanced with $y^2$, and the reduced cross
section $\sigma_r$ tends to \Fc $-$ \FLc for $y \rightarrow 1$. In the
quark-parton model, the longitudinal structure function is zero for
spin 1/2 quarks~\cite{calgro}. In QCD, parton radiation
processes~\cite{altmar} lead to non-zero values of \FLc.  Thus \FLc
contains information about the gluon distribution and about the strong
interaction dynamics which is complementary to that obtained from the
analysis of the scaling violations in \F. At low $Q^2$ the
longitudinal structure function is expected to be particularly
sensitive to higher-order corrections to DGLAP QCD~\cite{zijner, rob, bart}.

The longitudinal structure function can be extracted from the
inclusive cross section only in the region of large $y$.  An important
advantage of HERA, compared to fixed target DIS lepton-nucleon
experiments, is the wide range of $y$ values covered.  This allows the
behaviour of \Fc at low $y$ to be determined reliably and to be
extrapolated into the region of high $y$.  Two methods are used here
to perform an extraction of the longitudinal structure function.  For
larger $Q^2$, a NLO DGLAP QCD fit is used to extrapolate \Fc into the
high $y$ region, and \FLc is obtained with the ``extrapolation
method'' introduced previously~\cite{sfl}.  This fit uses only H1 data
in the restricted kinematic range $y<0.35$ and $Q^2 \geq 3.5$~\gv.
Details of this and other QCD fits to H1 and fixed target data are
presented in section \ref{qcdana}. At low $Q^2 < 10$\,\gv, the
behaviour of \Fc as a function of $\ln y$ is used in a new extraction
method~\cite{doris}, termed the ``derivative method'' as it is based
on the cross section derivative \pdsi.

\subsection{Cross Section Derivative
{\boldmath $(\partial \sigma_r / \partial \ln y)_{Q^2}$ }} \label{pdsi}
The derivative of the reduced cross section, taken at fixed $Q^2$,
is given by  
\begin{equation}
  \left(\frac {\partial \sigma_r }{ \partial \ln y}\right)_{Q^2}=
   \left(\frac {\partial F_2 }{ \partial \ln y}\right)_{Q^2}
   - F_L \cdot 2y^2 \cdot \frac{2-y}{Y_+^2}
   -\frac {\partial F_L }{ \partial \ln y} \cdot \frac{y^2}{Y_+}
       \label{dsig}
\end{equation}
  For $y \rightarrow 1$
the cross section derivative tends to the limit $( \partial F_2 /
\partial \ln y)_{Q^2}~ - ~2 \cdot F_L$, neglecting the contribution from the
derivative of $F_L$. At largest $y$ the \FLc contribution
dominates the derivative of the reduced cross section $\sigma_r$.  This is in
contrast to the influence of \FLc on $\sigma_r$ which is dominated by the
contribution of $F_2$ for all $y$.  A further advantage of
the derivative method is that it can be applied down to very low $Q^2
\simeq 1\,$\gv where a QCD description of \F is complicated due to
higher order and possible non-perturbative corrections.

\begin{figure}[tbp]
 \epsfig{file=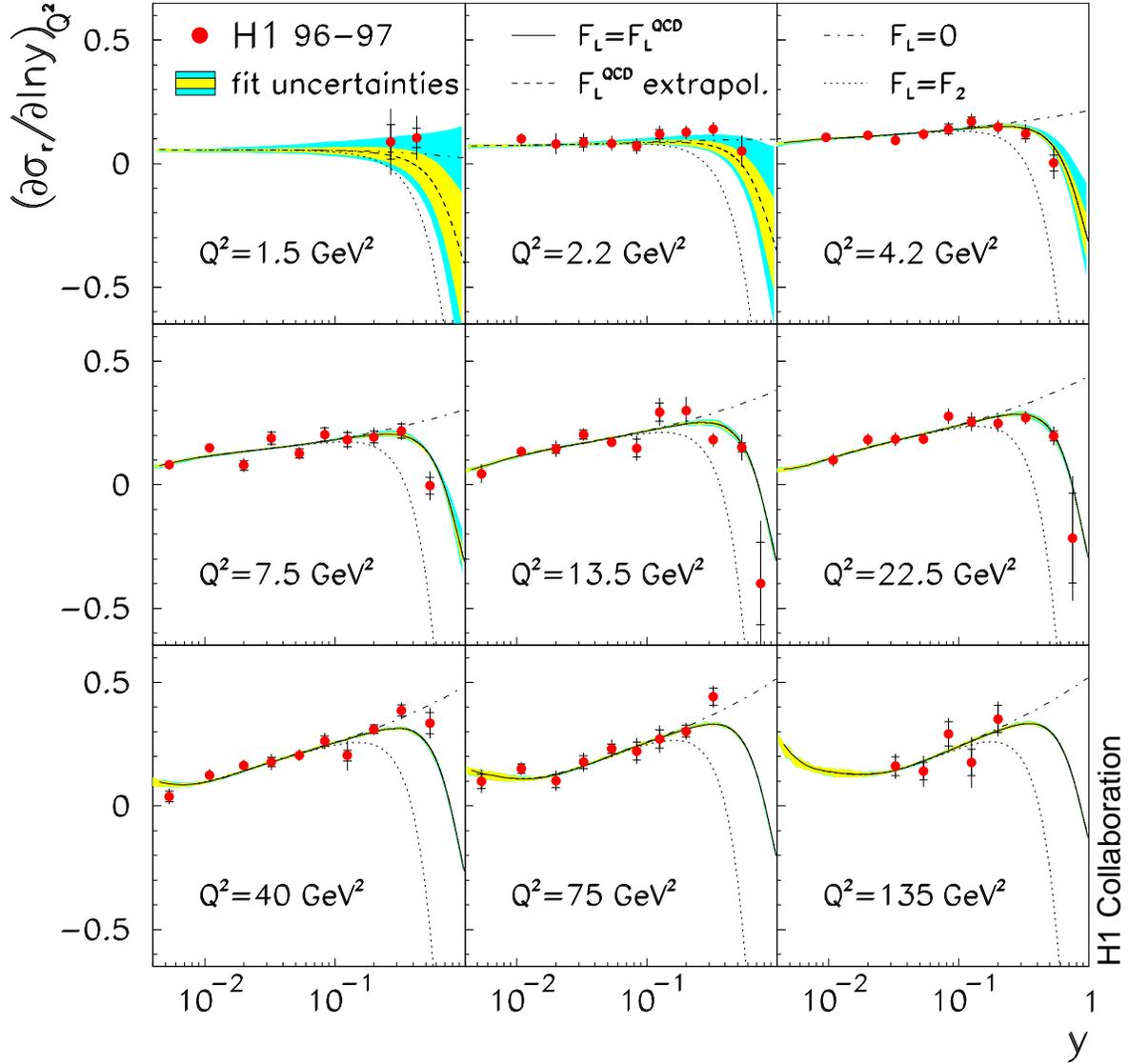,width=15.5cm}
  \caption{ \sl Measurement of the derivative 
    $\partial \sigma_r / \partial \ln y$  at fixed $Q^2$. The
    inner error bars represent the statistical errors and the total 
    error bars
    the statistical and systematic errors, added in
    quadrature. The curves represent the
    QCD fit result to the H1 data, for $y < 0.35 $ and $Q^2 \geq 3.5$\,\gv,
    calculated with different assumptions  about \FLc.
    The solid curves use the QCD prediction of \FLc,
    the dashed (dashed dotted) curves assume \FLc = \Fc (\FLc = 0).
    The inner error band is the experimental uncertainty of the
    fit result (section~\ref{h1fit}),
    the outer band represents the additional
    uncertainty due to the fit assumptions. 
    The fit results shown for  $Q^2 < 3.5$~\gv (dashed)  
    are obtained by backward extrapolation.
} \protect\label{dsigy}
\end{figure}

To obtain an accurate determination of \pdsi the data are rebinned in
$Q^2$ by combining data of two adjacent $Q^2$ intervals.  Differences
$\Delta \sigma_r$ are calculated between cross section points adjacent
in $y$ at fixed $Q^2$. A bin-centre correction is applied to obtain
the derivative at each $y$ point, which is chosen to be the average of
the two $y$ values of the cross section measurements used to calculate
the derivative. A full error analysis is performed in order to account
for the correlations of errors, which partially cancel. The two
adjacent data points of the derivative are anti-correlated since they
use the same cross section measurement with different sign. The cross
section derivatives are shown in figure~\ref{dsigy}, and the values
are given in tables~\ref{tabder1} and \ref{tabder2}. The measured
derivatives are well described by the QCD calculation
(section~\ref{h1fit}).

For low $Q^2$ and $y < 0.3$ the derivative is observed to be a linear
function of $\ln y$. The structure function \F, at fixed $Q^2$,
behaves like $x^{-\lambda} \propto y^{\lambda}$. At low $Q^2$ the
exponent $\lambda$ is observed to be small and the derivative is thus
expected to be approximately linear in $\ln y$. This approximation is
used to determine the longitudinal structure function at low $Q^2$.
For larger $Q^2$ the exponent $\lambda$ rises and  a
curvature is expected as can be seen in figure~\ref{dsigy}.
\subsection{Determination of \boldmath{$F_L$}} \label{fldet}
For the determination of \FLc for $Q^2 < 10$\,GeV$^2$ straight line
fits are performed in $\ln y$ to the derivative \pdsi for $y \leq
0.2$.  These straight lines describe the data well
(figure~\ref{dsigy}) and are extrapolated to estimate the contribution
of $\partial F_2 / \partial \ln y$ at high $y$. The uncertainty on
this extrapolation is included in the systematic errors of the \FLc
determination, taking into account the correlations of errors at low
$y$ with those at high $y$.  The extrapolations are compared with the
values obtained from the QCD analysis and very good agreement has been
found. The contribution of $\partial F_L / \partial \ln y$ to the
derivative (equation~\ref{dsig}) is neglected. The uncertainty of this
approximation is estimated using the derivative of \FLc as calculated
in QCD. It is taken as an additional uncertainty on the measurement,
and amounts to about a quarter of the systematic uncertainty of \FLc.

\begin{figure}[tbp]
 \epsfig{file=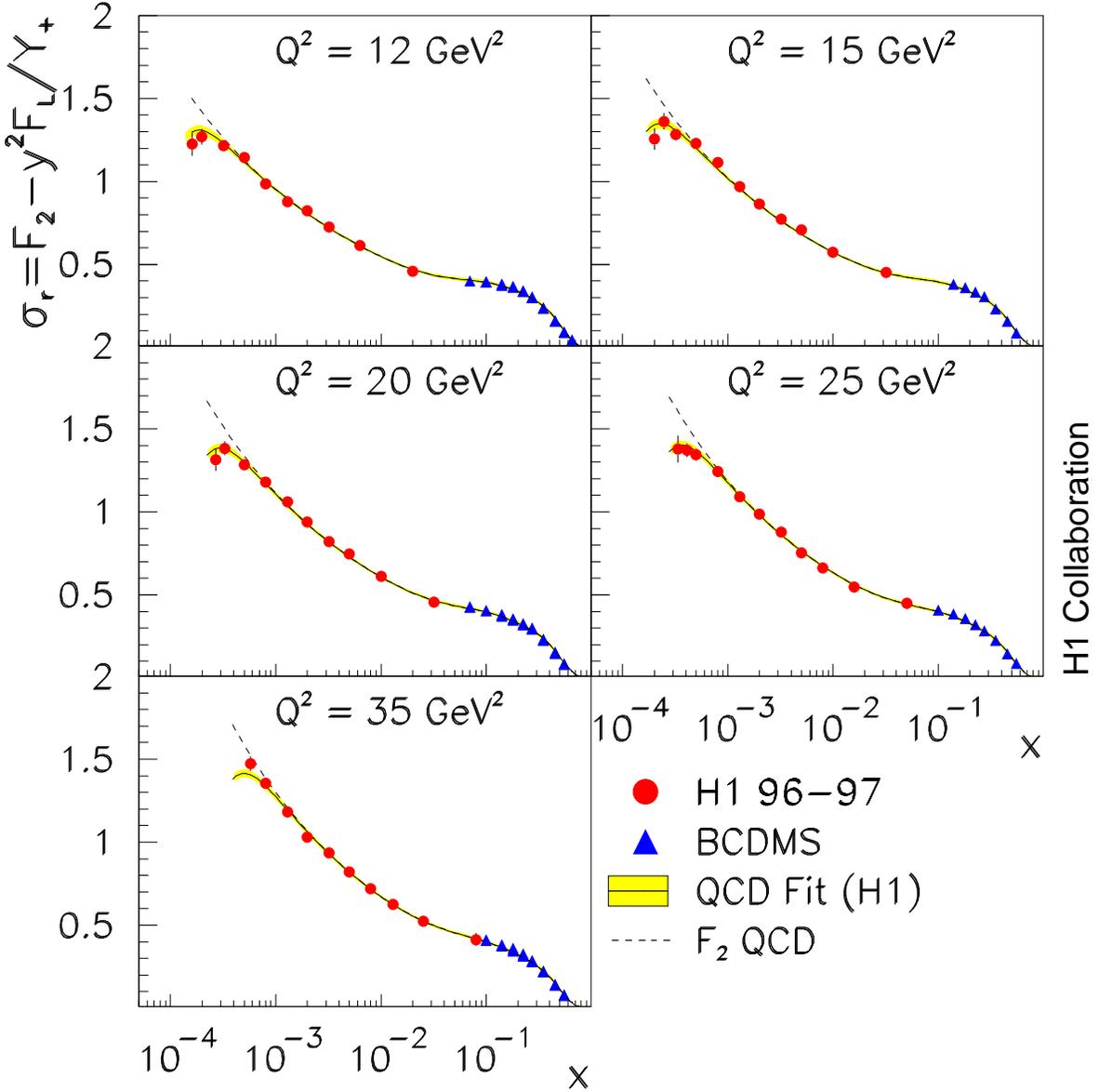,width=15.5cm}
  \caption{\sl
  Measurement of the reduced DIS scattering cross section
  (closed points). Triangles represent data from
    the BCDMS muon-proton scattering
 experiment. The curves represent a NLO QCD fit
    to the H1 data alone, using data with $y < 0.35$
     and $Q^2 \geq 3.5$~\gv. The dashed curves 
    show the $F_2$ structure function as
     determined with this fit.
   The error bands represent the  experimental and model
   uncertainty of the QCD fit.
} \protect\label{smstamp}
\end{figure}

For $Q^2 > 10$\,GeV$^2$, the NLO QCD fit of the H1 data for $y < 0.35$
is used to estimate \Fc in the high $y$ (i.e. small $x$) region. In
this $Q^2$ range, the extrapolation method is more accurate than the
derivative method. The derivative method is statistically limited at
maximum $y$ since this region is accessed only with a combined SPACAL
and track trigger. Figure~\ref{smstamp} compares the fit with the
measured cross section for those five $Q^2$ bins above 10~\gv which
access the high $y$ region.  The difference between the measured
$\sigma_r$ and the extrapolated \Fc is used to determine \FL as
described in~\cite{sfl}. Systematic errors, which are common to the
lower $y$ and the large $y$ region, are considered in the fit as
described in~\cite{zopa}.

The \FLc values obtained are presented in table~\ref{tabfl}.  The
uncertainties on the longitudinal structure function include several sources:
the statistical errors, uncorrelated systematic errors and correlated
systematic errors, resulting e.g. from the $y$ dependent amount of
subtracted photoproduction background. In addition, errors are
associated with the assumptions inherent to the extraction methods.
For the derivative method these errors are dominated by the uncertainty of
the straight line fit, and for the extrapolation method by the
variation of the smallest $Q^2$ of data used in the QCD fit, see
section~\ref{h1fit}. As can be seen in table~\ref{tabfl}, these errors
are smaller than the experimental systematic errors. In the region of
overlap, for $Q^2$ between 4~\gv and 15~\gv, the derivative method and
the extrapolation method give consistent results. Further details of
this analysis are described in~\cite{doris}.

\begin{figure}[tbp]
 \epsfig{file=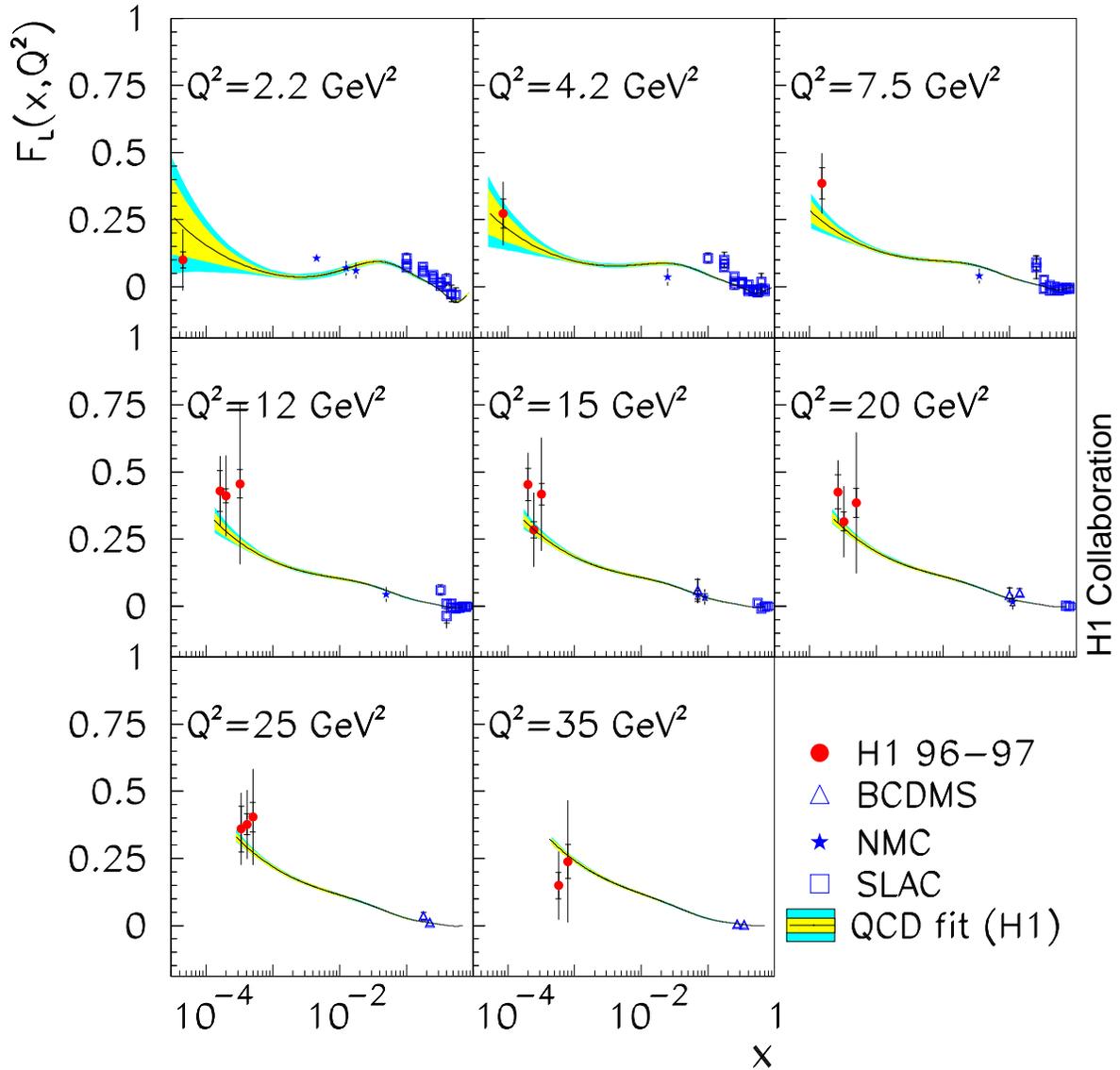, width=15.5cm}
  \caption{\sl The longitudinal structure function \FL 
    for different bins of $Q^2$  as obtained
    by H1 at low $x$, and by charged lepton-nucleon fixed
    target experiments at large $x$.
    The measurements for $Q^2 < 10$~\gv are
    determined with the derivative method while the points for larger
    $Q^2$ are due to the extrapolation method. The error on the
    data points is the total uncertainty of the determination
    of \FLc representing the statistical, the systematic and the
    model errors added in quadrature. The inner error
    bars show the statistical error. 
    The error bands are due to the
    experimental (inner) and model (outer) uncertainty of the
    calculation of $F_L$ using the NLO QCD fit to the H1 
    data for $y < 0.35$ and $Q^2 \geq  3.5$\,\gv.
     }
  \protect\label{figflu}
\end{figure}

The values for \FL, given in table~\ref{tabfl}, are consistent with
the previous determination of \FLc by the H1 collaboration~\cite{sfl},
but they are more precise and obtained in a wider kinematic range.
The H1 data extend the knowledge of the longitudinal structure
function beyond that obtained from fixed target lepton-proton
scattering experiments into the region of much lower $x$, see
figure~\ref{figflu}. The increase of \FL towards low $x$ is consistent
with the NLO QCD calculation (section~\ref{h1fit}), reflecting the rise
of the gluon momentum distribution towards low~$x$. A measurement of
the $x$ dependence of \FL can be performed independently of
assumptions about the behaviour of $F_2$ with a variation of the 
proton beam energy at HERA~\cite{lat}.
%
%
\section{Structure Function \boldmath{$F_2$} and  
Derivative
  \boldmath{$(\partial F_2 / \partial \ln Q^2)_x$}} \label{deris}
The proton structure function \F is obtained from the measured
reduced cross section using equation~\ref{sig} rewritten as
\begin{equation}
F_2 =  \sigma_r  \cdot \left(1- \frac{y^2}{Y_+} \cdot \frac{R}{1+R}\right)^{-1}. 
 \label{siglt}
\end{equation}  
The ratio $R=F_L/(F_2-F_L)$ is determined using the standard DGLAP QCD
fit to the H1 data (section~\ref{h1fit}), calculating \FLc to order
$\alpha_s^2$.  In order to reduce the dependence of the measurement on
\FLc, the \Fc extraction is limited to the range $y \leq 0.6$.  The
results for \F and the calculated values of \FL are given in
tables~\ref{tabsiga}-\ref{tabsigd}. This measurement is consistent
with, and improves upon, the previous results~\cite{sf94}, which were
obtained with a different backward apparatus.

\begin{figure}[tbp]
 \epsfig{file=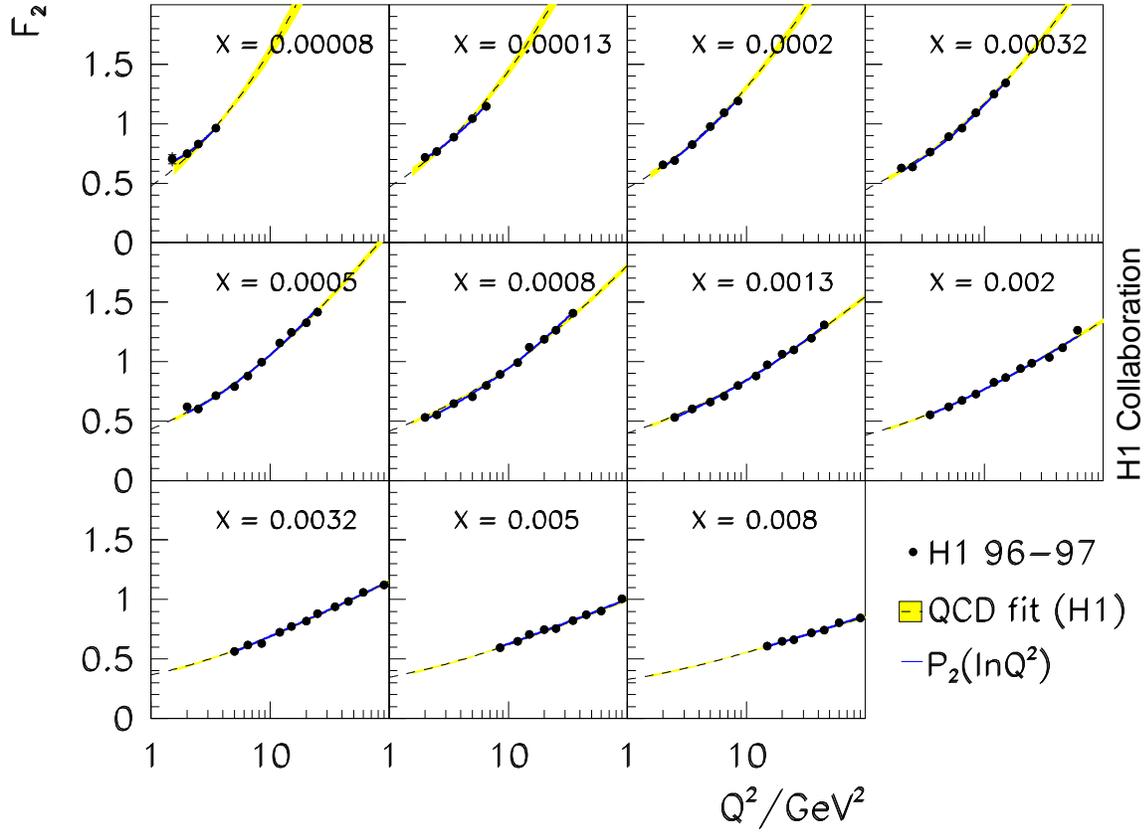,width=15cm}
  \caption{\sl Measurement of the proton structure function \F,
  plotted as  functions of $Q^2$
  in bins of $x$ (points). The solid lines
  represent fits to \Fc in bins of $x$  according to a polynomial
  $P_2 (x,Q^2) = a(x) + b(x) \ln Q^2 + c(x) (\ln Q^2)^2$. The dashed lines
  are obtained from
  the NLO QCD fit to the H1 data (section~\ref{h1fit}), for $Q^2 \geq 3.5$~\gv.
   The error bands are due to the experimental and
  model uncertainties in the QCD fit.
    } \protect\label{f2logq2}
\end{figure}

In figure~\ref{f2logq2} the measurement of the structure function \F
at low $x$ is shown as a function of $Q^2$. The data are well
described by the NLO QCD fit, as is discussed in detail in
section~\ref{h1fit}.  The $\ln\,Q^2$ dependence of \Fc is observed to
be non-linear. It can be well described by a quadratic expression
\begin{equation}
  P_2 (x,Q^2)= a(x) + b(x) \ln Q^2 + c(x) (\ln Q^2)^2,
       \label{p2}
\end{equation}
which nearly coincides with the QCD fit in the kinematic range of this
measurement.

The DGLAP evolution equations are governed by the derivative \pdff
taken at fixed $x$. Measurement of this derivative has long been
recognised as a powerful constraint on \xgc and \as \cite{wuki}.
In the low $x$ DIS region its behaviour is a
direct reflection of the behaviour of the gluon density~\cite{prytz}.
This quantity has also been studied in~\cite{bartels} in view of
possible non-linear gluon interaction effects~\cite{glr}.  A study of
the derivative \pdff at low $x$ was presented previously by the ZEUS
Collaboration~\cite{zeuspheno} where \F was assumed to depend linearly
on $\ln Q^2$. This approximation is inconsistent with the data
presented here, see figure~\ref{f2logq2}.

\begin{figure}[tbp]
 \epsfig{file=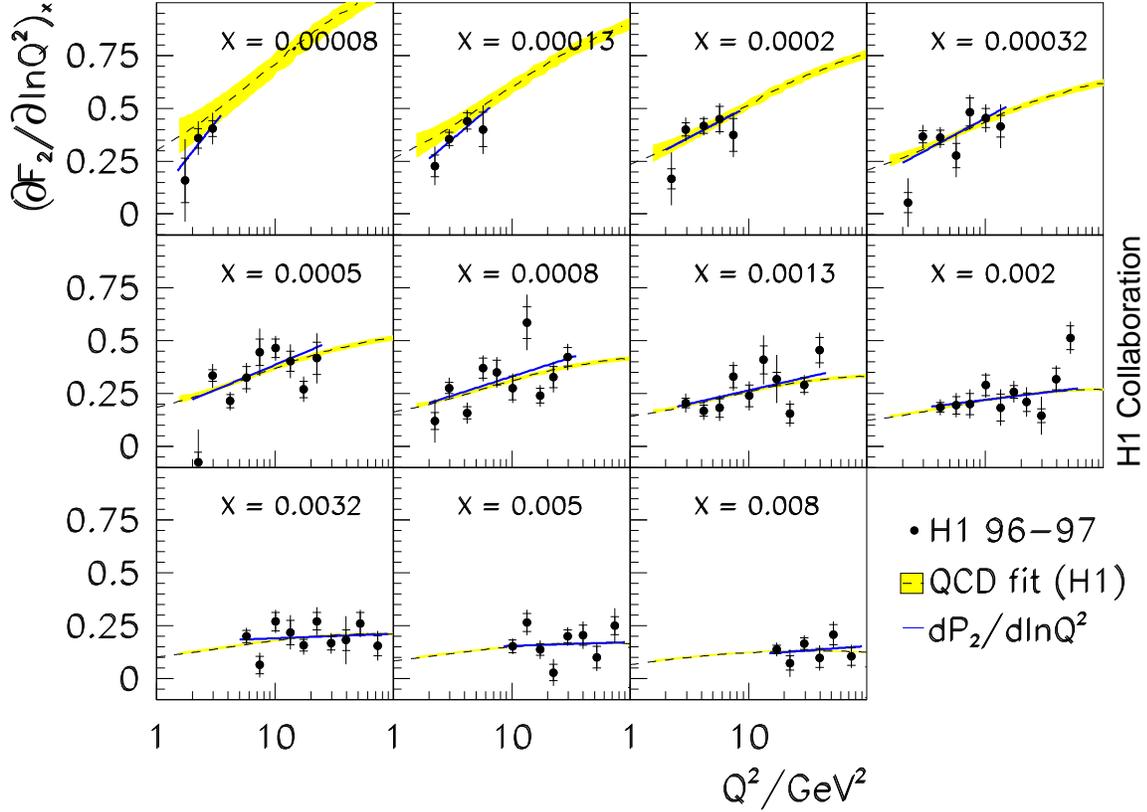,width=15cm}
  \caption{\sl
  Measurement of the partial
  derivative \pdff
  taken at fixed $x$ and plotted
  as  functions of $Q^2$. The error bars represent the
  quadratic sum of statistical and systematic 
  errors. The straight solid lines are given by the function
  $ b(x) + 2c(x) \ln Q^2$ determined in fits to \F at fixed $x$.
  The dashed lines represent the  derivatives as calculated with
  the QCD fit to the H1 data. The error bands are due to the 
  experimental and
  model uncertainties in the QCD fit which includes data
  for $Q^2 \geq 3.5$~\gv.
    } \protect\label{df21}
\end{figure}

\begin{figure}[tbp]
  \epsfig{file=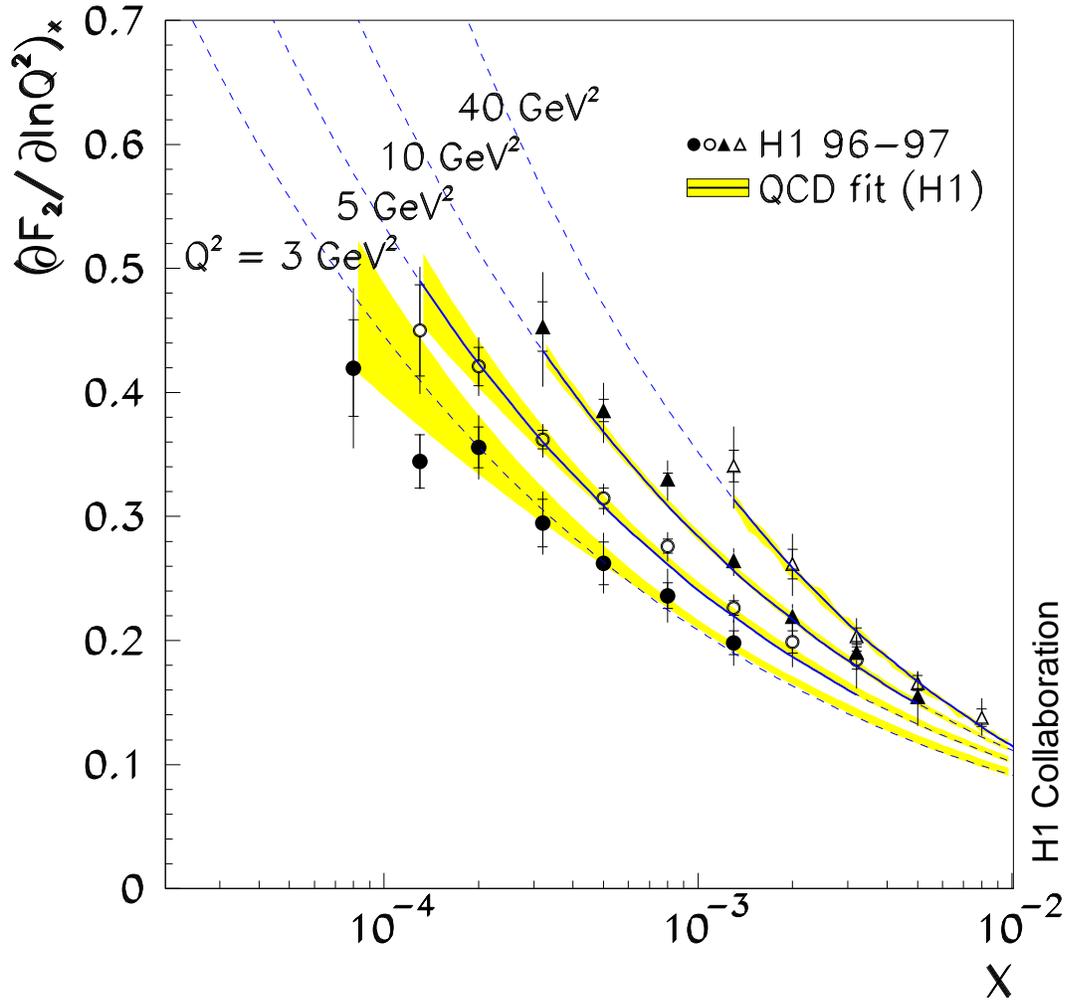,width=14cm}
  \caption{ \sl
The derivative \pdff  plotted
 as functions of $x$ for fixed $Q^2$,
for the H1 data  (symbols) and the QCD fit to
the H1 data, for $Q^2 \geq 3.5$~\gv (solid lines). 
 The dashed curves extrapolate this fit below $Q^2_{min}$
 and outside the range of $x$.
The error bands represent the model uncertainty of the
QCD analysis. 
    } \protect\label{df22}
\end{figure}

Using the procedure adopted for the derivatives \pdsi
(section~\ref{pdsi})
 the local derivatives \pdff are measured.
 The results are shown in figure
\ref{df21} for different $x$ as a function of $Q^2$, and the values
are quoted in tables~\ref{tabderq1},\ref{tabderq2} and \ref{tabderq3}.
For each bin of $x$ these derivatives can be described by the function
$b(x) + 2 \cdot c(x) \ln Q^2$ (solid lines).
Small deviations of \pdff from the linearity in $\ln Q^2$ occur in NLO
QCD (dashed lines).  Using the linear expression the derivatives are
calculated at fixed $Q^2 \geq 3$\,\gv and are shown as functions of $x$ in
figure~\ref{df22}.  The derivatives show a continuous rise towards low
$x$ for fixed $Q^2$ which is well described by the NLO DGLAP QCD to
the H1 data  (section~\ref{h1fit}).  The shape of \pdff reflects the
behaviour of the gluon distribution in the associated kinematic
range.
%
%
\section{QCD Analysis \label{qcdana}}
In this section the predictions of the DGLAP evolution equations in
NLO are confronted with the reduced differential cross section
measurement.  By comparison to the cross section data, the strong
coupling constant $\alpha _s$, and the shape and normalisation of the
gluon and quark distributions are determined.  This is done using a
$\chi ^2 $ minimisation procedure (fit) and a suitable choice of
parameterisations for the input parton distributions at an initial scale
$Q_0^2$.

Traditionally, this kind of analysis makes use of both lepton-proton
and lepton-deuteron data~\cite{NMCg, mv, botje, alek, andreas}
in order to separate the non-singlet and singlet evolution, and also to
determine the parton distributions of up and down quarks
simultaneously. The present analysis introduces a new parameterisation
of quark distributions which permits lepton-proton data to be analysed
alone. Thus the use of deuteron scattering data, which involves bound
state corrections and their uncertainties, is avoided.

Two complete analyses are performed, one with H1 data only to
determine the gluon distribution \xg at low $x$, and a second one in which
 the H1 data are combined with data from the BCDMS experiment in order to
simultaneously determine the strong coupling constant \amz
and the gluon distribution.

The most difficult aspect of these fits is an adequate treatment of
systematic errors, which lead to strong correlations among the data
points. The present analysis uses a sophisticated treatment of
systematic error correlations allowing their true effect on the
extracted quantities to be estimated by the fit~\cite{zopa}. This
procedure is used to identify data regions which are strongly affected
by correlated systematic errors. The present analysis is therefore
based on a minimum number of data sets in regions where their
systematic errors are well understood. Uncertainties due to physics
and analysis assumptions are estimated by a systematic exploration of
the parameter space.

The $Q^2$ evolution of parton distributions is the result of processes
of radiation from gluons which dominate the
scaling violations at small $x \le 0.1$, and from quarks which
dominate at large $x$.  The present H1 data allow the gluon
contribution to be well determined for fixed $\alpha_s$.  The strong
correlation between the gluon distribution with $\alpha _s$ can be
much reduced by using DIS data at large $x$ and low $Q^2$, in addition
to the H1 data.  Thus, the analysis is extended to include the precise
$\mu p$ data at large $x$ from the BCDMS collaboration.  This allows
the gluon distribution and $\alpha _s$ to be determined
simultaneously.
\subsection{Analysis Procedure\label{secanapro}}
In the quark-parton model, the proton structure function \F
is given by a sum of quark and anti-quark momentum distribution
functions
\begin{eqnarray}\label{f2q}
 F_2(x,Q^2) = x \sum_q{Q_q^2 \cdot [q(x,Q^2)+\bq(x,Q^2)}],
\end{eqnarray}
where $Q_q$ represents the electric charges of quarks.  In the present
analysis, the sum extends over up, down and strange ($u,~d,~s$)
quarks.  The  charm and beauty contributions are added
using NLO QCD calculations~\cite{lae2} in the on-mass shell
renormalisation scheme using $m_c~=~1.4$\,GeV
and $m_b = 4.5$\,GeV. At low $x$ about 20-30\% of the inclusive
cross section is due to charm production, dominated by the photon-gluon
fusion process.

The present QCD analysis, described in detail in \cite{rainer}, uses a
flavour decomposition of \F into two independent combinations of
 parton
distribution functions \V and \A, according to
\begin{equation}
  \label{f2va}
  F_2 =  \frac{1}{3}x V + \frac{11}{9}x A. 
\end{equation}
The $x$ dependences of \xgc, \Vc and \Ac are parameterised at an
initial scale $Q_0^2$, and a $\chi^2$ minimisation determines these
distributions and \as.  The function \Vc  is
defined by the valence-quark distributions, i.e.
\begin{equation}
  \label{vsimp}
  V = \frac{9}{4} u_v - \frac{3}{2} d_v.
\end{equation}
It is thus constrained by the relation
\begin{equation}
  \label{intV} 
 \int_0^1{V}dx=3,
\end{equation}
which is used in the fit procedure together with the momentum-sum
rule.  The function \Ac~contains the sea-quark distribution and a
small valence-quark correction. It is given as
\begin{equation}
  \label{asimp}
  A = \bu - \frac{1}{4}  ( u_v  - 2  d_v),
\end{equation}
and determines the low $x$ behaviour of \F.

These equations hold for a strange contribution $s+\bs = (\bu +
\bd)/2$ and flavour symmetry of the sea, $u_{sea}=\bu=d_{sea}=\bd$. 
 This ansatz is generalised in
Appendix~\ref{aflavour} to account for the
observed small deviations of the strange~\cite{ccfr} and
antiquark~\cite{handt} distributions from the conventional
assumptions about the sea. In the analysis target-mass
corrections~\cite{tarmas} are applied.

The analysis is performed in the $\overline{MS}$ renormalisation
scheme using the DGLAP evolution equations~\cite{dglap} in
NLO~\cite{furmanski}. Thus the formulae for \Fc given here are modified 
by replacing the sums over parton distributions by sums
over integrals of products of coefficient functions times parton
distributions.

The strong coupling constant is defined by the solution of the
renormalisation group equation to order $\alpha_s^3$,
\begin{equation}
\frac{da_s}{d{\rm ln}\mu_r^2}=-\beta_0a_s^2-\beta_1a_s^3,
 \label{renorm}
\end{equation}  
where $a_s=\alpha_s/4\pi$, $\mu_r$ is the renormalisation scale, and
the $\beta$ functions are defined in~\cite{furmanski}. The
longitudinal structure function is calculated to order $\alpha_s^2$.
The analysis uses an $x$ space program developed inside the H1
collaboration~\cite{zopa2}. This program has been checked in detail
against different evolution codes~\cite{bluvo, botje, anleiden}, and
very good agreement is found.

In the fit procedure, a $\chi^2$ function is minimised which is
defined in Appendix~\ref{achi2}. This definition takes into account
correlations of data points caused by systematic uncertainties. It is
desirable that the fit results depend neither significantly on the
functional form of the parameterisations which are used for the input
distributions, nor on the input scale $Q_0^2$ at which these are
defined. Thus, for each fit described below, a grid of nearly $10^3$
initial fit conditions is considered, with about tenfold variations of
each of the parameters $Q_0^2$, $Q^2_{min}$ and $\alpha_s$.  Here
$Q^2_{min}$ denotes the smallest $Q^2$ value of data included in the
fit.

The quality of the fits is studied in a statistical evaluation of the
parameter space for all data sets and parton distribution
parameterisations considered.  As described in Appendix~\ref{apara},
this leads to a best choice of
\begin{equation}
xq(x) = a_qx^{b_q}(1-x)^{c_q}
      [1+d_{q}\sqrt{x} +e_{q}x],
\label{eqparah1}
\end{equation}
for the parameterisations of the functions $V,~A$ and $xg$.
The standard value of $Q^2_0$ is 4~GeV$^2$, but it can be varied over a
reasonable range without significantly influencing the result. 
\subsection{Fit to  H1 Data and Determination of the Gluon Distribution \label{h1fit}}
The measurements of the reduced differential cross section presented
here are combined with recent data of the H1 collaboration~\cite{hiq}
from the same data taking period, which cover the large $x$ range at
high $Q^2 \geq 150$\,\gv.  A cut $Q^2 \leq 3000$~\gv is applied to
eliminate the region where electroweak interference effects are
important, which involve the structure function $xF_3$.  Since the H1
data have still limited precision at large $x$, the parameter $d_g$ in
the expression for the gluon distribution (equation~\ref{eqparah1}) is
superfluous, see Appendix~\ref{apara}.  For the fits to the H1 data
alone it is therefore set to $d_g=0$.

\begin{figure}[t]
 \epsfig{file=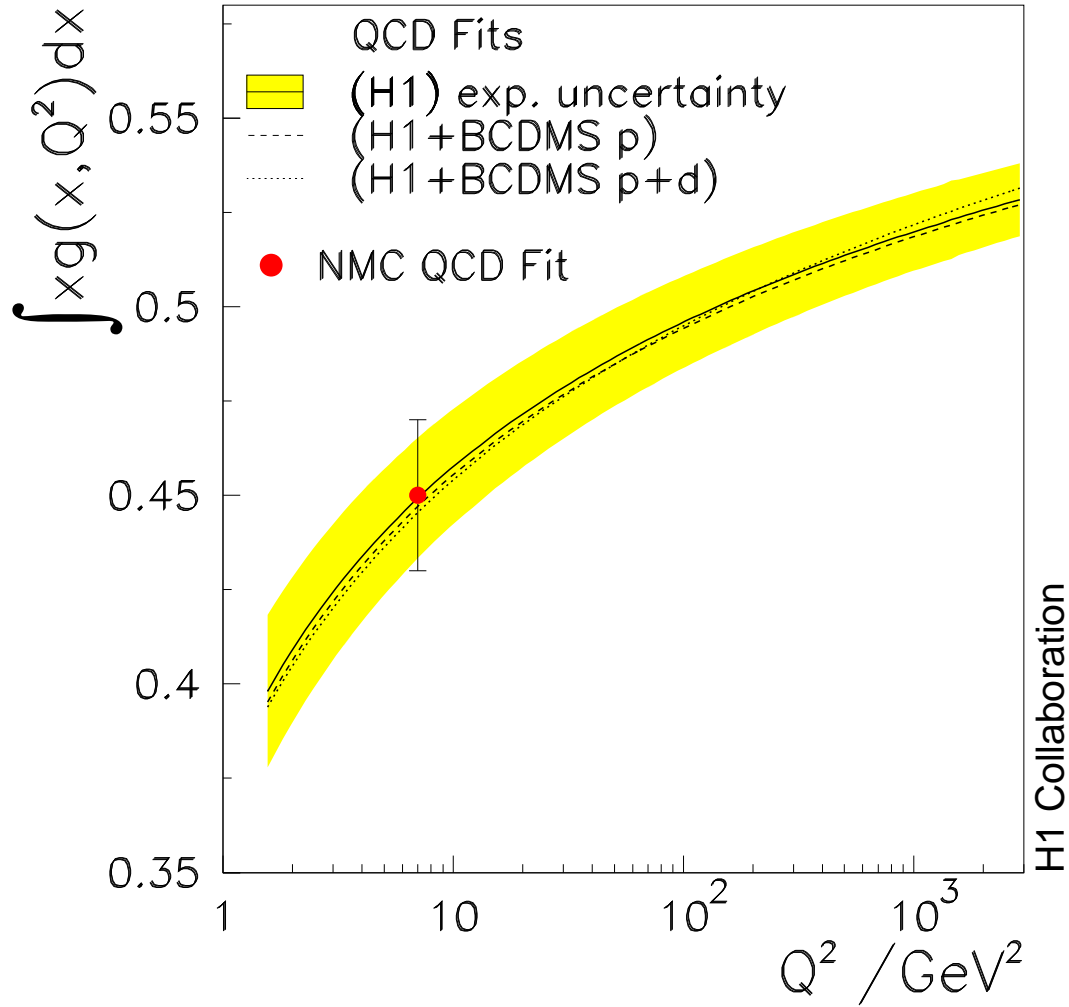,width=14cm}
  \caption{ \sl The fraction of the proton momentum carried by 
  gluons  as a function of 
   $Q^2$, obtained in different NLO DGLAP fits.
   Solid curve:  fit to H1 data alone;  
  dashed curve: fit to H1 and BCDMS proton data;
  dotted curve: fit to H1 $ep$ and BCDMS $\mu p$ and $\mu d$ data.
  The shaded error band represents the experimental uncertainty
  in the analysis of the H1 data alone. The solid point is due to
   a  QCD analysis by the NMC collaboration~\cite{NMCg}.}
  \protect\label{glumo}
\end{figure}

The standard fit assumes a fixed \amz = 0.115 and uses all H1 data for
$Q^2 \ge 3.5$~\gv.  In this range derivatives of $F_2$ with respect to
$\ln Q^2$ are measured and found to be described by the DGLAP
evolution equations (see section~\ref{deris}).  The momentum fraction
carried by the gluons is $0.43 \pm 0.02~(exp)$ at the input scale
$Q_0^2=4$~\gv where the error is due to the measurement uncertainties.
The variation of the gluon momentum fraction with $Q^2$ is shown in
figure~\ref{glumo}.  The result agrees with previous determinations at
$Q^2 = 7$~\gv~\cite{NMCg,cteqg}. The stability of this result has been
checked by adding $\mu p$ and also $\mu d$ data of the BCDMS
collaboration. As is shown in figure~\ref{glumo}, no significant
change is observed. The fit is also repeated without using the
constraint given by the momentum sum rule.  This fit determines the
integral $\int_0^1{x(\Sigma +g)}dx$ to be $1.016 \pm 0.017 (exp)$,
where $\Sigma$ is the singlet parton distribution function, see
Appendix~\ref{apara}.  This value is found to be nearly independent of
the minimum $Q^2$ value of the data included in the analysis.

\begin{figure}[t]
  \epsfig{file=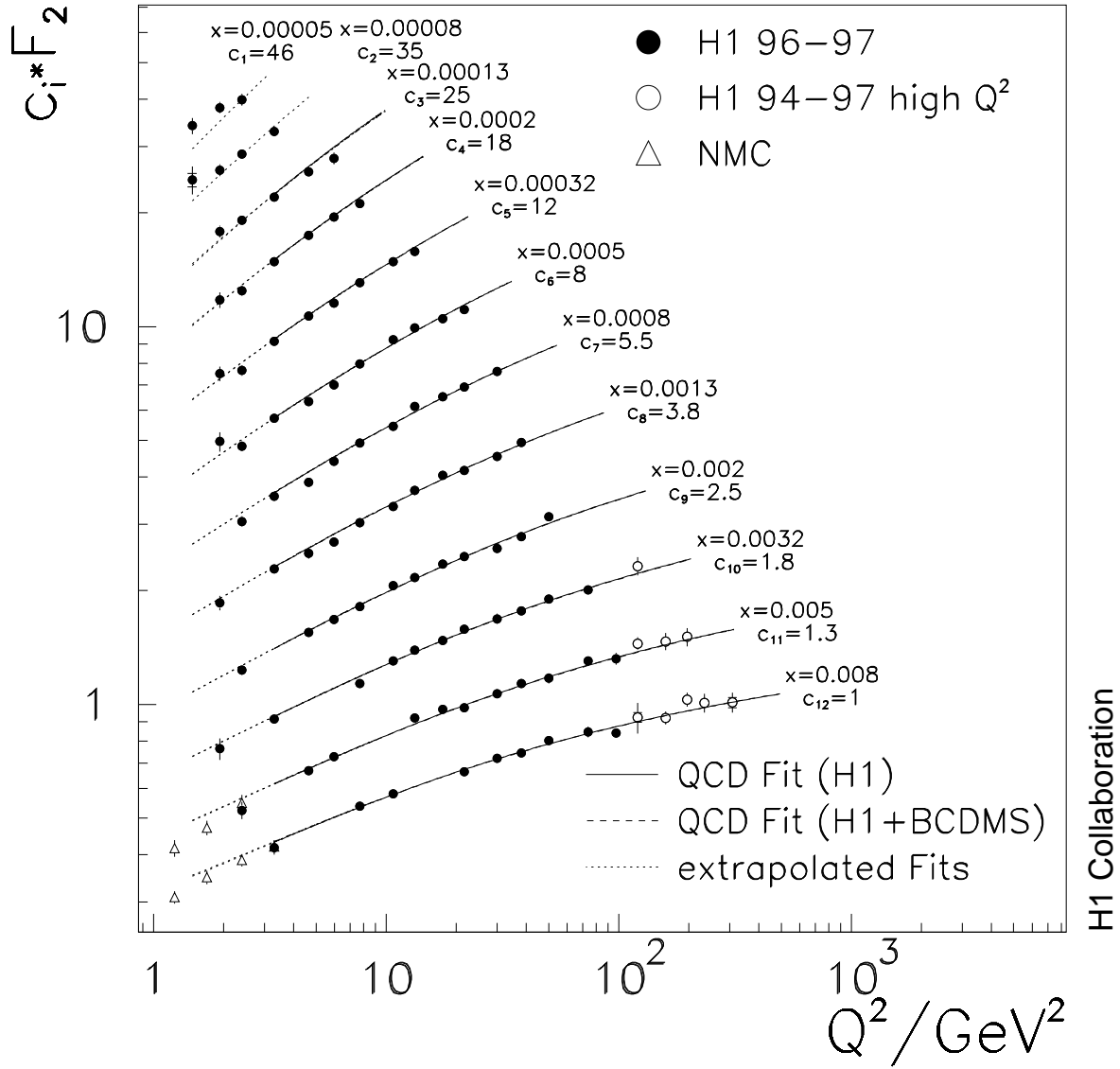,width=15.5cm}
  \caption{ \sl Measurements of the proton
    structure function \F  by the H1  and the NMC 
    experiments. Solid curves: NLO DGLAP QCD fit to the H1
    cross section data. Overlayed as dashed curves 
    are the results of the 
    QCD fit to the H1 $ep$ and
    BCDMS $\mu p$ data, for  $y_{\mu} > 0.3$,
    which are indistinguishable from
    those of the pure H1 fit.
   Dotted curves:  fit extrapolations
    at fixed $x$  into the region below $Q^2 = 3.5$~\gv.
    } \protect\label{f2qa}
\end{figure}

The structure function \F is extracted from the reduced cross section
data using the prediction of the fit for the longitudinal structure
function \FL.  The result is shown in figures \ref{f2qa} and
\ref{f2qb}. The data are compared to published $\mu p$ data of the
fixed target muon-proton scattering experiments BCDMS and NMC. The
solid lines give the result of the QCD fit with $Q^2_{min} = 3.5$~\gv
to the H1 data. This fit also describes the fixed target data in the
non-overlapping regions rather well, except for the data points at
$x=0.65$ where the fit curve is below the BCDMS data.  The H1 data at
this value of $x$ \cite{hiq} have a correlated systematic uncertainty
of 12\%, due to the energy scale error for the scattered positron,
which accomodates the observed difference.

\begin{figure}
 \epsfig{file=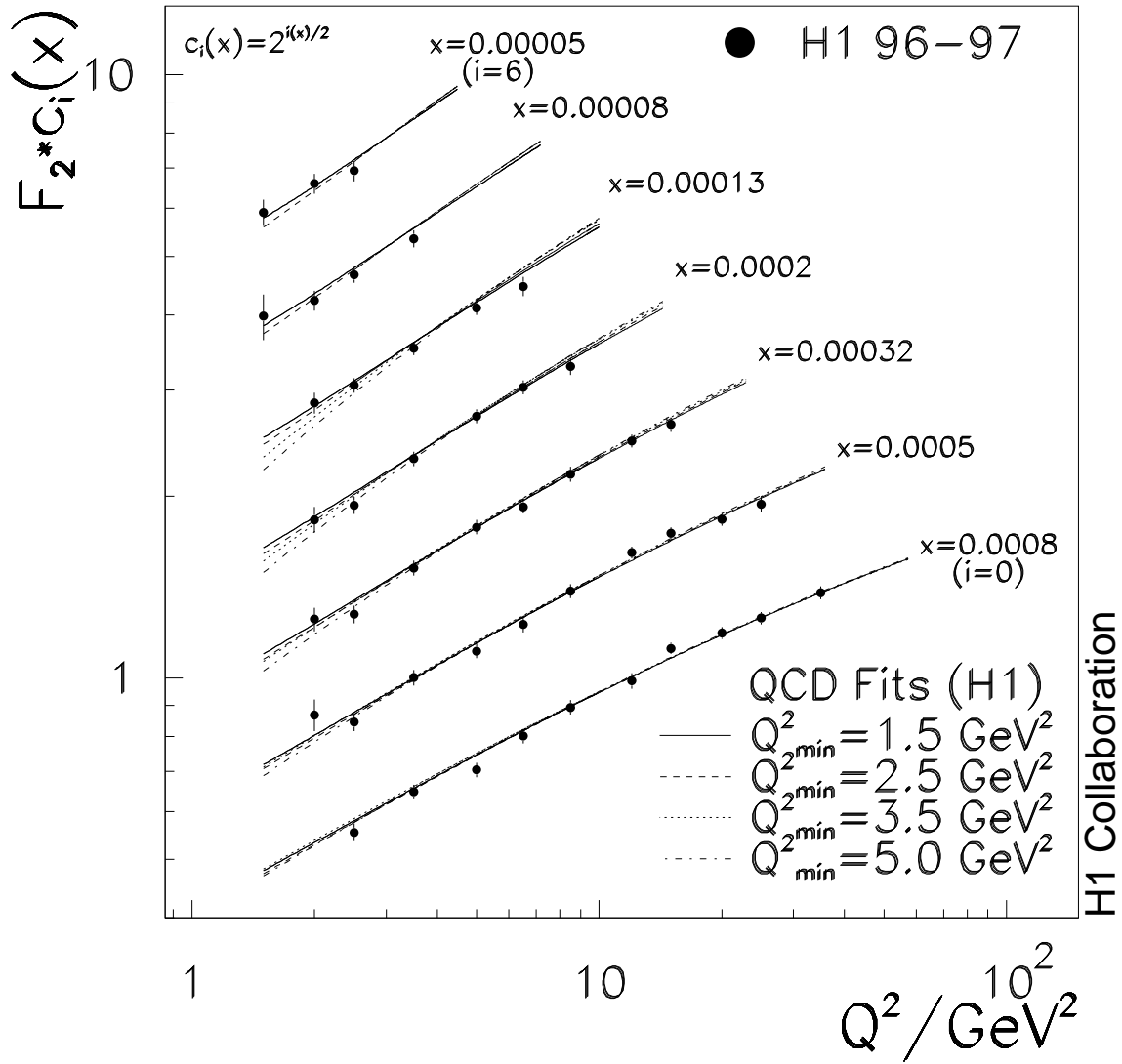,width=15cm} 
  \caption{ \sl
  Effect of the $Q^2_{min}$ cut 
  on  the structure function \Fc in  the DGLAP QCD fit to the H1
  data (points). The curves represent fits with different 
  minimum $Q^2$ values. The analysis uses $Q^2_{min} = 3.5$\,\gv
  as default.
  } 
  \protect\label{fsqmin}
\end{figure}

\begin{figure}
 \epsfig{file=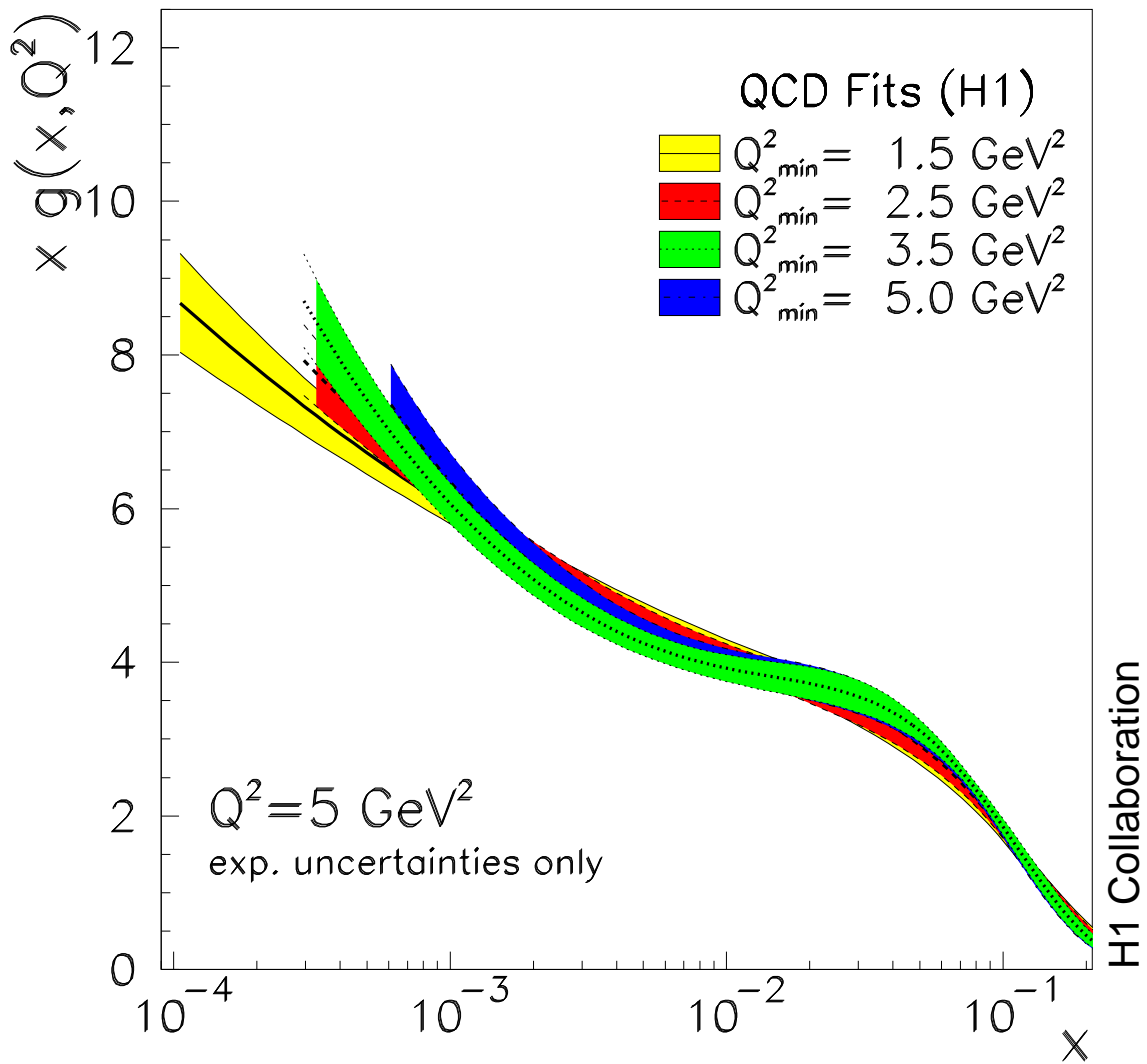,width=15cm}
  \caption{ \sl
  Effect of  the $Q^2_{min}$ cut, applied in
  the DGLAP QCD fit to the H1 data,  on 
  the gluon distribution at $Q^2=5$~\gv.
  The distributions are shown down to $x$ values corresponding to twice the
  minimum $x$ values of the data  which
  allow a $Q^2$ slope to be measured.}
  \protect\label{gsqmin}
\end{figure}

The $x$ range is restricted at small $x$ by the choice of $Q^2_{min}$.
An extension of the analysis to low values of $Q^2$ and $x$ is of
interest to study possible deviations from NLO DGLAP evolution.  The
dependence of the fit result on the chosen $Q^2_{min}$ is studied
systematically. Figure~\ref{fsqmin} shows the H1 $F_2$ data for $x \le
8 \cdot 10^{-4}$ together with the fit curves for different values of
$Q^2_{min}$. The fit with $Q^2_{min}=1.5$~\gv describes all the data
very well. If $Q^2_{min}$ is raised, the fit curves extrapolated below
$Q^2=Q^2_{min}$ tend to undershoot the data excluded from the fit.
The gluon distributions at $Q^2=5$~\gv obtained from these fits are
shown in figure~\ref{gsqmin} in the low $x$ range where the gluon
distribution is constrained\footnote{ The measurement of the slope
  \pdff requires at least two data points with different $Q^2$ above
  $Q^2_{min}$ for fixed x. Thus the minimum value $x_{min}$ at which
  this slope can be measured depends on $Q^2_{min}$.  It has been
  shown in~\cite{prytz} that the derivative \pdff determines the gluon
  distribution at a value of approximately $2x$. The gluon
  distributions in figure~\ref{gsqmin} are therefore shown only down
  to $x \simeq 2x_{min}$.}. They are consistent within the estimated
uncertainty in the overlapping regions. Extension of this study into
the region of $Q^2 \simeq 1$\,\gv is of interest. It requires
precision data in a large range of $x$. For such $Q^2$ values the
gluon distribution \xg, in twist 2 NLO QCD, is observed to
approximately vanish at low $x$ aquiring a valence-like shape.

\begin{figure}
 \epsfig{file=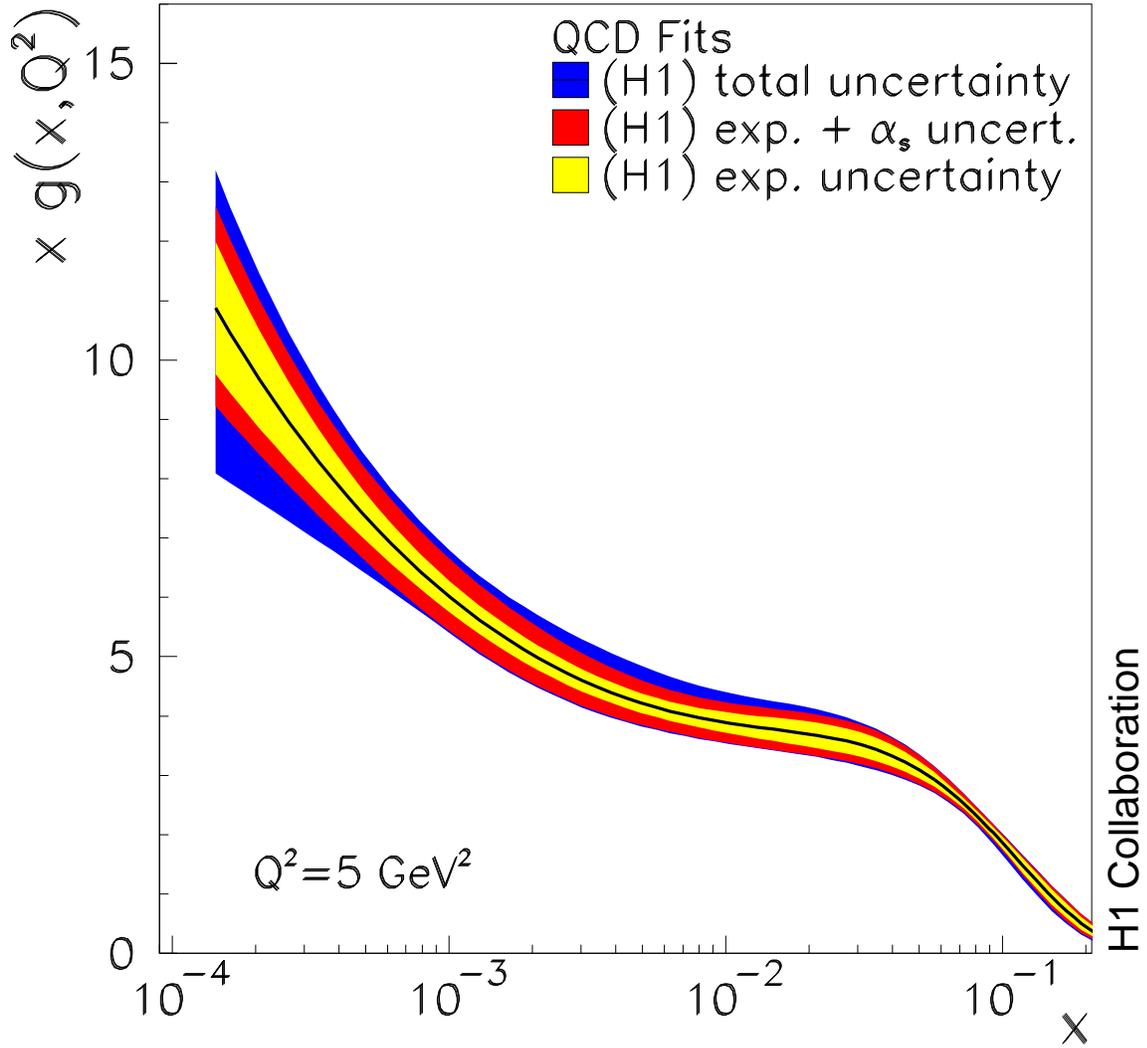,width=15cm}
  \caption{ \sl
  The gluon distribution \xg at  $Q^2=5$~\gv, determined in the NLO 
   DGLAP QCD fit to the H1 data. Inner error band: experimental
  uncertainty; middle error band: effect of the experimental error
  and of the \amz uncertainty of $\pm 0.0017$; outer error band:
  effect of experimental, \as and model uncertainties.
} 
  \protect\label{h1glu}
\end{figure}

The fit to the H1 data determines \amz to be $0.115$ with an
experimental error of $\pm 0.005$ and an optimum $\chi ^2$ of 180 for
224 degrees of freedom.  This is the first measurement of \as with
HERA inclusive cross section data alone.  The result for the gluon
distribution obtained from the H1 measurements is shown in
figure~\ref{h1glu}.  The innermost error band is due to the
experimental measurement uncertainty, which for $3 \cdot 10^{-4} \leq
x \leq 0.1$ is about 5\% and decreases to about 3\% at $Q^2 =
20$\,\gv.  The middle error band illustrates the effect of an \as
uncertainty of $\pm 0.0017$, which is derived in the fit to the
combined H1 and BCDMS data (see below). The outer error band includes
in addition the uncertainty of the QCD model as described
subsequently. For values of $x > 0.1$ this analysis is not able to
reliably determine the shape of the gluon distribution because in this
range the scaling violations are dominated by quark radiation rather
than gluon radiation. Yet, the integrated momentum fraction carried by
gluons at large $x$ is constrained by the momentum sum rule as
discussed above.  The fit determines this contribution to be
$\int_{0.1}^{1} xg(x,Q^2) dx = 0.13 \pm 0.04~(exp)$ for $Q^2=4$\,\gv.

\begin{figure}
 \epsfig{file=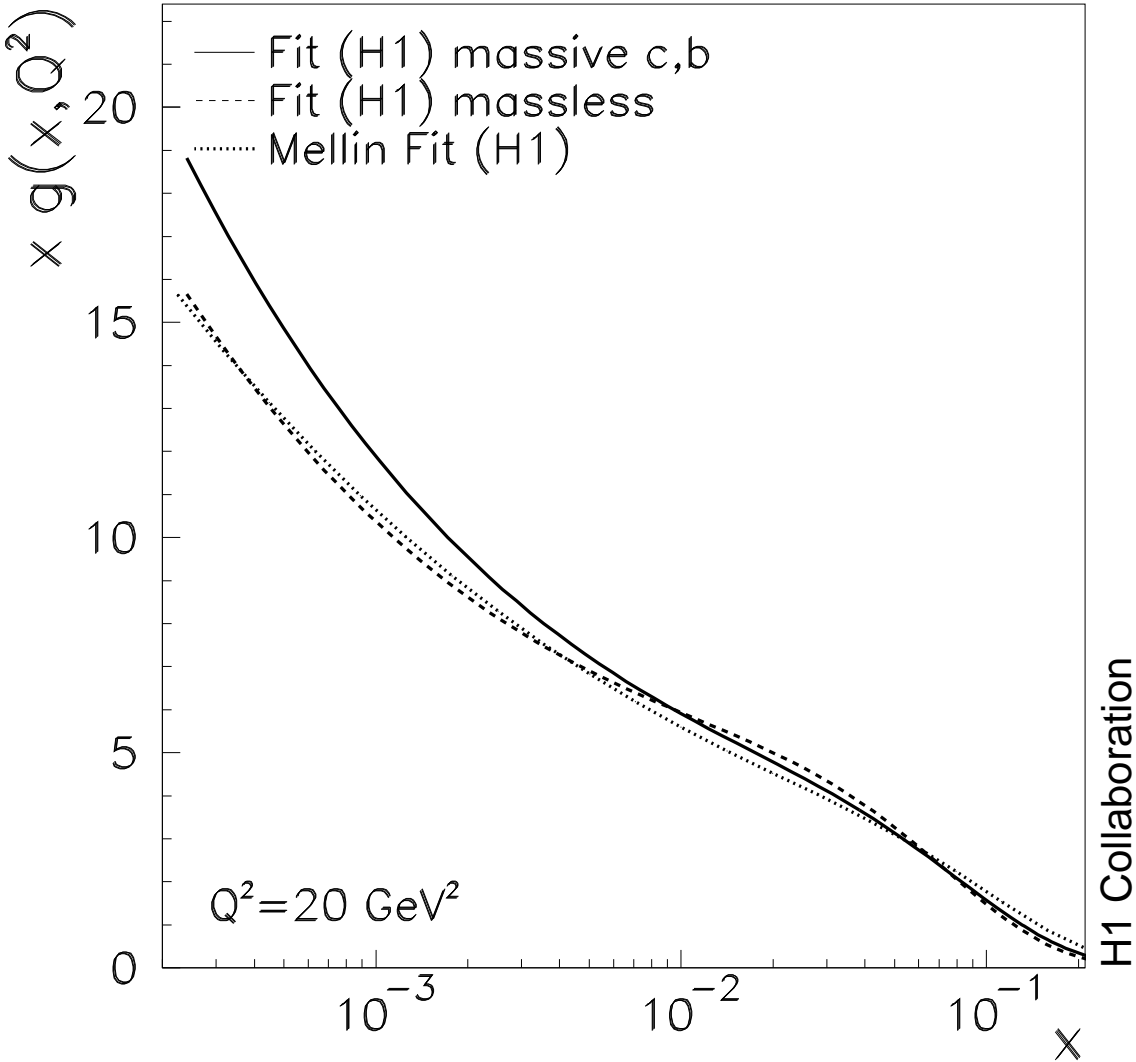,width=15cm}
  \caption{ \sl
Comparison of  gluon distributions 
 obtained in NLO DGLAP QCD fits to the H1 data,
  using different prescriptions: solid curve: standard fit using the
  massive heavy flavour scheme; dashed curve: fit in the massless
  scheme; dotted curve: fit in the massless
  scheme using a Mellin $n$  space program.}
  \protect\label{gmellin}
\end{figure}

The result for the gluon distribution depends on the theoretical
framework since \xgc  is not an observable. For example, if
the massive quark description for charm and beauty production is
replaced by the massless treatment of heavy quarks, the gluon
distribution changes as illustrated in figure~\ref{gmellin}. The gluon
distribution in the massless fit is about 15\% lower at small $x$
compared to the standard result.  A consistent cross check of this
massless fit result is obtained with a Mellin $n$ space
program~\cite{mellin}.

The analysis of the longitudinal structure function $F_L$ (section
5.2) uses a QCD fit to the reduced cross section for $y \leq 0.35$.
This fit follows exactly the same procedure as that described above.
It results in a $\chi^2$ of 151 for 180 degrees of freedom and agrees
very well with the full fit. In particular, the gluon distribution
obtained in this fit is nearly indistinguishable from $xg$ obtained in
the standard fit covering the full $y$ range which is sensitive to
\FLc. Thus, \xg appears to be determined by the scaling violations of
\F.
\subsection{Simultaneous Determination
 of  {$\bf \alpha_s(M_Z^2)$} and  the Gluon Distribution  \label{secas}}
\begin{figure}[tbp]
 \epsfig{file=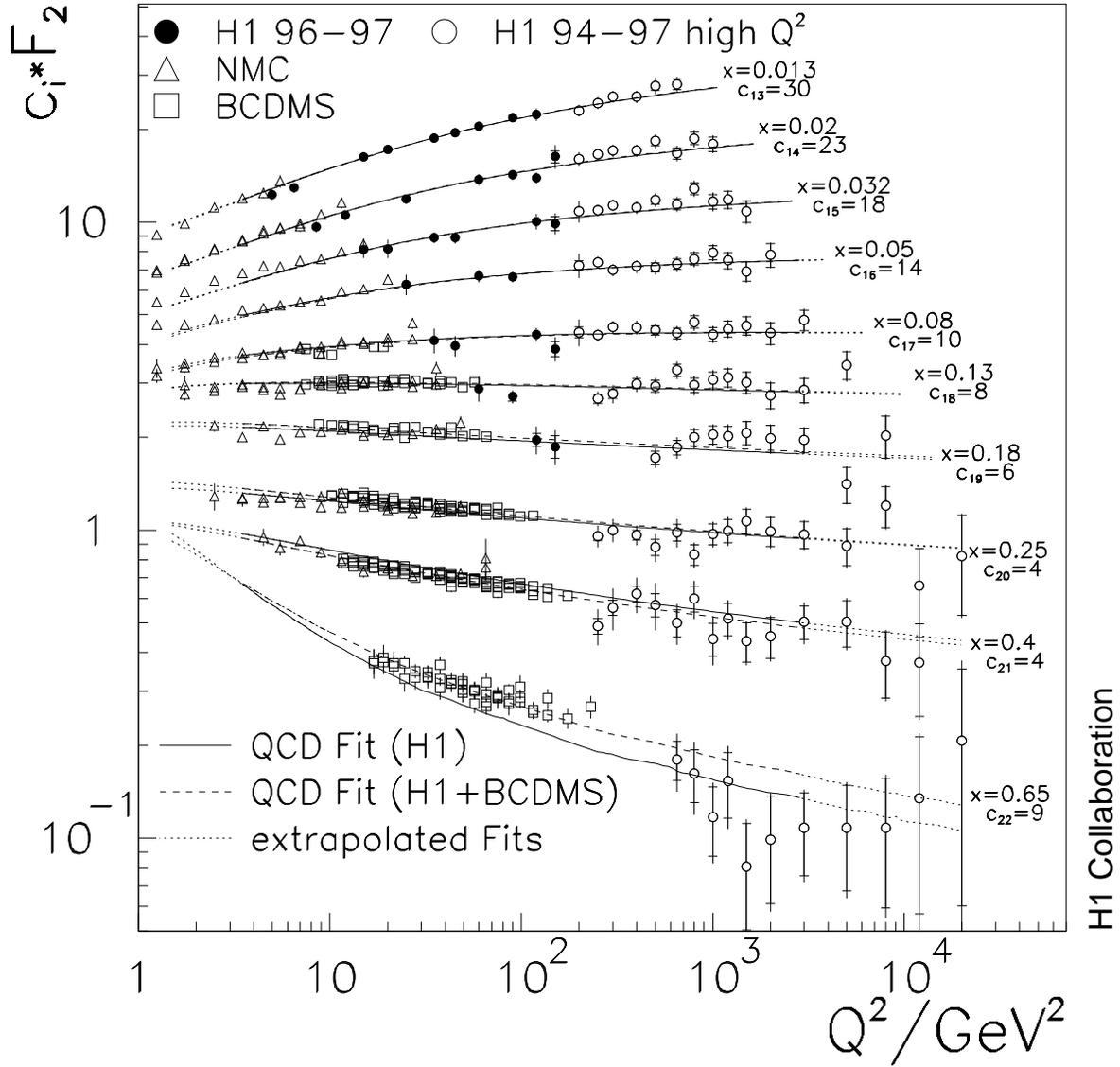,width=15.5cm}
  \caption{ \sl
    Measurements of the proton
    structure function \F
    by the H1 experiment and by fixed 
     target muon-proton scattering experiments.
    The error on the data points is the total mesurement
    uncertainty.
    The inner error bars represent the statistical
    error.  
     Solid curves: fit to the H1
     cross section data.  
    Dashed curves: fit to the H1 $ep$ and BCDMS $\mu p$ data,
    for $y_{\mu} >  0.3$.
  Dotted curves: extrapolations
    to  data  not used in the fit.
 } \protect\label{f2qb}
\end{figure}

%
%
The precision of the large $x$, high $Q^2$ H1 data~\cite{hiq} is not
sufficient to enable a competitive determination of \amz and of the
gluon distribution simultaneously from the H1 data alone. The most
precise measurement of the DIS inclusive cross section at large $x$
was obtained by the BCDMS $\mu p$ scattering experiment~\cite{BCDMS}
(figure~\ref{f2qb}). These data are therefore combined with the
H1 measurements.

In a first step, a fit is performed to the complete data sets.  The
correlated systematic errors of the data are fitted, together with the
other parameters.  Regions of data are identified in which the fit
causes large systematic shifts of the data points.  For the BCDMS data
in the range $y_{BCDMS} =y_{\mu} < 0.3$ the data points are shifted by
more than the quoted systematic error. The low $y$ region in this
experiment is particularly strongly affected by the energy scale
uncertainty of the scattered muon, which leads to correlated
systematic errors $\propto 1/y$. These become large at low $y$ for
each of the four data sets at different muon energy.  Note that the
low $y$ data of BCDMS differ from measurements of the $ep$ scattering
cross section at SLAC~\cite{slacd, mrs} in the region of overlap. In
this region the BCDMS data accuracy is dominated by systematic errors
while the SLAC measurement is statistically limited.  This also
suggests the presence of large systematic effects in the low $y$
region of the BCDMS data which were studied previously~\cite{ast}.
Therefore in all analyses only BCDMS data with $y_{\mu} > 0.3$ are
used.

The result of the QCD fit to the combined H1 and BCDMS data sets,
leaving \as as a free fit parameter, is shown in figures~\ref{f2qa}
and ~\ref{f2qb} as dashed curve. It describes the data very well with
a $\chi^2/$dof of 394/451. The fit to the H1 and BCDMS data is nearly
indistinguishable from the fit to the H1 data alone, except for the
two highest $x$ bins.  The parameterisations used in the fit to the H1
and BCDMS data are given in equation~\ref{eqparah1}, and the
parameters are summarised in table~\ref{h1fitpar}. The choice of these
particular shapes results from a detailed analysis of the behaviour of
the $\chi^2$ function, similar to the fit to the H1 data which is
described in Appendix~\ref{apara}.
\begin{table}[h]
  \begin{center}
    \begin{tabular}{|l|r|r|r|r|r|}
\hline
           &     a &     b &    c &    d  &    e   \\
\hline
    gluon  & 1.10  &  -0.247 & 17.5 &  -4.83  & 68.2 \\

\hline
     V      &  86.3 & 1.47  & 4.48 & -2.12 &  1.60 \\

\hline
     A       & 0.229  & -0.130 & 19.7 & -3.82 & 29.8 \\
\hline
    \end{tabular}
    \caption { \sl Parameters of the input distributions 
      $xq(x) = a_qx^{b_q}(1-x)^{c_q}[1+d_{q}\sqrt{x} +e_{q}x]$ for
      \xg, \V and \A at the initial scale $Q^2_0=4$~\gv using H1 and
      BCDMS data for $Q^2 \geq 3.5$~\gv and $y_{\mu} > 0.3$.  A fit
      with $d_g = e_g = 0$ yields $c_g=6.5$, not far from the
      dimensional counting rule expectation~\cite{cglue}, yet with a
      worsened $\chi^2$.}
    \label{h1fitpar}
  \end{center}
\end{table}

The \as value obtained in the NLO analysis of the H1 and BCDMS proton
data is
\begin{equation}
\alpha_s(M_Z^2) = 0.1150~~\pm~~0.0017~(exp)~~^{+~~0.0009}_{-~~0.0005}~(model).
 \label{eqalf}
\end{equation}  
In this combined fit both data sets consistently favour a value of
\amz~$\simeq 0.115$ with comparable accuracy. The first error
represents the experimental uncertainty of the data sets.  The second
error includes all uncertainties associated with the construction of
the QCD model for the measured cross section. These are summarised in
table~\ref{dalf}.  The strong coupling constant is defined here by the
solution of the renormalisation group equation to order $\alpha_s^3$.
In the double logarithmic approximation the value for \amz is
calculated to be lower by 0.0003.

\begin{figure}
  \epsfig{file=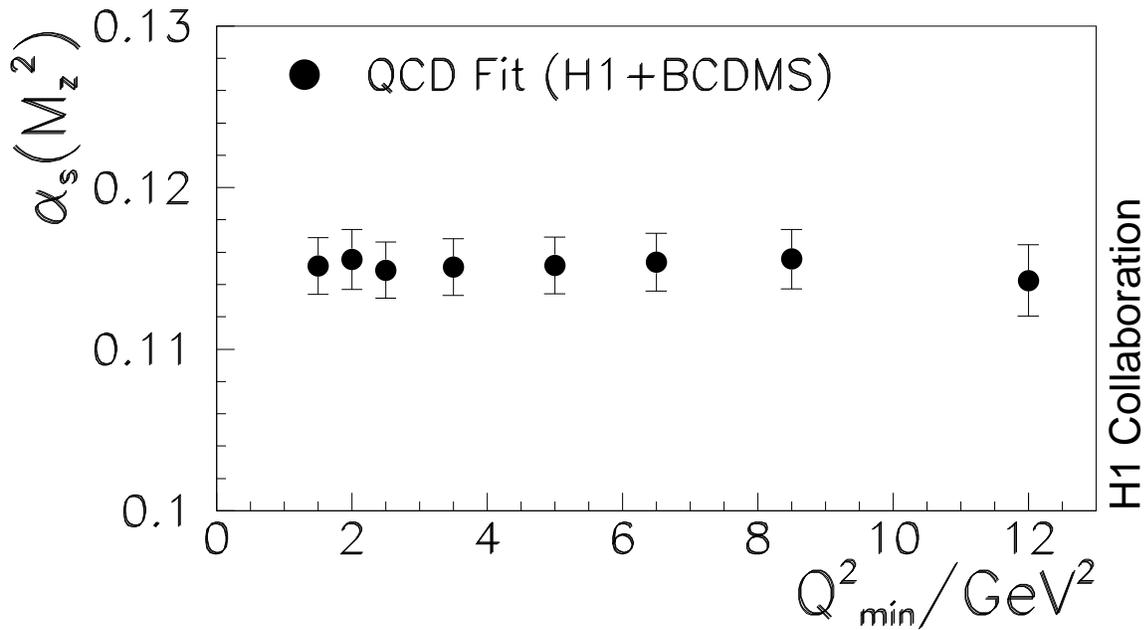,width=15cm}
  \caption{\sl
   Dependence of \amz obtained in fits to the H1 and BCDMS data 
   on the minimum $Q^2$ value used.
   The error bars denote the experimental uncertainty of
   \amz. Note that 
   the BCDMS data have an intrinsic $Q^2_{min}$
   of 7.5~\gv and are limited in this analysis to $y_{\mu} \geq 0.3$
   (see text). An
   increase of $Q^2_{min}$ 
   implies that the minimum $x$  rises correspondingly,
   i.e. from $x = 3.2 \cdot 10^{-5}$ at $Q^2_{min} = 1.5$\,\gv
   to  $x  = 8 \cdot 10^{-4}$ at $Q^2_{min} = 12$\,\gv.}
  \protect\label{alfcont}
\end{figure}

The value obtained for \amz is nearly independent of $Q_0^2$ and of
the chosen parameterisation for the large set of input distributions
considered in Appendix \ref{apara}. Residual effects are included in
the estimation of the systematic error on \as.  The dependence of \as
on $Q^2_{min}$ is shown in figure~\ref{alfcont}.  No systematic trend
is observed.  Note that the BCDMS data are limited to $Q^2 \geq
7.5$~\gv, such that a choice of $Q^2_{min}$ below this value affects
the low $x$ H1 data only.
\begin{table}[h]
  \begin{center}
    \begin{tabular}{|l|l|l|}
\hline
   analysis uncertainty & +$\delta$~\as &  -$\delta$~\as \\
\hline
    $ Q^2_{min} = 2$~\gv&    &  0.00002   \\
    $ Q^2_{min} = 5$~\gv      &  0.00016 &   \\ 
    parameterisations           &  0.00011 &  \\ 
    $ Q^2_{0} = 2.5$~\gv      & 0.00023  & \\
    $ Q^2_{0} =   6$~\gv      &  & 0.00018   \\
    $y_e <$ 0.35               &  0.00013 & \\
    $x     <$ 0.6                &  0.00033 & \\
    $y_{\mu} >$ 0.4            & 0.00025   &   \\
    $x     > 5 \cdot 10^{-4}$     &  0.00051 & \\
    uncertainty of $\bu - \bd$       & 0.00005 & 0.00005   \\
    strange quark contribution $\epsilon=0$  & 0.00010 & \\ 
    $ m_c + 0.1$\,GeV            & 0.00047 &  \\
    $ m_c - 0.1$\,GeV         &  & 0.00044   \\
    $ m_b + 0.2$\,GeV            & 0.00007 &  \\
    $ m_b - 0.2$\,GeV         &  & 0.00007   \\
\hline
total uncertainty & 0.00088   &  0.00048 \\
\hline
    \end{tabular}
    \caption{ \sl Contributions to the error of \amz
     in the analysis of H1 $ep$ and BCDMS $\mu p$
     data which are due to the selection of data and to the 
     fit assumptions.}
    \label{dalf}
  \end{center}
\end{table}

\begin{figure}[tbp]
  \epsfig{file=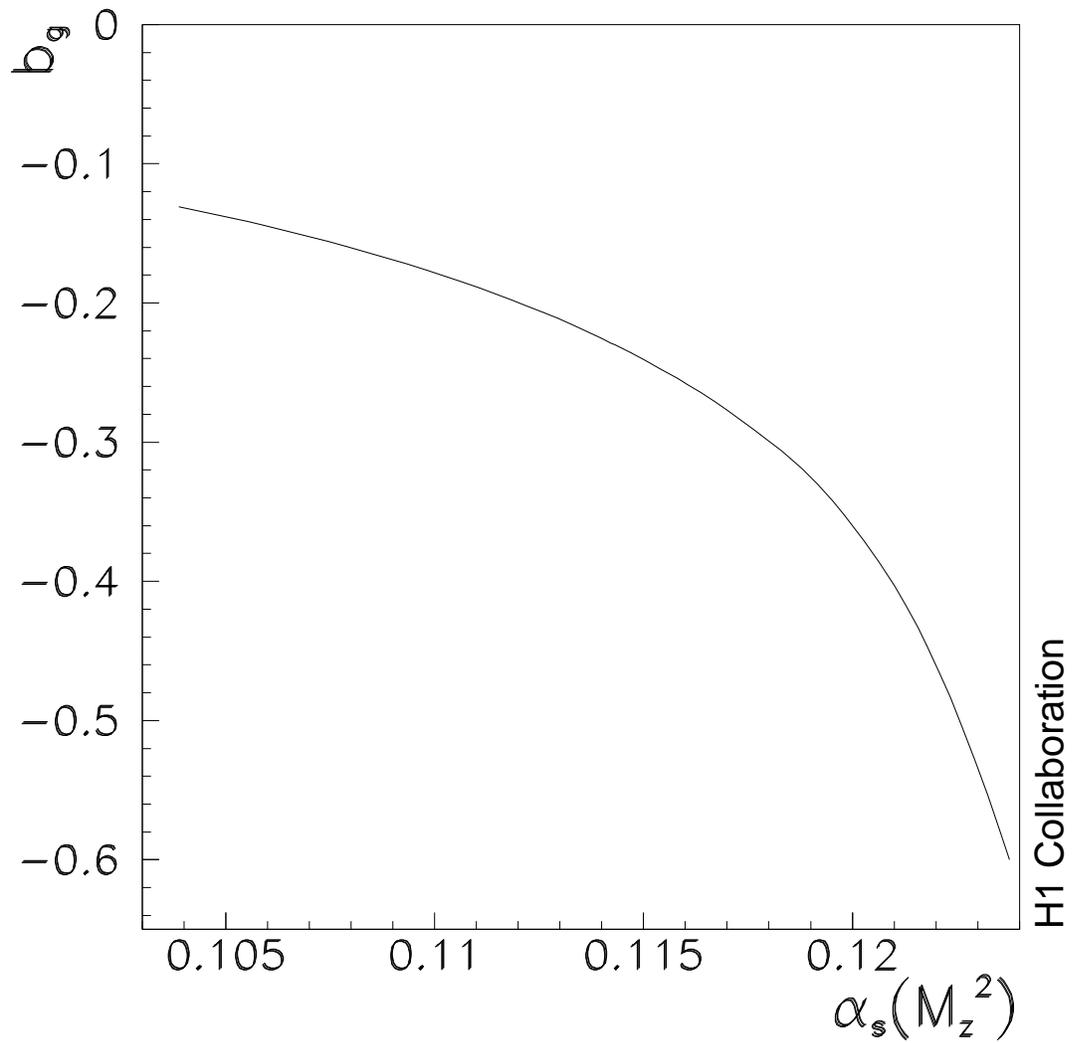,width=14cm}
  \caption{\sl The correlation of the gluon distribution parameter
  $b_g$  with $\alpha_s$ in the fit to
  H1 and BCDMS data. The parameter $b_g$
  governs the low $x$ behaviour of $ xg \propto x^{b_g}$.}
  \protect\label{bcalf}
\end{figure}

\begin{figure}[tbp]
 \begin{picture}(150,150)
 \put(0,0){
\epsfig{file=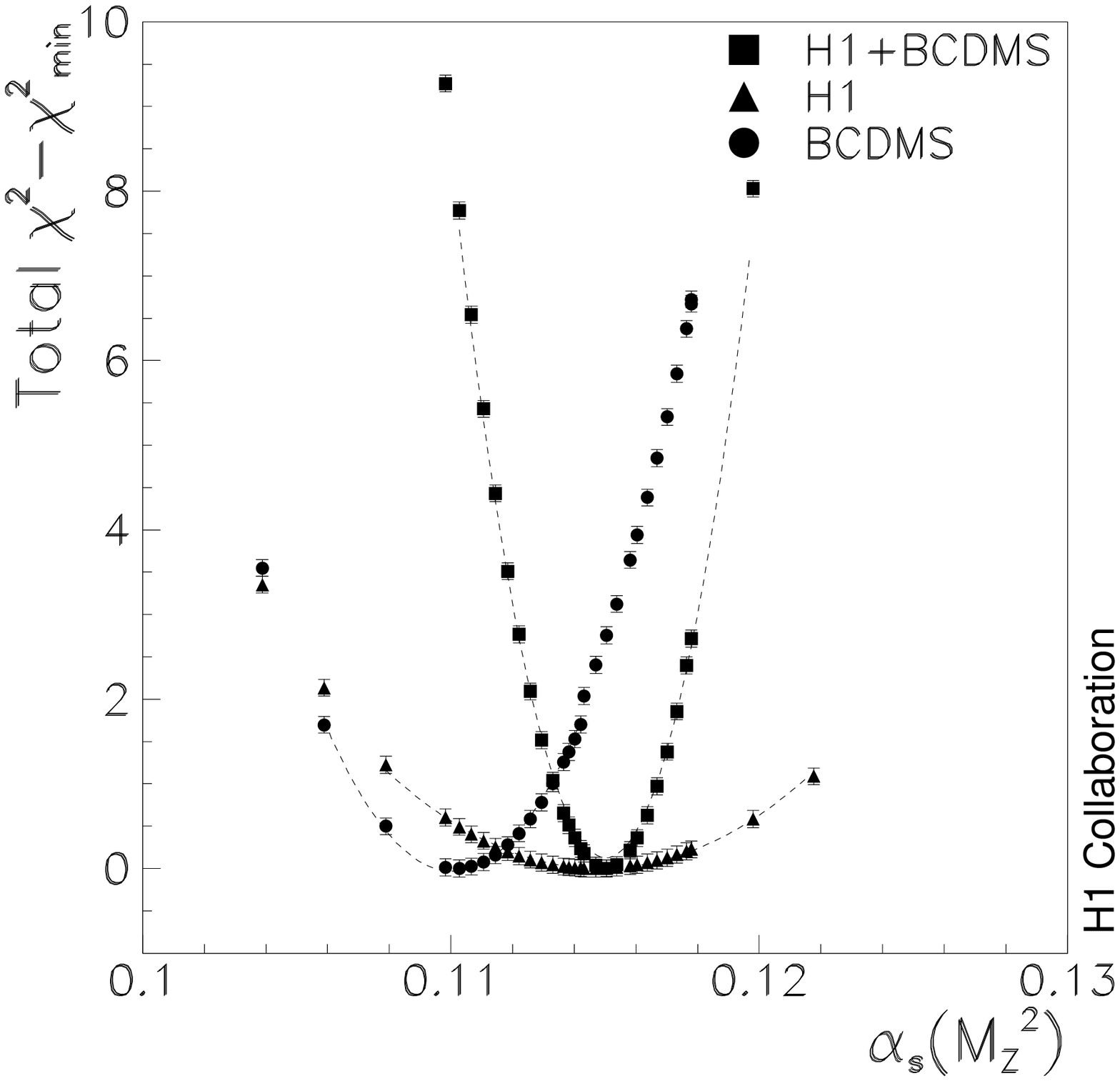,width=7.5cm}}
 \put(20,50){a)}
 \put(75,0){ 
\epsfig{file=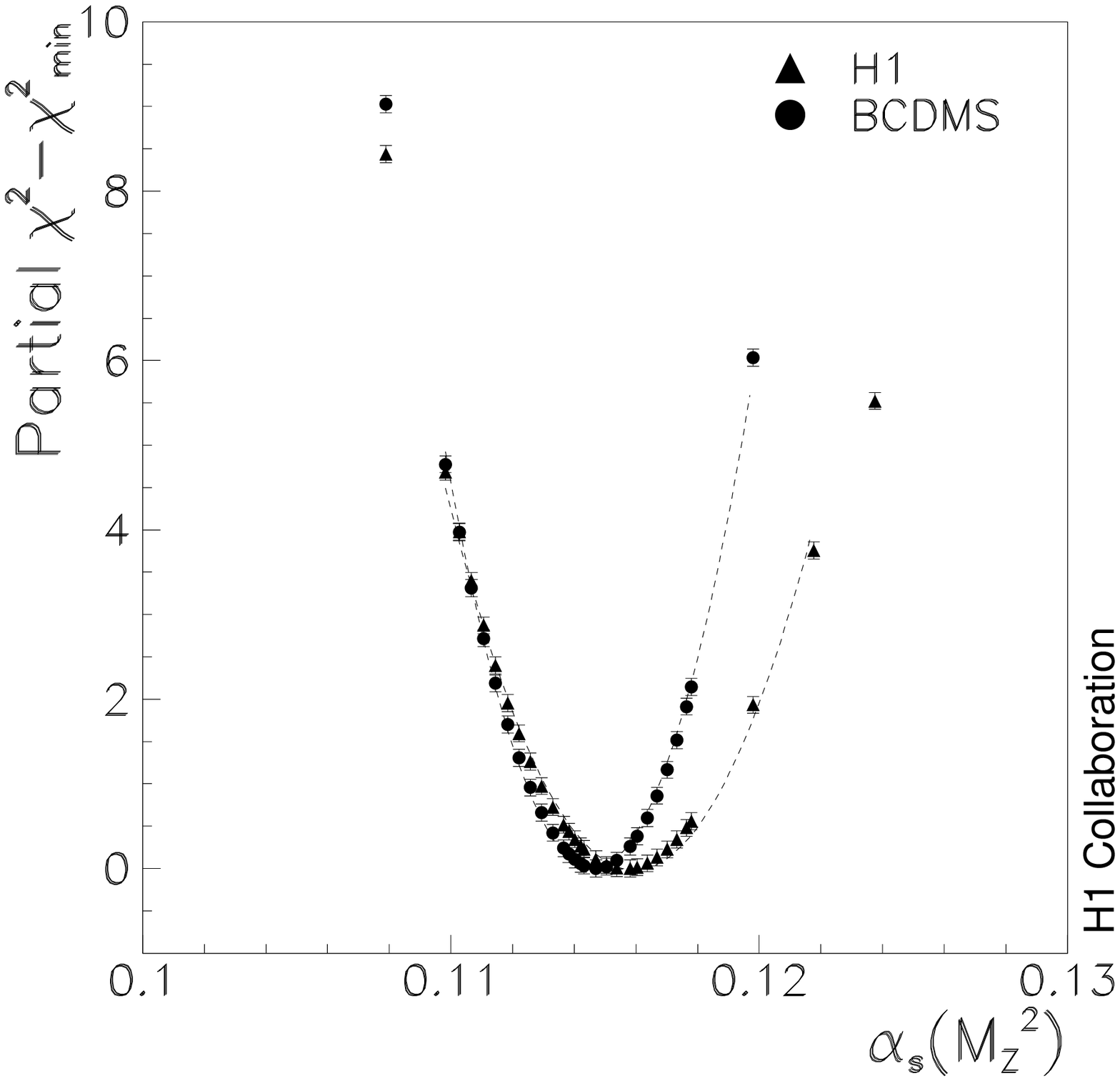,width=7.5cm}}
 \put(95,50){b)}
 \end{picture}
  \caption {\sl
   Determination of the strong coupling constant \amz in NLO
   DGLAP QCD.
   a) total $\chi^2$ for fits to the H1 $ep$ and BCDMS $\mu p$ data 
   ($y_{\mu} > 0.3$) separately and for the  fit using
   data of the two experiments combined; 
   b)  partial $\chi^2$ contributions of the 
   H1 and BCDMS proton data in the fit to determine $\alpha_s$
   using both experiments.}
  \protect\label{allalf}
\end{figure}

The combination of low $x$ data with high $x$ data constrains the
gluon distribution and \as.  A correlation is observed
(figure~\ref{bcalf}) between \as and the parameter $b_g$, which
governs the shape of the gluon distribution at low $x$
(equation~\ref{eqparah1}).  In a fit to the BCDMS data alone, for
$y_{\mu} > 0.3$ and using $xg = a x^b (1-x)^c$, a value of \amz~$=
0.111 \pm 0.003~(exp)$ is found~\footnote{The requirement $y_{\mu} >
  0.3$ causes \amz to increase by about 0.004 in the fit to the BCDMS
  data only, and by about 0.002 in the fit to the H1 and BCDMS data
  combined.}, and $b_g$ is positive. A positive value of $b_g$ implies
that $xg(x,Q_0^2)$ falls as $x$ decreases. An early \as
analysis~\cite{mv}, in the absence of detailed information about the
low $x$ behaviour of \xgc, assumed $b_g=0$.  A positive or zero value
of $b_g$, for $Q_0^2 \geq 4$\,\gv, is incompatible with analyses of
the HERA DIS data at low $x$. The fit to the BCDMS data, when
complemented with low $x$ H1 data, leads to a negative $b_g$
parameter, and therefore a larger value of \amz is obtained with the
BCDMS data than hitherto. The results of these fits to the H1 and
BCDMS data are shown in figure~\ref{allalf}a.  In the combined fit
both data sets give a consistent and comparable contribution to the
error on \as. This is illustrated in figure~\ref{allalf}b.

A rather large theoretical uncertainty of the NLO analysis results
from the choices of the renormalisation scale $\mu_r^2 = m_r \cdot
Q^2$ (equation~\ref{renorm}), and of the factorisation scale $\mu_f^2
= m_f \cdot Q^2$ which leads to scale dependent parton distributions.
In the $\overline{MS}$ scheme both scales are set equal to $Q^2$, i.e.
$m_r=m_f=1$.  In the absence of a clear theoretical prescription, the
effect of both scales on \as is estimated by varying the scale factors
$m_r$ and $m_f$ between 0.25 and 4.  The results are summarized in
table~\ref{scales}.
\begin{table}[h]
  \begin{center}
    \begin{tabular}{|l|l|l|l|l|}
\hline
          &     $m_r = 0.25$      &   $m_r = 1$    &     $m_r =  4$     \\
\hline
 $m_f = 0.25 $     &  $-0.0038 $     & $-0.0001 $    &   $  +0.0043 $   \\
 $m_f = 1 $        &  $-0.0055 $    &  $      --     $    &   $  +0.0047 $   \\
 $m_f = 4 $       &   $ -- $    &  $+0.0005 $    &   $  +0.0063 $   \\
\hline
    \end{tabular}
    \caption{ \sl Dependence of  \amz on the renormalisation
     and factorisation scales $m_f$ and $m_r$, respectively,
     expressed as the difference of \amz obtained for 
     scales different from one and the central value of \amz=0.1150.
    The combination $m_f=4$ and $m_r=0.25$
    is abandoned since the splitting function term $\propto \ln{(m_r/m_f)}^2$ 
    becomes negative at low $Q^2$ which causes a huge
    increase of $\chi^2$.} 
    \label{scales}
  \end{center}
\end{table}
In agreement with previous studies~\cite{andreas} it is found that the
renormalisation scale causes a much larger uncertainty on \amz than
the factorisation scale. Depending on which set of $m_r$ and $m_f$ is
chosen, the obtained $\chi^2$ differs by several units. This suggests
that the assumed variation of the scales is too large. These scales,
however, are not considered to represent physics quantities which may
be determined in the minimisation procedure. The estimated overall
uncertainty of about 0.005 on \amz is much larger than the
experimental error. It is expected to be significantly reduced when
next-to-NLO calculations become available~\cite{nnlo, anleiden}.
Recently an \as analysis of moments of structure functions, measured
in charged lepton-nucleon scattering, was presented extending to NNLO
QCD~\cite{yndu}.

The stability of the fit results is checked further with respect
to possible changes in the analysis procedure:
\begin{itemize}
\item{
 The value of \amz  increases by 0.0005
 if the correlation due to systematic errors is neglected, i.e.
 if the correlated systematic error parameters are not part of the
  minimisation (see Appendix~\ref{achi2}).}
\item{
In the present analysis, the relative normalisations of the
data sets are left free.  The change imposed by the fit to
the BCDMS data is about $-1.5$\% within a total normalisation
uncertainty of 3\%. The H1 data are moved by about 1\% within the
experimental error of 1.7\%.  Thus the selected 
H1 and BCDMS data are compatible with each other.
If the fit is repeated with all normalisations
fixed then $\chi^2$ increases by 26, and \amz increases by 0.0005.}
\item{
 If the BCDMS data is replaced by data of the NMC collaboration \cite{nmc},
 imposing the low  $Q^2$ limit of the BCDMS data, a consistent value
 of   \amz$ = 0.116 \pm 0.003~(exp)$ is obtained.}
\item{If the heavy flavour treatment is changed and a massless,
 four flavour fit performed, \amz is enlarged by +0.0003.}
\item{ The addition of the BCDMS deuteron target data, with $y_{\mu} > 0.3$,
    to the H1 and BCDMS proton data yields \amz = 0.1158 $\pm 0.0016~(exp)$,
 i.e. \amz increases by 0.0008. In this analysis nuclear
    corrections~\cite{melni} are applied, and the conventional flavour
    decomposition into valence and sea quarks is used.}
\end{itemize}

\begin{figure}[tbp]
 \epsfig{file=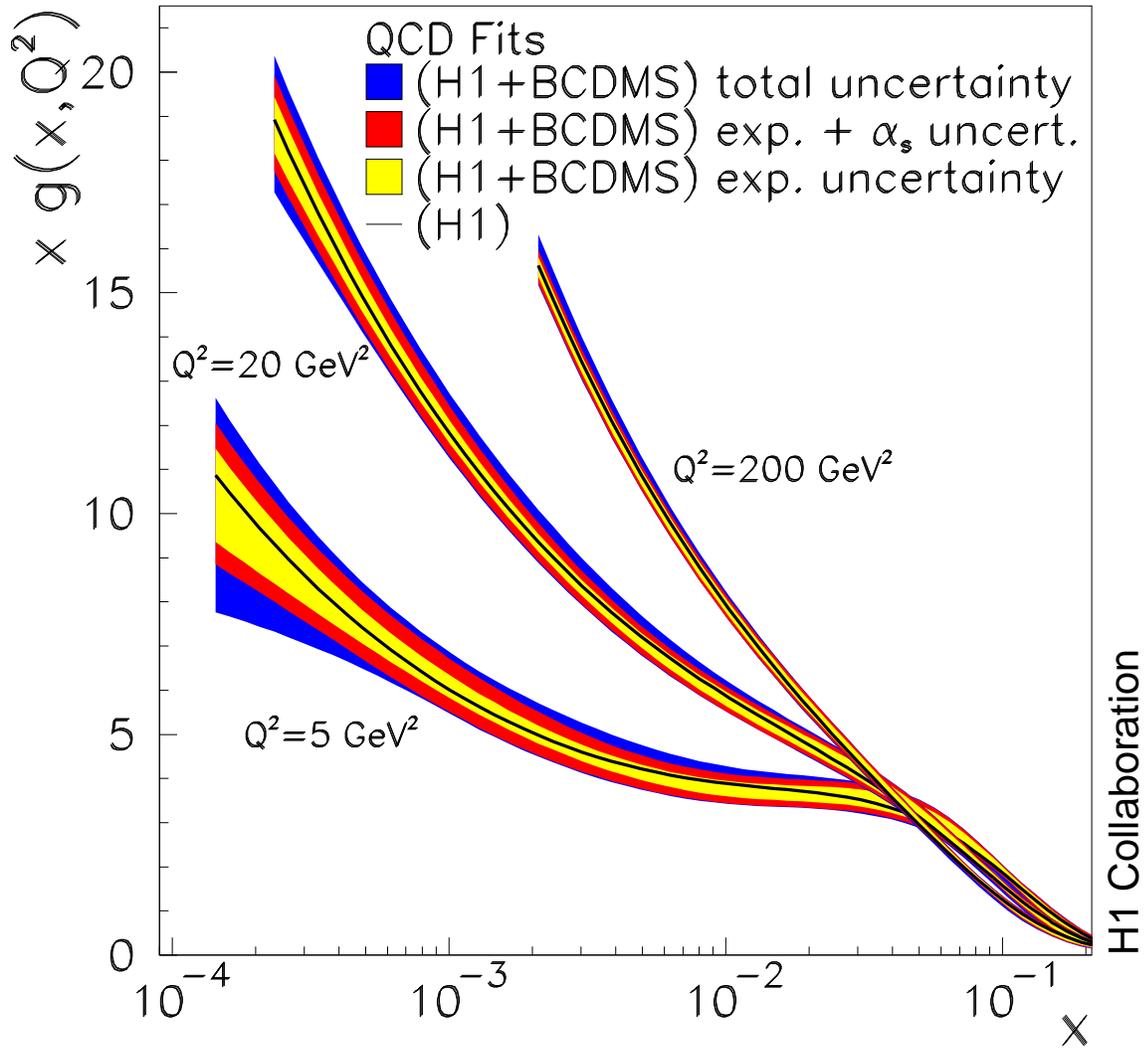,width=15cm}
  \caption{ \sl
  Gluon distribution
  resulting from the NLO DGLAP QCD fit 
    to H1 
   $ep$ and  BCDMS $\mu p$ cross section data
   in the massive heavy flavour scheme.
   The innermost error bands represent the experimental
   error for fixed \amz=0.1150. The middle error bands include in addition
   the contribution due to the simultaneous fit of \as.
   The outer error bands also include  the  
   uncertainties related to the QCD model and data range.
   The solid lines inside the error band
   represent the gluon distribution obtained in the fit
   to the H1 data alone. }
  \protect\label{h1gluon}
\end{figure}

The gluon distribution from the fit to the H1 and the BCDMS proton
cross section data is shown in figure~\ref{h1gluon} for $Q^2=5$\,\gv.
The inner error band represents the experimental uncertainty of the
determination of $xg$ for \as fixed. This fit, however, simultaneously
determines \xg and \as, which leads to a small increase of the
experimental error of \xgc as is illustrated by the middle error band.
The full error band includes in addition the uncertainties connected
with the fit ansatz, as listed in table~\ref{dalf} for the
determination of \amz. For the low $x$ behaviour of \xg these are
dominated by the choice of $Q^2_{min}$, as is discussed in
section~\ref{h1fit}. The gluon distribution from the combined fit is
shown also for two higher $Q^2$ values, 20 and 200\,\gv.  The DGLAP
evolution leads to a gluon distribution which rises dramatically at
small $x$ with increasing $Q^2$ (figure~\ref{h1gluon}). The inner
solid line illustrates the behaviour of \xg, as determined with the H1
data alone, which is seen to be in very good agreement. The fits with
deuteron data or NMC data lead to very similar gluon distributions.
This analysis determines \xgc from the scaling violations of \Fc. It
is more accurate but  consistent with determinations of the
gluon distribution by the H1 experiment in charm~\cite{gcharm} and
deep-inelastic dijet ~\cite{dijet} production.
\section{Summary}
A new measurement of the deep-inelastic positron-proton scattering
cross section is presented for squared four-momentum transfers $1.5
\leq~Q^2~\leq 150{~ \rm GeV^2}$ and Bjorken-$x$ values $3 \cdot
10^{-5}~\leq x~\leq 0.2$ which is more precise than previous
measurements in this kinematic range.  The statistical accuracy of the
present inclusive cross section measurement is better than 1\%, for a
large part of the kinematic region. The systematic uncertainty has
decreased to about 3\%, apart from the edges of the covered range.
This is due to improved detectors in the backward region used for the
identification and measurement of the scattered positrons at low
$Q^2$.  These are an electromagnetic calorimeter (SPACAL) with very
good spatial and energy resolutions, a drift chamber (BDC) and a
silicon tracker (BST).

The kinematic range is extended down to $y = 0.004$ such that the
present data overlap kinematically with the measurements of
muon-proton scattering experiments. This is achieved with higher
statistics, improved event vertex reconstruction and calibration of
the forward parts of the calorimeter. The present data agree with the
$\mu p$ data in the region of overlap within the accuracy of about
7\%.  The kinematic range is extended also up to $y = 0.82$ using
track reconstruction in front of the SPACAL in an extended angular
range.

The cross section measurement is used to determine derivatives with
respect to $\ln y$ and to $\ln Q^2$ as functions of $x$ and $Q^2$.
The partial derivative \pdff is measured in the full $x, Q^2$ range of
this measurement. When considered as a function of $x$ at fixed $Q^2$,
for $3 \le Q^2 \le 40$~\gv, it is observed to rise continuously
towards low $x$ in agreement with QCD.

The partial derivative of the reduced cross section $(\partial
\sigma_r / \partial \ln y)_{Q^2}$ is used to extract \FLc at low $Q^2
< 10$\,\gv. This is complemented by a determination of \FLc at $Q^2 >
10$\,\gv using the difference between the measured reduced
cross section, $\sigma_r$, and \Fc calculated from an extrapolation of
a NLO QCD fit to low $y$ data.  Thus the longitudinal structure
function \FLc at low $x$ is measured more precisely than hitherto, and
in a larger $Q^2$ range.

A detailed, systematic analysis is presented of the structure function
data using the DGLAP evolution equations in NLO. The salient features
of this analysis are a new formalism for charged lepton-proton
scattering and a comprehensive study of the influence of model
parameters and parton distribution parameterisations on the results.
These are obtained from a minimum number of data sets with special
emphasis on the treatment of correlated systematic errors.

The scaling violations of \Fc, the behaviour of the derivatives and of
the longitudinal structure function at low $x$ are found to agree with
DGLAP QCD. The present precise low $x$ data, when combined with high
$Q^2$ data of H1 from the same running period, determine the shape of
the gluon distribution at small $x$. The gluon distribution \xg is
determined at $Q^2=20$\,\gv to an experimental accuracy of about 3\%
in the kinematic range $3 \cdot 10^{-4} \leq x \leq 0.1$.

A simultaneous determination of the gluon distribution and of \as is
obtained by combining the low\,$x$ data of H1 with $\mu p$ scattering
data of the BCDMS collaboration at large $x$.  A value of the coupling
constant \amz = $0.1150 \pm 0.0017
(exp)^{+~~0.0009}_{-~~0.0005}(model)$ is obtained.  The value of \as
changes by about 0.005, much more than the experimental error, if the
renormalisation scale is allowed to vary by a factor of four, and to a
lesser extent if the factorisation scale is changed by the same
amount. This uncertainty is expected to be reduced significantly in
next-to-NLO perturbation theory.  
\vspace{1cm}

{\bf Acknowledgements}                                                         
\normalsize   
\noindent We are very grateful to the HERA machine group whose
outstanding efforts made this experiment possible. We acknowledge the
support of the DESY technical staff. We appreciate the substantial
effort of the engineers and technicians who constructed and maintain
the detector. We thank the funding agencies for financial support of
this experiment.  We wish to thank the DESY directorate for the
support and hospitality extended to the non-DESY members of the
collaboration.  We thank J.~Bl\"umlein, M.~Botje, W.~van~Neerven,
R.~Roberts, and W.-K.~Tung for interesting discussions on the QCD
interpretation of this data. We are particularly grateful to A.~Vogt
for his cooperative effort in the analysis of higher-order effects on
\as and to R.~Engel for help in the photoproduction background
simulation. 
\newpage
\begin{appendix}
\section{Details of the QCD Analysis}
\subsection{Flavour Decomposition of {\boldmath $F_2$} \label{aflavour}}
The structure function \Fc can be written as
\begin{eqnarray}\label{f2ud}
 F_2 =  \frac{4}{9} \cdot xU + \frac{1}{9} \cdot xD,
\end{eqnarray}
with $U=u+\bu$ and $D=d+\bd+s+\bs$ (equation~\ref{f2q}).
 A modified projection yields
\begin{eqnarray}\label{f2sns}
 F_2 =  \frac{2}{9} \cdot x \Sigma + \frac{1}{3} \cdot x \Delta.
\end{eqnarray}
The sum $\Sigma=U+D$ defines a singlet combination of quark distributions
which has a $Q^2$ evolution coupled to the gluon
momentum distribution \xgc. The difference $\Delta=(2U-D)/3$ defines a
non-singlet distribution which evolves independently of \xgc. Thus 
\Fc is defined by two independent quark distribution functions. 

In this analysis two specific functions \V and \A  are chosen which
are related to $U$ and $D$ according to
\begin{eqnarray}\label{UVA}
 U = \frac{2}{3} V +2 A 
\end{eqnarray}
and
\begin{eqnarray}\label{DVA}
 D = \frac{1}{3} V + 3 A.
\end{eqnarray}
The inverse relations  defining \Vc and \Ac are
\begin{eqnarray}\label{V}
 V = \frac{3}{4} (3U-2D) = \frac{9}{4} u_v -  \frac{3}{2} d_v
 +  \frac{9}{2} \bu - 3 (\bd + \bs)  
\end{eqnarray}
and
\begin{eqnarray}\label{A}
 A = \frac{1}{4} (2D-U) = \bd + \bs -\frac{1}{2} \bu
 -  \frac{1}{4} u_v +  \frac{1}{2} d_v,
\end{eqnarray}
which for the conventional assumption $\bu = \bd = 2\bs$ lead to the
relations presented in the introduction of the QCD analysis, see
section~\ref{secanapro}.  In this approximation the $V$ distribution
vanishes for small $x < 0.01$. The behaviour for large $x$ is defined
by $u_v$.  For small $x$ the function $A$ is given by the sea
distribution $A \simeq \bu$. 

Recent measurements of Drell-Yan muon pair production at the
Tevatron~\cite{handt} have established a difference between the $\bu$
and $\bd$ distributions. Charged current neutrino-nucleon experiments
determined the relative amount of strange quarks in the nucleon sea to
be
\begin{eqnarray}\label{ssbar}
 s +\bs = (\frac{1}{2} + \epsilon) \cdot (\bu + \bd),
\end{eqnarray} 
with a recent value of $\epsilon = -0.08$~\cite{nutev}.  These results
lead to modifications of the simple assumptions on the sea\footnote{
  The evolution of $s+\bs$ in DGLAP QCD is found to yield a linear
  dependence of $\epsilon$ on $ \ln Q^2$ which is used to extrapolate
  the NuTeV result~\cite{nutev}, obtained at 16\,\gv, to $Q^2 =
  Q^2_0$.}. They have been accounted for by modifying
equation~\ref{DVA} according to
\begin{eqnarray}\label{dmodi}
  D  = \frac{1}{3} V + k A, 
\end{eqnarray} 
which, using equation~\ref{UVA}, results in
\begin{eqnarray}\label{Vmod}
 V = \frac{3}{2} \cdot \frac{1}{k-1} (kU-2D) 
\end{eqnarray}
and $\Sigma = V + A \cdot (2+k)$.
Choosing $k=3+2\epsilon$ can be shown to remove the strange
contribution to the function $V$ yielding 
\begin{eqnarray}\label{Vmodq}
 V = \frac{3}{4} \cdot \frac{1}{1+\epsilon} [(3+2\epsilon) u_V
            -2d_V + (5+2 \epsilon)(\bu-\bd)], 
\end{eqnarray}
which coincides with equation~\ref{vsimp} for $\epsilon=0$ and
$\bu=\bd$.  Because the integral $\delta = \int (\bu-\bd) dx$ is
finite~\footnote{The most accurate measurement of $\int_0^1 (\bu-\bd)
  dx$ has been performed by the E866/NuSea Collaboration~\cite{handt}
  which obtained a value of $-0.118 \pm 0.011$ at $ \langle Q^2
  \rangle= 54$\,GeV$^2$.}, this choice of $k$ allows the counting rule
constraint (equation~\ref{intV}) to be maintained as
\begin{eqnarray}\label{Vintmod}
 \int_0^1{V}dx = 3 + \delta \cdot \frac{3}{4} \cdot
                \frac{5+2\epsilon}{1+\epsilon} = v(\epsilon,\delta).
\end{eqnarray}
If this constraint is released in a fit to the H1 data,
a value of $\int{V}dx = 2.24 \pm 0.13 (exp)$ is obtained instead 
of about 2.5 following from equation~\ref{Vintmod}.    
The modified expression for the $A$ function
in terms of quark distributions becomes
\begin{eqnarray}\label{Aqmod}
 A =  \frac{1}{4} \cdot \frac{1}{1+\epsilon}[
                4\bu -(u_V - 2 d_V)
               -5 (\bu-\bd) + 2 \epsilon (\bu+\bd)]. 
\end{eqnarray} 
For the naive assumptions $\epsilon = 0$ and $\bu = \bd$ this yields the
approximate relation~\ref{asimp} and $A \simeq \bu$ at low $x < 0.1$.
In the analysis these generalised expressions are used for $V$, its 
integral and $A$.
\subsection{Definition of Minimisation Procedure \label{achi2}}
The $\chi^2$ is computed as
\begin{eqnarray}\label{lechi2}
\chi^2
&=&\sum_{exp} \sum_{dat}
 \frac{[{\sigma_r}^{dat}_{exp}-
{\sigma_r}^{ fit}\times(1-\nu_{exp}\delta_{exp} 
-\sum_k\delta^{dat}_k(s^{exp}_k))
]^2}
{\sigma_{dat,sta}^2+\sigma_{dat,unc}^2}\nonumber\\
&+&\sum_{exp}\nu_{exp}^2
+\sum_{exp}\sum_k (s^{exp}_k)^2. 
\end{eqnarray}
The first two sums run over the data ($dat$) of the various
experiments ($exp$).  $\delta_{exp}$ is the relative overall
normalisation uncertainty.  $\sigma_{dat,sta}$ and $\sigma_{dat,unc}$
are the statistical error and the uncorrelated systematic error,
respectively, corresponding to the datum $dat$.  $\nu_{exp}$ is the
number of standard deviations corresponding to the overall
normalisation of the experimental sample $exp$.
$\delta^{dat}_k(s^{exp}_k)$ is the relative shift of the datum $dat$
induced by a change by $s^{exp}_k$ standard deviations of the $k^{th}$
correlated systematic uncertainty source of the experiment $exp$.
\subsection{Parameterisations \label{apara}}
As explained above, three parton distributions ($xg$, $V$ and $A$) are
necessary to describe the proton structure function \F and its $Q^2$
evolution.  The following general type of parameterisation is used
\begin{equation}
xq = a_qx^{b_q}(1-x)^{c_q}
      [1+d_{q}\sqrt{x} +e_{q}x+f_{q}{x}^2]
\label{eqpara}
\end{equation}
for $q = g,~V$~and $A$.  An attempt is made to describe these
functions with the least number of parameters in the brackets of
equation~\ref{eqpara}. All distributions are first calculated using
recent parameterisations of parton distributions, by
GRV98~\cite{grv98}, MRS99~\cite{mrs} and CTEQ5~\cite{cteq}. The
resulting functions are fitted using the expressions of
equation~\ref{eqpara} in order to obtain initial information about how
the new linear combinations of parton distributions $V$ and $A$ are
possibly parameterised best. All global analysis distributions require
the presence of $d_V$ and $e_V$ but allow $f_V$ to be set to zero.
This defines the parameterisation of $V$ which mainly is a combination
of valence quark distributions, see equation~\ref{Vmodq}. For $xg$
and $A$, however, different parameter combinations are tested in a
systematic way using fits to data.

\begin{figure}[tbp]
 \epsfig{file=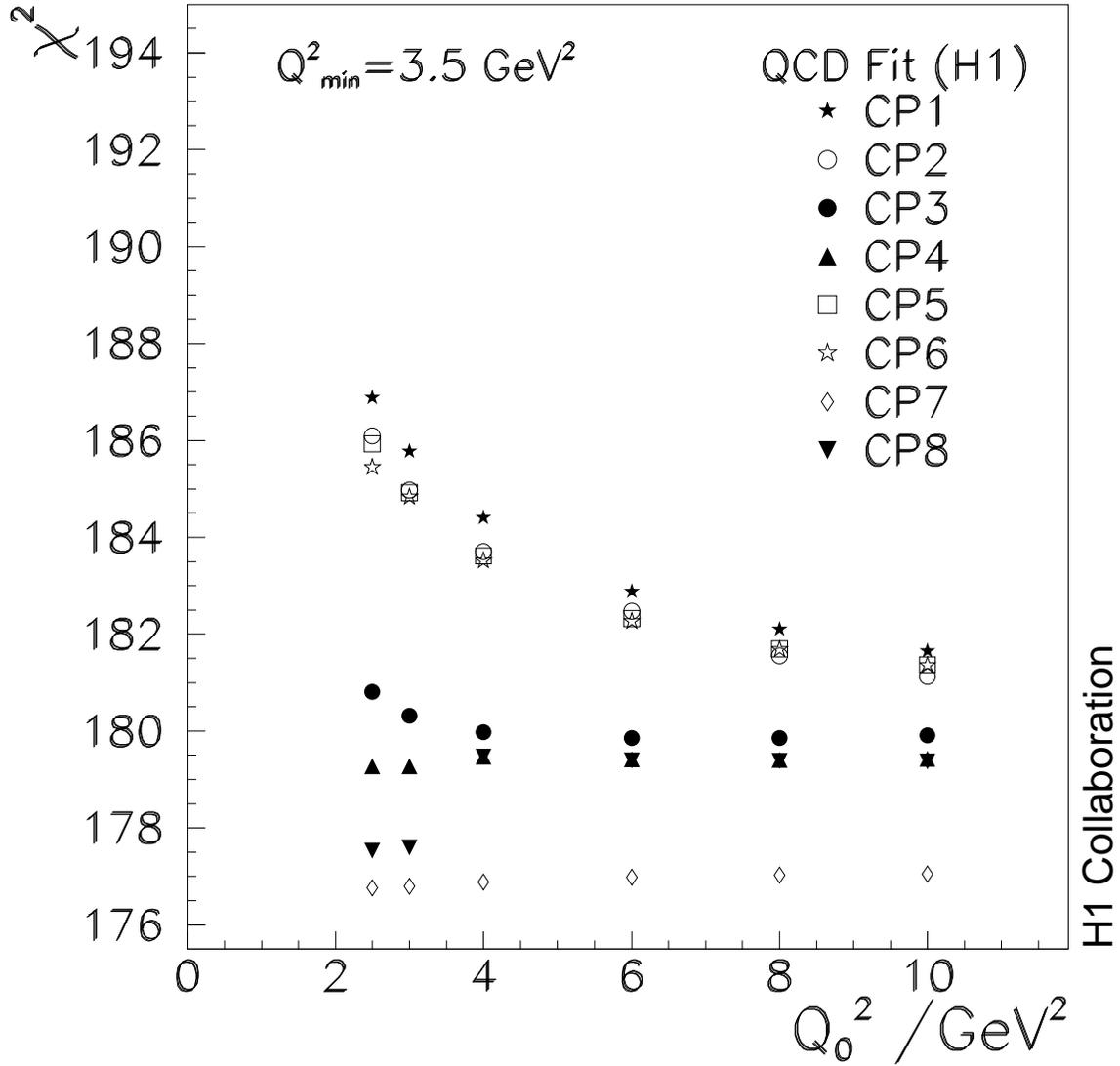,width=15cm}
  \caption{ \sl Dependence of
   $\chi^2$ on the  initial scale parameter 
   $Q^2_0$ for different parameterisations of the parton distributions
   $xg$ and $A$ (appendix~\ref{apara}, table~\ref{fs})  in the NLO QCD
   fit to the H1 data.}
  \protect\label{fh1para}
\end{figure}

The choice of a set of parameterisations is guided by the desire for a
weak dependence of the $\chi^2$ function on the initial scale $Q_0^2$,
and by the observed saturation of the $\chi^2$ when the number of
parameters becomes too large. This is demonstrated in
figure~\ref{fh1para} for the fit to H1 data alone. The functions
without a term $\propto \sqrt{x}$ in the $A$ distribution, see
table~\ref{fs}, yield a steady decrease of $\chi^2$ with $Q_0^2$.

Stability is observed for $Q_0^2 \geq 4$~\gv for the other
parameterisations.  Three of them have a similar $\chi^2$.  For the H1
fit the parameterisation CP3 is chosen.  The functions CP4 and CP8
have one more parameter but only one unit of $\chi^2$ is gained which
points to saturation of the parameter list. Although for CP7 a
somewhat better $\chi^2$ is found, this parameterisation is not
considered. It yields a too large gluon momentum as compared to the
other fits performed, including those using H1 and BCDMS data. These
all agree for the gluon momentum fraction among each other and also
with a previous analysis by the NMC collaboration~\cite{NMCg}. In the
CP7 fit to the H1 data apparently too many parameters are assigned to
describe the quark distributions at large $x$ since this leads also to
a distorted $V$ distribution.
\begin{table}[h] \centering 
\begin{tabular}{|l|c|c|}
\hline
  type  &  gluon & A  \\
\hline
 CP1     &  $1+ex$  &  $1+ex$ \\ 
 CP2    & $1+ d\sqrt{x} +ex $ & $1 +ex $ \\ 
\hline
 CP3     & $1+ex $  & $1+ d\sqrt{x} +ex $ \\
 CP4     & $1+ d\sqrt{x} +ex $ & $1+ d\sqrt{x} +ex$ \\ 
\hline
 CP5     & $1 +ex $ & $1 +ex+f{x}^2$ \\ 
 CP6     & $1+ d\sqrt{x} +ex $ & $1 +ex+f{x}^2$ \\
\hline
 CP7     & $1 +ex $ & $1+ d\sqrt{x} +ex+f{x}^2$ \\
 CP8     & $1+ d\sqrt{x} +ex $ & $1+ d\sqrt{x} +ex+f{x}^2$ \\
\hline
\end{tabular}
\caption{\label{fs}
  \sl Types of parameterisations of the $xg$ and $A$ distributions
  at the initial scale $Q_0^2$.}
\end{table}

The choice of parameterisation depends on the data set considered.  In
a similar study for the fit to H1 and BCDMS data, the parameterisation
CP4 is chosen.  Use of parameterisations with a high $x$ term
$(1+\alpha x^{\beta})$, as introduced by the CTEQ
collaboration~\cite{cteq}, worsens the $\chi^2$ by eleven units and
has thus not been considered further.
\end{appendix}
%

%
%
%
%
\newpage
%
%
\newpage
\newpage
\newpage
\clearpage
\newpage
%
\newpage
\newpage
%
\newpage
\newpage
%
\newpage
%
\newpage
\newpage
\newpage
%
\newpage
%
\newpage
%
%
\newpage
%
%
\newpage
%
\clearpage
%
\newpage
%
%
\newpage
\newpage
%
\newpage
%

%
%
%
%
\begin{table}[h] \centering 
\begin{tabular}{|c|l|c|r|r|r|r|r|r|r|}
\hline
$Q^2/GeV^2$ & $x$ & $y$ & $\sigma_r$ & $R$ & $F_2$ & $\delta_{tot}$ & $\delta_{sta}$ & $\delta_{unc}$ & $\delta_{cor}$   \\
\hline
   1.5 & 0.0000320 & 0.518 & 0.786 & 0.346 & 0.832 & 5.5 & 1.3 & 3.5 & 4.0 \\
   1.5 & 0.0000500 & 0.331 & 0.739 & 0.290 & 0.752 & 4.9 & 1.9 & 3.9 & 2.3 \\
   1.5 & 0.0000800 & 0.207 & 0.698 & 0.242 & 0.702 & 8.6 & 4.2 & 6.8 & 3.2 \\
   2.0 & 0.0000327 & 0.675 & 0.805 &   --  &   --  & 7.4 & 1.5 & 6.3 & 3.6 \\
   2.0 & 0.0000500 & 0.442 & 0.823 & 0.278 & 0.851 & 3.5 & 0.9 & 2.8 & 1.9 \\
   2.0 & 0.0000800 & 0.276 & 0.740 & 0.242 & 0.748 & 3.5 & 1.0 & 2.7 & 2.0 \\
   2.0 & 0.000130  & 0.170 & 0.714 & 0.209 & 0.716 & 3.7 & 1.3 & 2.9 & 1.8 \\
   2.0 & 0.000200  & 0.111 & 0.653 & 0.183 & 0.654 & 4.7 & 1.2 & 3.0 & 3.4 \\
   2.0 & 0.000320  & 0.069 & 0.625 & 0.159 & 0.626 & 4.4 & 1.4 & 3.1 & 2.8 \\
   2.0 & 0.000500  & 0.044 & 0.620 & 0.139 & 0.620 & 5.8 & 1.5 & 3.3 & 4.6 \\
   2.0 & 0.00100   & 0.022 & 0.512 & 0.115 & 0.513 & 4.5 & 1.2 & 3.0 & 3.2 \\
   2.0 & 0.00320   & 0.007 & 0.424 & 0.112 & 0.424 & 6.6 & 1.5 & 4.3 & 4.7 \\
   2.5 & 0.0000409 & 0.675 & 0.899 &   --  &   --  & 7.4 & 1.6 & 6.2 & 3.6 \\
   2.5 & 0.0000500 & 0.552 & 0.859 & 0.276 & 0.909 & 3.7 & 1.3 & 2.2 & 2.7 \\
   2.5 & 0.0000800 & 0.345 & 0.814 & 0.246 & 0.828 & 2.6 & 0.9 & 1.8 & 1.7 \\
   2.5 & 0.000130  & 0.212 & 0.763 & 0.219 & 0.767 & 2.5 & 0.9 & 1.6 & 1.7 \\
   2.5 & 0.000200  & 0.138 & 0.690 & 0.198 & 0.691 & 3.0 & 1.0 & 2.7 & 1.1 \\
   2.5 & 0.000320  & 0.086 & 0.637 & 0.177 & 0.638 & 3.1 & 1.0 & 2.7 & 1.4 \\
   2.5 & 0.000500  & 0.055 & 0.603 & 0.161 & 0.603 & 3.0 & 1.0 & 2.7 & 0.9 \\
   2.5 & 0.000800  & 0.035 & 0.555 & 0.147 & 0.555 & 3.1 & 1.1 & 2.7 & 1.1 \\
   2.5 & 0.00158   & 0.018 & 0.516 & 0.137 & 0.516 & 2.9 & 0.8 & 2.6 & 1.1 \\
   2.5 & 0.00500   & 0.005 & 0.403 & 0.167 & 0.403 & 5.2 & 1.0 & 3.9 & 3.3 \\
   3.5 & 0.0000573 & 0.675 & 0.897 &   --  &   --  & 7.0 & 2.1 & 6.2 & 2.6 \\
   3.5 & 0.0000800 & 0.483 & 0.928 & 0.254 & 0.964 & 2.9 & 1.1 & 1.8 & 2.1 \\
   3.5 & 0.000130  & 0.297 & 0.876 & 0.233 & 0.886 & 2.2 & 0.9 & 1.5 & 1.4 \\
   3.5 & 0.000200  & 0.193 & 0.822 & 0.216 & 0.826 & 2.3 & 0.9 & 1.5 & 1.5 \\
   3.5 & 0.000320  & 0.121 & 0.760 & 0.201 & 0.761 & 2.6 & 1.0 & 1.6 & 1.7 \\
   3.5 & 0.000500  & 0.078 & 0.715 & 0.188 & 0.716 & 2.7 & 1.0 & 1.6 & 1.9 \\
   3.5 & 0.000800  & 0.048 & 0.647 & 0.178 & 0.647 & 2.4 & 1.0 & 1.7 & 1.4 \\
   3.5 & 0.00130   & 0.030 & 0.601 & 0.173 & 0.601 & 2.6 & 1.0 & 1.7 & 1.7 \\
   3.5 & 0.00251   & 0.015 & 0.532 & 0.176 & 0.532 & 2.1 & 0.8 & 1.5 & 1.2 \\
   3.5 & 0.00800   & 0.005 & 0.418 & 0.236 & 0.418 & 4.2 & 0.9 & 3.3 & 2.4 \\
\hline
\end{tabular}
\caption{\label{tabsiga}
\sl 
The H1 measurement of the reduced deep-inelastic cross section
 with  data taken in a dedicated
low $Q^2$ trigger run in 1997.
 For $y < 0.6$ the structure function \Fc is  extracted
 using the quoted values of $R$, derived 
from a QCD fit to the H1 cross section data. 
 Fractional cross section errors are quoted in \%.
The total error ($\delta_{tot}$) is the quadratic sum of the uncorrelated
($\delta_{unc}$), the correlated ($\delta_{cor}$)  
and the experimental statistical error ($\delta_{sta}$).
}
\end{table}
\newpage
%
%
\begin{table}[h] \centering 
\begin{tabular}{|c|l|c|r|r|r|r|r|r|r|}
\hline
$Q^2/GeV^2$ & $x$ & $y$ & $\sigma_r$ & $R$ & $F_2$ & $\delta_{tot}$ & $\delta_{sta}$ & $\delta_{unc}$ & $\delta_{cor}$   \\
\hline
   5.0 & 0.0000818 & 0.675 & 1.019 &   --  &   --  & 6.6 & 2.1 & 4.8 & 3.9 \\
   5.0 & 0.000130  & 0.425 & 1.015 & 0.245 & 1.043 & 2.4 & 1.1 & 1.7 & 1.4 \\
   5.0 & 0.000200  & 0.276 & 0.965 & 0.232 & 0.974 & 2.3 & 1.0 & 1.4 & 1.5 \\
   5.0 & 0.000320  & 0.173 & 0.887 & 0.220 & 0.890 & 2.4 & 1.1 & 1.6 & 1.5 \\
   5.0 & 0.000500  & 0.111 & 0.791 & 0.210 & 0.792 & 2.3 & 1.1 & 1.6 & 1.4 \\
   5.0 & 0.000800  & 0.069 & 0.703 & 0.202 & 0.704 & 2.5 & 1.1 & 1.6 & 1.5 \\
   5.0 & 0.00130   & 0.043 & 0.661 & 0.198 & 0.661 & 3.0 & 1.1 & 1.6 & 2.3 \\
   5.0 & 0.00200   & 0.028 & 0.621 & 0.199 & 0.621 & 2.5 & 1.1 & 1.6 & 1.5 \\
   5.0 & 0.00398   & 0.014 & 0.538 & 0.213 & 0.538 & 2.3 & 0.9 & 1.5 & 1.5 \\
   5.0 & 0.0130    & 0.004 & 0.410 & 0.281 & 0.410 & 3.9 & 1.0 & 3.2 & 2.1 \\
   6.5 & 0.000130  & 0.552 & 1.089 & 0.252 & 1.148 & 3.5 & 1.6 & 1.7 & 2.6 \\
   6.5 & 0.000200  & 0.359 & 1.073 & 0.241 & 1.092 & 2.3 & 1.1 & 1.6 & 1.3 \\
   6.5 & 0.000320  & 0.224 & 0.957 & 0.230 & 0.963 & 2.1 & 1.2 & 1.4 & 1.2 \\
   6.5 & 0.000500  & 0.144 & 0.875 & 0.222 & 0.877 & 2.7 & 1.2 & 1.5 & 1.8 \\
   6.5 & 0.000800  & 0.090 & 0.800 & 0.215 & 0.801 & 2.5 & 1.2 & 1.5 & 1.5 \\
   6.5 & 0.00130   & 0.055 & 0.708 & 0.211 & 0.708 & 2.5 & 1.2 & 1.5 & 1.4 \\
   6.5 & 0.00200   & 0.036 & 0.672 & 0.212 & 0.672 & 2.7 & 1.3 & 1.6 & 1.7 \\
   6.5 & 0.00398   & 0.018 & 0.587 & 0.223 & 0.587 & 2.3 & 0.9 & 1.4 & 1.6 \\
   6.5 & 0.0130    & 0.005 & 0.432 & 0.273 & 0.432 & 3.6 & 1.0 & 3.2 & 1.4 \\
   8.5 & 0.000139  & 0.675 & 1.097 &   --  &   --  & 4.9 & 2.1 & 1.8 & 4.1 \\
   8.5 & 0.000200  & 0.470 & 1.152 & 0.247 & 1.193 & 2.9 & 1.4 & 1.6 & 2.0 \\
   8.5 & 0.000320  & 0.293 & 1.080 & 0.238 & 1.092 & 2.5 & 1.3 & 1.5 & 1.6 \\
   8.5 & 0.000500  & 0.188 & 0.992 & 0.231 & 0.996 & 2.3 & 1.3 & 1.4 & 1.3 \\
   8.5 & 0.000800  & 0.118 & 0.893 & 0.225 & 0.894 & 2.5 & 1.4 & 1.5 & 1.4 \\
   8.5 & 0.00130   & 0.072 & 0.797 & 0.222 & 0.797 & 2.8 & 1.4 & 1.6 & 1.8 \\
   8.5 & 0.00200   & 0.047 & 0.725 & 0.222 & 0.725 & 2.8 & 1.4 & 1.6 & 1.8 \\
   8.5 & 0.00320   & 0.029 & 0.632 & 0.226 & 0.632 & 2.8 & 1.5 & 1.6 & 1.8 \\
   8.5 & 0.00631   & 0.015 & 0.565 & 0.242 & 0.565 & 2.3 & 1.1 & 1.5 & 1.3 \\
   8.5 & 0.0200    & 0.005 & 0.419 & 0.268 & 0.419 & 4.2 & 1.2 & 3.2 & 2.4 \\
  12.0 & 0.000800  & 0.166 & 0.986 & 0.233 & 0.989 & 2.7 & 1.6 & 1.6 & 1.5 \\
  12.0 & 0.00130   & 0.102 & 0.878 & 0.230 & 0.879 & 2.3 & 1.6 & 1.6 & 0.6 \\
  12.0 & 0.00200   & 0.066 & 0.825 & 0.229 & 0.825 & 2.7 & 1.6 & 1.6 & 1.5 \\
  12.0 & 0.00320   & 0.041 & 0.725 & 0.231 & 0.725 & 2.9 & 1.6 & 1.6 & 1.7 \\
  12.0 & 0.00631   & 0.021 & 0.613 & 0.241 & 0.613 & 2.4 & 1.2 & 1.5 & 1.5 \\
  12.0 & 0.0200    & 0.007 & 0.459 & 0.249 & 0.459 & 4.0 & 1.4 & 3.2 & 1.9 \\

\hline
\end{tabular}
\caption{\label{tabsigb}
\sl 
The H1 measurement of the reduced deep-inelastic cross section
 with  data taken in a dedicated
low $Q^2$ trigger run in 1997.
 For $y < 0.6$ the structure function \Fc is  extracted
 using the quoted values of $R$, derived 
from a QCD fit to the H1 cross section data. 
 Fractional cross section errors are quoted in \%.
}
\end{table}
%
%
\newpage
\begin{table}[h] \centering 
\begin{tabular}{|c|l|c|r|r|r|r|r|r|r|}
\hline
$Q^2/GeV^2$ & $x$ & $y$ & $\sigma_r$ & $R$ & $F_2$ & $\delta_{tot}$ & $\delta_{sta}$ & $\delta_{unc}$ & $\delta_{cor}$   \\
\hline
  12.0 & 0.000161  & 0.825 & 1.226 &   --  &   --  & 5.8 & 4.1 & 3.8 & 1.8 \\
  12.0 & 0.000197  & 0.675 & 1.269 &   --  &   --  & 3.5 & 0.9 & 2.1 & 2.7 \\
  12.0 & 0.000320  & 0.415 & 1.217 & 0.245 & 1.249 & 2.0 & 0.6 & 1.7 & 1.0 \\
  12.0 & 0.000500  & 0.266 & 1.146 & 0.239 & 1.156 & 1.8 & 0.7 & 1.4 & 1.0 \\
  15.0 & 0.000201  & 0.825 & 1.255 &   --  &   --  & 5.2 & 3.2 & 3.6 & 1.9 \\
  15.0 & 0.000246  & 0.675 & 1.361 &   --  &   --  & 3.3 & 0.9 & 2.1 & 2.4 \\
  15.0 & 0.000320  & 0.519 & 1.283 & 0.249 & 1.342 & 2.4 & 0.7 & 1.9 & 1.4 \\
  15.0 & 0.000500  & 0.332 & 1.228 & 0.243 & 1.247 & 1.9 & 0.6 & 1.6 & 0.7 \\
  15.0 & 0.000800  & 0.208 & 1.115 & 0.238 & 1.121 & 1.7 & 0.6 & 1.4 & 0.8 \\
  15.0 & 0.00130   & 0.127 & 0.969 & 0.234 & 0.971 & 2.9 & 0.6 & 1.4 & 2.4 \\
  15.0 & 0.00200   & 0.083 & 0.865 & 0.232 & 0.866 & 2.5 & 0.6 & 1.5 & 2.0 \\
  15.0 & 0.00320   & 0.052 & 0.774 & 0.234 & 0.774 & 2.5 & 0.7 & 1.5 & 1.9 \\
  15.0 & 0.00500   & 0.033 & 0.708 & 0.237 & 0.708 & 2.4 & 0.8 & 1.5 & 1.7 \\
  15.0 & 0.0100    & 0.017 & 0.575 & 0.244 & 0.575 & 2.7 & 0.7 & 1.4 & 2.2 \\
  15.0 & 0.0320    & 0.005 & 0.453 & 0.211 & 0.453 & 6.4 & 1.0 & 3.3 & 5.5 \\
  20.0 & 0.000268  & 0.825 & 1.313 &   --  &   --  & 5.2 & 3.2 & 3.6 & 1.8 \\
  20.0 & 0.000328  & 0.675 & 1.383 &   --  &   --  & 2.7 & 1.0 & 2.0 & 1.5 \\
  20.0 & 0.000500  & 0.443 & 1.285 & 0.246 & 1.324 & 2.0 & 0.6 & 1.6 & 0.9 \\
  20.0 & 0.000800  & 0.277 & 1.178 & 0.241 & 1.190 & 1.8 & 0.6 & 1.4 & 1.0 \\
  20.0 & 0.00130   & 0.170 & 1.059 & 0.237 & 1.062 & 1.8 & 0.6 & 1.4 & 1.0 \\
  20.0 & 0.00200   & 0.111 & 0.939 & 0.235 & 0.940 & 2.8 & 0.7 & 1.4 & 2.3 \\
  20.0 & 0.00320   & 0.069 & 0.819 & 0.234 & 0.819 & 2.3 & 0.7 & 1.5 & 1.7 \\
  20.0 & 0.00500   & 0.044 & 0.747 & 0.235 & 0.747 & 2.2 & 0.8 & 1.5 & 1.3 \\
  20.0 & 0.0100    & 0.022 & 0.610 & 0.238 & 0.610 & 2.4 & 0.7 & 1.4 & 1.8 \\
  20.0 & 0.0320    & 0.007 & 0.455 & 0.198 & 0.455 & 6.8 & 1.0 & 3.3 & 5.9 \\
  25.0 & 0.000335  & 0.825 & 1.379 &   --  &   --  & 5.9 & 4.1 & 3.9 & 1.8 \\
  25.0 & 0.000410  & 0.675 & 1.371 &   --  &   --  & 2.6 & 1.2 & 2.0 & 1.2 \\
  25.0 & 0.000500  & 0.553 & 1.345 & 0.248 & 1.417 & 2.4 & 1.0 & 1.8 & 1.2 \\
  25.0 & 0.000800  & 0.346 & 1.242 & 0.243 & 1.263 & 1.9 & 0.7 & 1.6 & 0.9 \\
  25.0 & 0.00130   & 0.213 & 1.091 & 0.238 & 1.097 & 1.8 & 0.7 & 1.4 & 0.9 \\
  25.0 & 0.00200   & 0.138 & 0.985 & 0.236 & 0.987 & 2.9 & 0.8 & 1.4 & 2.4 \\
  25.0 & 0.00320   & 0.086 & 0.879 & 0.234 & 0.880 & 2.8 & 0.8 & 1.5 & 2.2 \\
  25.0 & 0.00500   & 0.055 & 0.754 & 0.234 & 0.754 & 2.4 & 0.9 & 1.5 & 1.6 \\
  25.0 & 0.00800   & 0.034 & 0.663 & 0.234 & 0.663 & 2.5 & 0.9 & 1.5 & 1.8 \\
  25.0 & 0.0158    & 0.018 & 0.547 & 0.226 & 0.547 & 3.7 & 0.9 & 1.5 & 3.3 \\
  25.0 & 0.0500    & 0.005 & 0.447 & 0.148 & 0.447 & 7.5 & 1.3 & 3.3 & 6.6 \\
  35.0 & 0.000574  & 0.675 & 1.473 &   --  &   --  & 2.7 & 1.4 & 2.0 & 1.2 \\
  35.0 & 0.000800  & 0.484 & 1.354 & 0.244 & 1.405 & 2.2 & 0.9 & 1.7 & 1.1 \\
  35.0 & 0.00130   & 0.298 & 1.181 & 0.239 & 1.195 & 1.8 & 0.8 & 1.4 & 0.8 \\
  35.0 & 0.00200   & 0.194 & 1.031 & 0.235 & 1.035 & 1.8 & 0.8 & 1.4 & 0.8 \\
\hline
\end{tabular}
\caption{\label{tabsigc}
\sl 
The H1 measurement of the reduced deep-inelastic cross section
 with  data taken in 1996/97.
 For $y < 0.6$ the structure function \Fc is  extracted
 using the quoted values of $R$, derived 
from a QCD fit to the H1 cross section data. 
 Fractional cross section errors are quoted in \%.
}
\end{table}
%
%
\newpage
\begin{table}[h] \centering 
\begin{tabular}{|c|l|c|r|r|r|r|r|r|r|}
\hline
$Q^2/GeV^2$ & $x$ & $y$ & $\sigma_r$ & $R$ & $F_2$ & $\delta_{tot}$ & $\delta_{sta}$ & $\delta_{unc}$ & $\delta_{cor}$   \\
\hline
  35.0 & 0.00320   & 0.121 & 0.935 & 0.233 & 0.936 & 3.1 & 0.9 & 1.5 & 2.5 \\
  35.0 & 0.00500   & 0.077 & 0.821 & 0.231 & 0.821 & 2.7 & 0.9 & 1.5 & 2.0 \\
  35.0 & 0.00800   & 0.048 & 0.719 & 0.228 & 0.719 & 2.4 & 1.0 & 1.5 & 1.6 \\
  35.0 & 0.0130    & 0.030 & 0.625 & 0.222 & 0.625 & 3.2 & 1.1 & 1.6 & 2.6 \\
  35.0 & 0.0251    & 0.015 & 0.524 & 0.195 & 0.524 & 4.1 & 1.1 & 1.6 & 3.6 \\
  35.0 & 0.0800    & 0.005 & 0.413 & 0.095 & 0.413 & 9.2 & 1.8 & 3.5 & 8.3 \\
  45.0 & 0.00130   & 0.383 & 1.282 & 0.238 & 1.309 & 1.9 & 0.9 & 1.7 & 0.3 \\
  45.0 & 0.00200   & 0.249 & 1.107 & 0.234 & 1.115 & 1.8 & 0.9 & 1.4 & 0.6 \\
  45.0 & 0.00320   & 0.156 & 0.979 & 0.231 & 0.982 & 1.8 & 0.9 & 1.4 & 0.7 \\
  45.0 & 0.00500   & 0.099 & 0.872 & 0.228 & 0.873 & 2.8 & 1.1 & 1.5 & 2.1 \\
  45.0 & 0.00800   & 0.062 & 0.743 & 0.224 & 0.743 & 2.5 & 1.1 & 1.5 & 1.6 \\
  45.0 & 0.0130    & 0.038 & 0.649 & 0.215 & 0.649 & 2.8 & 1.3 & 1.6 & 2.0 \\
  45.0 & 0.0251    & 0.020 & 0.525 & 0.187 & 0.525 & 4.3 & 1.2 & 1.6 & 3.8 \\
  45.0 & 0.0800    & 0.006 & 0.396 & 0.091 & 0.396 & 7.6 & 2.1 & 3.5 & 6.4 \\
  60.0 & 0.00200   & 0.332 & 1.245 & 0.231 & 1.263 & 2.1 & 1.0 & 1.7 & 0.7 \\
  60.0 & 0.00320   & 0.208 & 1.052 & 0.227 & 1.057 & 1.9 & 1.1 & 1.4 & 0.7 \\
  60.0 & 0.00500   & 0.133 & 0.900 & 0.223 & 0.902 & 3.1 & 1.2 & 1.6 & 2.3 \\
  60.0 & 0.00800   & 0.083 & 0.803 & 0.218 & 0.804 & 2.8 & 1.3 & 1.6 & 1.9 \\
  60.0 & 0.0130    & 0.051 & 0.683 & 0.208 & 0.683 & 2.9 & 1.4 & 1.7 & 2.0 \\
  60.0 & 0.0200    & 0.033 & 0.597 & 0.192 & 0.597 & 4.1 & 1.7 & 1.8 & 3.2 \\
  60.0 & 0.0398    & 0.017 & 0.506 & 0.145 & 0.506 & 4.7 & 1.8 & 1.9 & 3.9 \\
  60.0 & 0.130     & 0.005 & 0.360 & 0.057 & 0.360 & 9.4 & 3.0 & 3.8 & 8.1 \\
  90.0 & 0.00320   & 0.311 & 1.107 & 0.221 & 1.120 & 2.6 & 1.4 & 1.8 & 1.2 \\
  90.0 & 0.00500   & 0.199 & 0.999 & 0.216 & 1.004 & 2.2 & 1.3 & 1.5 & 0.9 \\
  90.0 & 0.00800   & 0.124 & 0.845 & 0.209 & 0.846 & 3.3 & 1.5 & 1.7 & 2.5 \\
  90.0 & 0.0130    & 0.076 & 0.728 & 0.197 & 0.728 & 2.8 & 1.6 & 1.8 & 1.4 \\
  90.0 & 0.0200    & 0.050 & 0.618 & 0.180 & 0.618 & 3.9 & 1.8 & 1.9 & 2.8 \\
  90.0 & 0.0398    & 0.025 & 0.506 & 0.135 & 0.506 & 3.8 & 2.0 & 1.9 & 2.7 \\
  90.0 & 0.130     & 0.008 & 0.339 & 0.053 & 0.339 & 4.8 & 3.3 & 2.0 & 2.9 \\
 120.0 & 0.00500   & 0.266 & 1.011 & 0.210 & 1.019 & 3.7 & 2.4 & 2.1 & 1.9 \\
 120.0 & 0.00800   & 0.166 & 0.839 & 0.202 & 0.841 & 3.0 & 1.9 & 1.8 & 1.4 \\
 120.0 & 0.0130    & 0.102 & 0.744 & 0.190 & 0.745 & 4.6 & 2.1 & 2.0 & 3.6 \\
 120.0 & 0.0200    & 0.066 & 0.604 & 0.173 & 0.605 & 4.0 & 2.3 & 2.1 & 2.5 \\
 120.0 & 0.0320    & 0.041 & 0.558 & 0.145 & 0.558 & 5.8 & 2.8 & 2.5 & 4.5 \\
 120.0 & 0.0631    & 0.021 & 0.462 & 0.094 & 0.462 & 4.9 & 3.0 & 2.7 & 2.9 \\
 120.0 & 0.200     & 0.007 & 0.312 & 0.037 & 0.312 &10.6 & 4.8 & 4.8 & 8.2 \\
 150.0 & 0.0200    & 0.083 & 0.709 & 0.167 & 0.709 & 8.9 & 4.4 & 3.8 & 6.6 \\
 150.0 & 0.0320    & 0.052 & 0.550 & 0.140 & 0.550 & 8.0 & 5.1 & 4.1 & 4.7 \\
 150.0 & 0.0631    & 0.026 & 0.418 & 0.090 & 0.418 & 9.0 & 5.7 & 4.4 & 5.4 \\
 150.0 & 0.200     & 0.008 & 0.296 & 0.036 & 0.296 &12.9 & 8.4 & 7.1 & 6.7 \\

\hline
\end{tabular}
\caption{\label{tabsigd}
\sl 
The H1 measurement of the reduced deep-inelastic cross section
 with  data taken in 1996/97.
 For $y < 0.6$ the structure function \Fc is  extracted
 using the quoted values of $R$, derived 
from a QCD fit to the H1 cross section data. 
 Fractional cross section errors are quoted in \%.
}
\end{table}
\newpage
%
%
\begin{table}[h] \centering 
\begin{tabular}{|c|l|r|c|r|r|r|}
\hline
$Q^2/GeV^2$ & $x$ & $y$ & $ (\partial \sigma_r / \partial \ln y)_{Q^2}$ & $\Delta_{sta}$ & $\Delta_{sys}$ & $\Delta_{tot}$    \\
\hline
    1.5& 0.000039& 0.425& 0.104& 0.039& 0.080& 0.089\\
    1.5& 0.000062& 0.269& 0.088& 0.070& 0.114& 0.134\\
    2.2& 0.000045& 0.538& 0.051& 0.017& 0.061& 0.064\\
    2.2& 0.000075& 0.325& 0.139& 0.015& 0.025& 0.029\\
    2.2& 0.000122& 0.200& 0.126& 0.015& 0.024& 0.028\\
    2.2& 0.000194& 0.125& 0.118& 0.018& 0.031& 0.036\\
    2.2& 0.000291& 0.083& 0.070& 0.018& 0.026& 0.032\\
    2.2& 0.000454& 0.054& 0.082& 0.014& 0.026& 0.030\\
    2.2& 0.000748& 0.032& 0.086& 0.018& 0.033& 0.038\\
    2.2& 0.00122 & 0.020& 0.081& 0.016& 0.039& 0.042\\
    2.2& 0.00224 & 0.011& 0.100& 0.007& 0.023& 0.024\\
    4.2& 0.000086& 0.538& 0.002& 0.031& 0.057& 0.065\\
    4.2& 0.000143& 0.325& 0.122& 0.020& 0.033& 0.039\\
    4.2& 0.000232& 0.200& 0.148& 0.017& 0.023& 0.029\\
    4.2& 0.000371& 0.125& 0.170& 0.020& 0.027& 0.034\\
    4.2& 0.000556& 0.083& 0.140& 0.019& 0.019& 0.027\\
    4.2& 0.000867& 0.054& 0.118& 0.014& 0.018& 0.023\\
    4.2& 0.00143 & 0.032& 0.095& 0.014& 0.016& 0.022\\
    4.2& 0.00232 & 0.020& 0.114& 0.011& 0.013& 0.017\\
    4.2& 0.00488 & 0.009& 0.106& 0.004& 0.013& 0.013\\
    7.5& 0.000154& 0.538& -0.004& 0.034& 0.049& 0.059\\
    7.5& 0.000255& 0.325& 0.218& 0.028& 0.024& 0.037\\
    7.5& 0.000414& 0.200& 0.193& 0.024& 0.031& 0.039\\
    7.5& 0.000663& 0.125& 0.182& 0.028& 0.027& 0.039\\
    7.5& 0.000992& 0.083& 0.203& 0.026& 0.021& 0.034\\
    7.5& 0.00155 & 0.054& 0.127& 0.019& 0.018& 0.026\\
    7.5& 0.00255 & 0.032& 0.189& 0.024& 0.021& 0.032\\
    7.5& 0.00414 & 0.020& 0.079& 0.019& 0.021& 0.029\\
    7.5& 0.00764 & 0.011& 0.149& 0.008& 0.016& 0.018\\
    7.5& 0.0155  & 0.005& 0.081& 0.013& 0.012& 0.018\\
   13.5& 0.000199& 0.748& -0.399& 0.167& 0.191& 0.253\\
   13.5& 0.000277& 0.538& 0.151& 0.018& 0.051& 0.054\\
   13.5& 0.000459& 0.325& 0.182& 0.015& 0.024& 0.028\\
   13.5& 0.000746& 0.200& 0.299& 0.016& 0.054& 0.056\\
   13.5& 0.00119 & 0.125& 0.295& 0.037& 0.043& 0.057\\
   13.5& 0.00179 & 0.083& 0.148& 0.037& 0.035& 0.051\\
   13.5& 0.00279 & 0.054& 0.173& 0.012& 0.014& 0.018\\
   13.5& 0.00459 & 0.032& 0.203& 0.019& 0.017& 0.025\\
   13.5& 0.00746 & 0.020& 0.145& 0.018& 0.027& 0.032\\
   13.5& 0.0137  & 0.011& 0.135& 0.010& 0.015& 0.018\\
   13.5& 0.0279  & 0.005& 0.045& 0.015& 0.035& 0.038\\
\hline
\end{tabular}
\caption{\label{tabder1}
  \sl The H1 measurement of the cross section derivative $(\partial \sigma_r /
  \partial \ln y)_{Q^2} = -(\partial \sigma_r / \partial \ln x)_{Q^2}$
  calculated at fixed $Q^2$.
  The errors are given in absolute values. The data below 13.5\,\gv
  belong to the special low $Q^2$ run taken in 1997.
  The data at larger $Q^2$ were taken in 1996/97.}
\end{table}
\newpage
\begin{table}[h] \centering 
\begin{tabular}{|c|l|r|c|r|r|r|}
\hline
$Q^2/GeV^2$ & $x$ & $y$ & $ (\partial \sigma_r / \partial \ln y)_{Q^2}$ & $\Delta_{sta}$ & $\Delta_{sys}$ & $\Delta_{tot}$    \\
\hline
   22.5& 0.000332& 0.748& -0.216& 0.181& 0.176& 0.253\\
   22.5& 0.000462& 0.538& 0.198& 0.020& 0.032& 0.037\\
   22.5& 0.000765& 0.325& 0.272& 0.017& 0.021& 0.027\\
   22.5& 0.00124 & 0.200& 0.249& 0.014& 0.032& 0.035\\
   22.5& 0.00199 & 0.125& 0.256& 0.017& 0.034& 0.038\\
   22.5& 0.00298 & 0.083& 0.277& 0.016& 0.026& 0.031\\
   22.5& 0.00465 & 0.054& 0.184& 0.012& 0.016& 0.020\\
   22.5& 0.00765 & 0.032& 0.185& 0.013& 0.024& 0.027\\
   22.5& 0.0124  & 0.020& 0.182& 0.012& 0.021& 0.025\\
   22.5& 0.0229  & 0.011& 0.099& 0.007& 0.025& 0.027\\
   40.0& 0.000822& 0.538& 0.335& 0.042& 0.045& 0.062\\
   40.0& 0.00136 & 0.325& 0.386& 0.023& 0.026& 0.034\\
   40.0& 0.00221 & 0.200& 0.311& 0.017& 0.014& 0.022\\
   40.0& 0.00354 & 0.125& 0.204& 0.021& 0.058& 0.062\\
   40.0& 0.00529 & 0.083& 0.263& 0.021& 0.026& 0.033\\
   40.0& 0.00826 & 0.054& 0.206& 0.015& 0.019& 0.024\\
   40.0& 0.0136  & 0.032& 0.179& 0.019& 0.027& 0.033\\
   40.0& 0.0221  & 0.020& 0.163& 0.016& 0.017& 0.024\\
   40.0& 0.0407  & 0.011& 0.124& 0.012& 0.023& 0.026\\
   40.0& 0.0826  & 0.005& 0.039& 0.022& 0.024& 0.032\\
   75.0& 0.00255 & 0.325& 0.442& 0.035& 0.029& 0.045\\
   75.0& 0.00414 & 0.200& 0.303& 0.024& 0.024& 0.034\\
   75.0& 0.00663 & 0.125& 0.271& 0.037& 0.042& 0.056\\
   75.0& 0.00992 & 0.083& 0.222& 0.036& 0.038& 0.053\\
   75.0& 0.0155  & 0.054& 0.232& 0.018& 0.031& 0.036\\
   75.0& 0.0255  & 0.032& 0.177& 0.026& 0.028& 0.038\\
   75.0& 0.0414  & 0.020& 0.102& 0.027& 0.033& 0.042\\
   75.0& 0.0764  & 0.011& 0.152& 0.019& 0.019& 0.027\\
   75.0& 0.155   & 0.005& 0.099& 0.029& 0.038& 0.048\\
  135.0& 0.00746 & 0.200& 0.352& 0.055& 0.045& 0.071\\
  135.0& 0.0119  & 0.125& 0.176& 0.054& 0.088& 0.103\\
  135.0& 0.0179  & 0.083& 0.292& 0.049& 0.041& 0.064\\
  135.0& 0.0279  & 0.054& 0.140& 0.035& 0.052& 0.063\\
  135.0& 0.0459  & 0.032& 0.161& 0.039& 0.032& 0.050\\

\hline
\end{tabular}
\caption{\label{tabder2}
  \sl 
The H1 measurement of the cross section derivative $(\partial \sigma_r /
  \partial \ln y)_{Q^2} = -(\partial \sigma_r / \partial \ln x)_{Q^2}$
 calculated at fixed $Q^2$. $\Delta_{sta}$ 
denotes the uncertainty due to the data statistics.
  The uncertainties are given in absolute values.}
\end{table}
\newpage
%
%
\begin{table}[h] \centering 
\begin{tabular}{|c|c|c|c|c|c|c|c|}
\hline
$Q^2/GeV^2$ & $x$ & $y$ &  \FL & $\delta_{sta}$ & $\delta_{sys}$ & $\delta_{met}$ & $\delta_{tot}$    \\
\hline
 2.2 & 0.000045 & 0.538 & 0.100 & 0.030 & 0.107 & 0.025 & 0.114 \\
 4.2 & 0.000086 & 0.538 & 0.273 & 0.055 & 0.101 & 0.027 & 0.118 \\
 7.5 & 0.000154 & 0.538 & 0.385 & 0.058 & 0.088 & 0.039 & 0.112 \\
12.0 & 0.000161 & 0.825 & 0.429 & 0.076 & 0.095 & 0.045 & 0.130 \\
12.0 & 0.000197 & 0.675 & 0.411 & 0.027 & 0.136 & 0.058 & 0.150 \\
12.0 & 0.000320 & 0.415 & 0.456 & 0.054 & 0.279 & 0.096 & 0.300 \\
15.0 & 0.000201 & 0.825 & 0.453 & 0.061 & 0.092 & 0.042 & 0.118 \\
15.0 & 0.000246 & 0.675 & 0.285 & 0.030 & 0.124 & 0.053 & 0.138 \\
15.0 & 0.000320 & 0.519 & 0.417 & 0.040 & 0.194 & 0.069 & 0.210 \\
20.0 & 0.000268 & 0.825 & 0.426 & 0.064 & 0.093 & 0.036 & 0.118 \\
20.0 & 0.000328 & 0.675 & 0.315 & 0.035 & 0.119 & 0.045 & 0.132 \\
20.0 & 0.000500 & 0.443 & 0.385 & 0.055 & 0.249 & 0.064 & 0.263 \\
25.0 & 0.000335 & 0.825 & 0.360 & 0.085 & 0.098 & 0.030 & 0.134 \\
25.0 & 0.000410 & 0.675 & 0.377 & 0.039 & 0.116 & 0.038 & 0.128 \\
25.0 & 0.000500 & 0.553 & 0.404 & 0.055 & 0.164 & 0.047 & 0.179 \\
35.0 & 0.000574 & 0.675 & 0.149 & 0.049 & 0.115 & 0.029 & 0.128 \\
35.0 & 0.000800 & 0.484 & 0.239 & 0.064 & 0.213 & 0.047 & 0.228 \\
        
%
\hline
\end{tabular}
\caption{\label{tabfl}
  \sl The H1 determination of the longitudinal structure function \FL.
 The errors are given in absolute, $\delta_{sta}$ 
representing the experimental statistics. The systematic errors
consider all contributions from
correlated and uncorrelated systematic error sources. $\delta_{met}$
is due to the uncertainties connected with the representation
of \Fc in the derivative method, for $Q^2 < 10$~\gv, and in the
QCD extrapolation method, for $Q^2 > 10$~\gv.
}
\end{table}
%
%
\begin{table}[h] \centering 
\begin{tabular}{|c|r|r|c|r|r|r|}
\hline
$Q^2/GeV^2$ & $x$ & $y$ & $ (\partial F_2 / \partial \ln Q^2)_x$ & $\Delta_{sta}$ & $\Delta_{sys}$ & $\Delta_{tot}$    \\
\hline
   1.7 & 0.00005 & 0.383 & 0.342 & 0.055 & 0.097 & 0.112 \\
   2.2 & 0.00005 & 0.495 & 0.261 & 0.062 & 0.094 & 0.113 \\
   1.7 & 0.00008 & 0.240 & 0.159 & 0.106 & 0.164 & 0.196 \\
   2.2 & 0.00008 & 0.309 & 0.359 & 0.046 & 0.064 & 0.079 \\
   3.0 & 0.00008 & 0.409 & 0.403 & 0.037 & 0.066 & 0.076 \\
   2.2 & 0.00013 & 0.190 & 0.228 & 0.052 & 0.075 & 0.091 \\
   3.0 & 0.00013 & 0.252 & 0.355 & 0.031 & 0.034 & 0.046 \\
   4.2 & 0.00013 & 0.356 & 0.439 & 0.039 & 0.036 & 0.052 \\
   5.7 & 0.00013 & 0.485 & 0.399 & 0.081 & 0.081 & 0.115 \\
   2.2 & 0.00020 & 0.124 & 0.166 & 0.046 & 0.115 & 0.124 \\
   3.0 & 0.00020 & 0.164 & 0.400 & 0.030 & 0.048 & 0.057 \\
   4.2 & 0.00020 & 0.232 & 0.417 & 0.035 & 0.029 & 0.046 \\
   5.7 & 0.00020 & 0.316 & 0.450 & 0.060 & 0.043 & 0.074 \\
   7.4 & 0.00020 & 0.411 & 0.374 & 0.078 & 0.069 & 0.104 \\
   2.2 & 0.00032 & 0.077 & 0.054 & 0.047 & 0.105 & 0.115 \\
   3.0 & 0.00032 & 0.102 & 0.366 & 0.029 & 0.047 & 0.055 \\
   4.2 & 0.00032 & 0.145 & 0.362 & 0.035 & 0.036 & 0.050 \\
   5.7 & 0.00032 & 0.197 & 0.276 & 0.057 & 0.084 & 0.102 \\
   7.4 & 0.00032 & 0.257 & 0.482 & 0.067 & 0.049 & 0.083 \\
  10.1 & 0.00032 & 0.349 & 0.453 & 0.045 & 0.048 & 0.066 \\
  13.4 & 0.00032 & 0.464 & 0.415 & 0.052 & 0.090 & 0.104 \\
   2.2 & 0.00050 & 0.050 &-0.076 & 0.049 & 0.148 & 0.156 \\
   3.0 & 0.00050 & 0.065 & 0.334 & 0.027 & 0.048 & 0.056 \\
   4.2 & 0.00050 & 0.093 & 0.214 & 0.031 & 0.032 & 0.045 \\
   5.7 & 0.00050 & 0.126 & 0.325 & 0.052 & 0.049 & 0.071 \\
   7.4 & 0.00050 & 0.165 & 0.445 & 0.063 & 0.093 & 0.113 \\
  10.1 & 0.00050 & 0.224 & 0.464 & 0.044 & 0.039 & 0.059 \\
  13.4 & 0.00050 & 0.297 & 0.403 & 0.047 & 0.064 & 0.080 \\
  17.3 & 0.00050 & 0.383 & 0.270 & 0.039 & 0.046 & 0.060 \\
  22.4 & 0.00050 & 0.495 & 0.417 & 0.076 & 0.091 & 0.119 \\
   2.2 & 0.00080 & 0.031 & 0.119 & 0.039 & 0.094 & 0.102 \\
   3.0 & 0.00080 & 0.041 & 0.275 & 0.026 & 0.041 & 0.049 \\
   4.2 & 0.00080 & 0.058 & 0.158 & 0.028 & 0.041 & 0.050 \\
   5.7 & 0.00080 & 0.079 & 0.369 & 0.048 & 0.040 & 0.062 \\
   7.4 & 0.00080 & 0.103 & 0.349 & 0.059 & 0.043 & 0.073 \\
  10.1 & 0.00080 & 0.140 & 0.276 & 0.057 & 0.041 & 0.070 \\
  13.4 & 0.00080 & 0.186 & 0.586 & 0.075 & 0.108 & 0.131 \\
  17.3 & 0.00080 & 0.240 & 0.240 & 0.034 & 0.038 & 0.051 \\
  22.4 & 0.00080 & 0.309 & 0.328 & 0.050 & 0.052 & 0.072 \\
  29.6 & 0.00080 & 0.409 & 0.423 & 0.045 & 0.044 & 0.062 \\
\hline
\end{tabular}
\caption{\label{tabderq1}
  \sl The H1 Measurement of the derivative \pdff.
$\Delta_{sta}$ denotes the uncertainty due to the data statistics.
The uncertainties are given in absolute values.}
\end{table}
%
\begin{table}[h] \centering 
\begin{tabular}{|c|r|r|c|r|r|r|}
\hline
$Q^2/GeV^2$ & $x$ & $y$ & $ (\partial F_2 / \partial \ln Q^2)_x $ & $\Delta_{sta} $ & $\Delta_{sys}$ & $\Delta_{tot}$    \\
\hline
   3.0 & 0.0013 & 0.025 & 0.205 & 0.023 & 0.040 & 0.046 \\
   4.2 & 0.0013 & 0.036 & 0.168 & 0.027 & 0.033 & 0.043 \\
   5.7 & 0.0013 & 0.049 & 0.181 & 0.044 & 0.042 & 0.061 \\
   7.4 & 0.0013 & 0.063 & 0.330 & 0.053 & 0.044 & 0.069 \\
  10.1 & 0.0013 & 0.086 & 0.239 & 0.051 & 0.047 & 0.069 \\
  13.4 & 0.0013 & 0.114 & 0.409 & 0.067 & 0.096 & 0.117 \\
  17.3 & 0.0013 & 0.147 & 0.318 & 0.032 & 0.110 & 0.114 \\
  22.4 & 0.0013 & 0.190 & 0.154 & 0.045 & 0.043 & 0.062 \\
  29.6 & 0.0013 & 0.252 & 0.291 & 0.035 & 0.032 & 0.048 \\
  39.7 & 0.0013 & 0.338 & 0.454 & 0.060 & 0.058 & 0.084 \\
   4.2 & 0.0020 & 0.023 & 0.185 & 0.023 & 0.029 & 0.037 \\
   5.7 & 0.0020 & 0.032 & 0.194 & 0.042 & 0.039 & 0.058 \\
   7.4 & 0.0020 & 0.041 & 0.200 & 0.049 & 0.048 & 0.069 \\
  10.1 & 0.0020 & 0.056 & 0.289 & 0.048 & 0.034 & 0.059 \\
  13.4 & 0.0020 & 0.074 & 0.183 & 0.064 & 0.058 & 0.086 \\
  17.3 & 0.0020 & 0.096 & 0.258 & 0.030 & 0.041 & 0.050 \\
  22.4 & 0.0020 & 0.124 & 0.209 & 0.044 & 0.054 & 0.070 \\
  29.6 & 0.0020 & 0.164 & 0.145 & 0.033 & 0.085 & 0.091 \\
  39.7 & 0.0020 & 0.220 & 0.318 & 0.050 & 0.040 & 0.064 \\
  52.0 & 0.0020 & 0.288 & 0.513 & 0.057 & 0.052 & 0.077 \\
   5.7 & 0.0032 & 0.020 & 0.200 & 0.029 & 0.027 & 0.040 \\
   7.4 & 0.0032 & 0.026 & 0.064 & 0.041 & 0.044 & 0.060 \\
  10.1 & 0.0032 & 0.035 & 0.270 & 0.043 & 0.034 & 0.055 \\
  13.4 & 0.0032 & 0.046 & 0.219 & 0.058 & 0.056 & 0.081 \\
  17.3 & 0.0032 & 0.060 & 0.157 & 0.028 & 0.033 & 0.044 \\
  22.4 & 0.0032 & 0.077 & 0.271 & 0.041 & 0.053 & 0.067 \\
  29.6 & 0.0032 & 0.102 & 0.167 & 0.032 & 0.036 & 0.048 \\
  39.7 & 0.0032 & 0.137 & 0.182 & 0.050 & 0.103 & 0.114 \\
  52.0 & 0.0032 & 0.180 & 0.261 & 0.051 & 0.041 & 0.065 \\
  73.5 & 0.0032 & 0.254 & 0.156 & 0.047 & 0.054 & 0.071 \\
  10.1 & 0.0050 & 0.022 & 0.153 & 0.030 & 0.022 & 0.038 \\
  13.4 & 0.0050 & 0.030 & 0.265 & 0.043 & 0.043 & 0.061 \\
  17.3 & 0.0050 & 0.038 & 0.137 & 0.027 & 0.033 & 0.043 \\
  22.4 & 0.0050 & 0.050 & 0.029 & 0.039 & 0.050 & 0.064 \\
  29.6 & 0.0050 & 0.065 & 0.200 & 0.030 & 0.032 & 0.044 \\
  39.7 & 0.0050 & 0.088 & 0.206 & 0.048 & 0.045 & 0.066 \\
  52.0 & 0.0050 & 0.115 & 0.101 & 0.051 & 0.042 & 0.066 \\
  73.5 & 0.0050 & 0.163 & 0.251 & 0.042 & 0.072 & 0.083 \\
 103.9 & 0.0050 & 0.230 & 0.054 & 0.095 & 0.089 & 0.130 \\

\hline
\end{tabular}
\caption{\label{tabderq2}
  \sl  The H1 measurement of the derivative \pdff.
 $\Delta_{sta}$ 
denotes the uncertainty due to the data statistics.
  The uncertainties are given in absolute values.}
\end{table}
%
\begin{table}[h] \centering 
\begin{tabular}{|c|r|r|c|r|r|r|}
\hline
$Q^2/GeV^2$ & $x$ & $y$ & $ (\partial F_2 / \partial \ln Q^2)_x $ & $\Delta_{sta}$ & $\Delta_{sys}$ & $\Delta_{tot}$    \\
\hline
  17.3 & 0.008 & 0.024 & 0.138 & 0.021 & 0.025 & 0.033 \\
  22.4 & 0.008 & 0.031 & 0.074 & 0.034 & 0.056 & 0.065 \\
  29.6 & 0.008 & 0.041 & 0.166 & 0.028 & 0.029 & 0.040 \\
  39.7 & 0.008 & 0.055 & 0.098 & 0.044 & 0.039 & 0.059 \\
  52.0 & 0.008 & 0.072 & 0.209 & 0.047 & 0.039 & 0.061 \\
  73.5 & 0.008 & 0.102 & 0.105 & 0.041 & 0.037 & 0.055 \\
 103.9 & 0.008 & 0.144 &-0.017 & 0.072 & 0.098 & 0.122 \\
  29.6 & 0.013 & 0.025 & 0.148 & 0.026 & 0.027 & 0.037 \\
  39.7 & 0.013 & 0.034 & 0.095 & 0.044 & 0.046 & 0.063 \\
  52.0 & 0.013 & 0.044 & 0.117 & 0.045 & 0.036 & 0.057 \\
  73.5 & 0.013 & 0.063 & 0.113 & 0.038 & 0.037 & 0.053 \\
 103.9 & 0.013 & 0.088 & 0.058 & 0.068 & 0.089 & 0.112 \\
  39.7 & 0.020 & 0.022 & 0.014 & 0.037 & 0.033 & 0.050 \\
  52.0 & 0.020 & 0.029 & 0.136 & 0.043 & 0.042 & 0.060 \\
  73.5 & 0.020 & 0.041 & 0.053 & 0.037 & 0.030 & 0.048 \\
 103.9 & 0.020 & 0.058 &-0.047 & 0.062 & 0.057 & 0.084 \\
 134.2 & 0.020 & 0.074 & 0.463 & 0.153 & 0.221 & 0.268 \\
  73.5 & 0.032 & 0.025 & 0.008 & 0.035 & 0.045 & 0.057 \\
 103.9 & 0.032 & 0.036 & 0.074 & 0.065 & 0.070 & 0.096 \\
 134.2 & 0.032 & 0.046 &-0.038 & 0.141 & 0.143 & 0.201 \\
 134.2 & 0.050 & 0.030 &-0.205 & 0.131 & 0.116 & 0.175 \\
\hline
\end{tabular}
\caption{\label{tabderq3}
  \sl 
The H1 measurement of the derivative $\partial F_2 /
  \partial \ln Q^2$.
$\Delta_{sta}$ denotes the uncertainty due to the data statistics.
The uncertainties are given in absolute values.}
\end{table}
\end{document}